\documentclass[rmp,aps,twocolumn,groupedaddress]{revtex4-1}
\usepackage{bm,amsmath}
\usepackage{bbm}
\usepackage{hyperref} 
\usepackage{color}
\usepackage{amssymb}
\hypersetup{
    colorlinks,
    citecolor=blue,
    filecolor=blue,
    linkcolor=blue,
    urlcolor=blue,          
    pdfborder = {0 0 0},        
    linktocpage
}
\usepackage{graphics}
\usepackage[dvips]{graphicx}
\usepackage{float,graphicx}
\usepackage{epstopdf}
\newcommand{\unit}{1\!\!1}
\newcommand{\beq}{\begin{equation}}
\newcommand{\eeq}{\end{equation}}
\newcommand{\beqa}{\begin{eqnarray}}
\newcommand{\eeqa}{\end{eqnarray}}

\def\be{\begin{equation}}

\def\ee{\end{equation}}

\def\ss{{\bf s}}

\def\eeff{\scriptsize\textmd{eff}}

\def\mmin{\scriptsize\textmd{min}}
\def\mmax{\scriptsize\textmd{max}}

\def\Hhc{\textmd{H.c.}}

\def\ddet{\textmd{det}}

\def\iint{\scriptsize\textmd{int}}
\def\ssign{\textmd{sign}}

\def\ddos{\scriptsize\textmd{DOS}}
\def\Ttr{\textmd{Tr}}
\def\ttot{\scriptsize\textmd{tot}}
\def\ddiss{\scriptsize\textmd{diss}}
\def\pper{\scriptsize\textmd{pe}}

\def\eel{\textmd{el}}

\def\ccor{\scriptsize\textmd{corr}}
\def\eeff{\scriptsize\textmd{eff}}

\def\Rre{\textmd{Re}}
\def\Iim{\textmd{Im}}
\def\aad{\scriptsize\textmd{ad}}

\def\ddet{\textmd{det}}
\def\eeq{\scriptsize\textmd{eq}}

\def\sst{\scriptsize\textmd{st}}
\def\Ttr{\textmd{Tr}}

\def\be{\begin{equation}}
\def\ee{\end{equation}}
\def\bea{\begin{eqnarray}}
\def\eea{\end{eqnarray}}

\begin{document}

\title[Dynamics of non-Markovian open quantum systems]{Dynamics of non-Markovian open quantum systems}
\author{In\'es de Vega}
\affiliation{Department of Physics and Arnold Sommerfeld Center for Theoretical Physics, Ludwig-Maximilians-Universit{\"a}t M{\"u}nchen, Theresienstr. 37, 80333 Munich, Germany}
\author{Daniel Alonso}
\affiliation{Instituto de Estudios Avanzados (IUdEA) and Departamento de F\'{\i}sica, Universidad de La Laguna,
La Laguna 38204, Tenerife, Spain}


\begin{abstract}

Open quantum systems (OQSs) cannot always be described with the Markov approximation, which requires a large separation of system and environment time scales. Here, we give an overview of some of the most important techniques available to tackle the dynamics of an OQS beyond the Markov approximation. Some of these techniques, such as master equations, Heisenberg equations and stochastic methods, are based on solving the reduced OQS dynamics, while others, such as path integral Monte Carlo or chain mapping approaches, are based on solving the dynamics of the full system. We emphasize the physical interpretation and derivation of the various approaches, explore how they are connected and examine how different methods may be suitable for solving different problems. 
\end{abstract}
\maketitle

\tableofcontents

\section{Introduction}

In most realistic situations, a quantum system shall be considered as an $open$ quantum system (OQS), coupled to an environment that induces decoherence and dissipation. 
The dynamics of an OQS can be described, in many cases, with a Markov approximation, which assumes that the environment recovers instantly from the interaction, leading to a continuous flow of information from the system to the environment. 

However, our increasing capability to fabricate new materials and to observe and control quantum systems at different times, length scales and energy ranges is constantly revealing new scenarios where dissipation and decoherence play a fundamental role. In many of these scenarios, a large separation between system and environment time scales can no longer be assumed, leading to non-Markovian behavior and eventually a back-flow of information from the environment into the system. It is therefore crucial to develop an accurate but efficient description of the system-environment interaction that goes beyond the Markov approximation. 


The main goal of the theory of OQSs is to avoid having to integrate the full system, comprising both the OQS itself and its environment, by describing the dynamics of the open system in its reduced Hilbert space. As it will be discussed in Sec. \ref{concepts}, the structure of the system-environment initial state is fundamental to determine the evolution for the reduced density matrix of the OQS, $\rho_s(t)$, defined by tracing out the environment degrees of freedom from the full system density matrix.
To compute such evolution, many different {\bf master equations} have been proposed. In particular, within the Markov approximation, master equations can often be arranged in the well-known Lindblad form \cite{kossakowski1972, lindblad1976, gorini1976}, which preserves complete positivity of the OQS dynamics. This equation is sometimes referred to as the Lindblad-Kossakowski equation. However, as discussed in Sec. \ref{ME}, master equations beyond the Markov approximation have also been derived by considering different approximations and methods. 

An alternative to master equations is to consider {\bf stochastic Schr{\"o}dinger equations (SSE)} \cite{zoller1997,diosi1998,gaspard1999b,stockburger2002,alonso2005,piilo2008}, discussed in Sec. \ref{SSE}. SSE enable the calculation of all the dynamical quantities of a non-Markovian OQS by evolving a state vector within its reduced Hilbert space. This state vector may depend on one or two noises whose statistical properties encode the relevant environmental information influencing the state vector dynamics. The reduced density matrix or the multiple-time correlations of the system observables can then be obtained as a Monte Carlo average over an ensemble of projectors of such stochastic trajectories. 
The closely related {\bf path-integral and quantum Monte Carlo methods} conform a broad and active area of research that we do not intend to cover exhaustively in this review. The interested reader can go for instance to the excellent reviews by \textcite{pollet2012} and \textcite{gull2011} that discuss quantum Monte Carlo applications in the fields of ultra-cold gases and quantum impurity models respectively. Besides that, the path integral representation is also the basis of different analytical derivations and approximations, that lead to Heisenberg, stochastic, and master equations similar to the ones covered in this review. Three of the most important approaches of this type are discussed in Sec. \ref{path}, namely the noninteracting blip approximation, the stochastic Liouville von-Neumann equation, and the hierarchical equations of motion.

As discussed in Sec. \ref{HR}, the {\bf Heisenberg representation}, standard for describing the evolution of quantum operators, can also be extended to tackle OQS dynamics, as already shown by \textcite{ackerhalt1973,wodkiewicz1976,kimble1977}. It allows one to introduce the well-known input-output formalism, first derived by \textcite{yurke1984,gardiner1985,barchielli1986,barchielli1987} (see also \cite{quantumnoise}) under the Markov approximation. As discussed, the input-output formalism was recently extended to non-Markovian systems by \textcite{diosi2012} in the context of stochastic Schr\"odinger equations, and  by \textcite{zhang2012} in the context of non-Markovian cascaded networks. 
Furthermore, when no approximation is considered, the multiple-time correlations of OQS observables follow a hierarchical structure when no approximation is considered: quantum mean values depend on two-time correlations, and in general $N$ time correlations depend on N+$1$ correlations. To truncate such a hierarchy, either a Markov, a semiclassical or a weak coupling approximation has to be assumed. 
The Heisenberg approach is particularly advantageous for many-body OQSs, where the dimension, $d$, of the systems Hilbert space grows exponentially with the number of particles. The reason is that it allows for an effective reduction of the problem dimension, by considering a semiclassical truncation of correlations involving multiple-particle operators. 

To describe OQSs, a second possibility is to integrate the degrees of freedom of the total system. This is a difficult task, due to the large number of degrees of freedom of the environment. In this regard, a judicious selection of the relevant states of the full system is of primary importance; for instance, in the context of electron-phonon interaction, this can be done by discarding states with low probability, as in the density matrix approach \cite{zhang1998b}, or by considering as relevant only those states generated during the evolution, as in the variational approach \cite{bonca1999,vidmar2010} (see also \citet{computationalmany} for a review on exact diagonalization methods). Another alternative is to map the original problem of a system coupled to a set of environment harmonic oscillators into a one-dimensional structure, where the system is coupled to a chain of transformed oscillators \cite{bulla2008,prior2010}. Either in the original star configuration, or in the chain form, the system can be solved with numerical renormalization group (NRG) \cite{bulla2008}, or with time-dependent density matrix renormalization group (t-DMRG) or matrix product states techniques \cite{white1992,white1998,scholl2011}. Some of these ideas are briefly discussed in Sec. \ref{unitary}.

The first two sections, which discuss models and scales of the problem (Sec. \ref{characterization}) as well as the main concepts of the theory of OQS (Sec. \ref{concepts}), are meant to give an overview of the subject. In contrast, Sec. \ref{ME}, \ref{SSE}, \ref{path}, and \ref{HR}, discuss different methods for solving the dynamics of OQS that are to some degree independent from each others, and therefore can be read independently. Also, while most of the derivations of master equations, SSE and Heisenberg equations rely on a perturbative expansion, the path integral related derivations discussed in Sec. \ref{path}, do not rely on such type of expansions, and therefore in principle they do not share this limitation. Finally, Sec. \ref{Exact} discusses some exactly solvable models. 

Before ending the introduction, we clarify our working use of the wording \textit{non-Markovian}. Here, we refer as \textit{Markovian} those derivations that are based on assuming a vanishing environment correlation time, \textit{i.e.} a Markov approximation (discussed in Secs. \ref{ch1sec7} and \ref{bornmarkov}). Similarly, we denominate as \textit{non-Markovian} those derivations that are not based on using the Markov approximation, and thus are in principle able to capture the non-Markovian behavior that could occur in some parameter regimes. Importantly, a different question is whether the resulting dynamics is indeed Markovian or non-Markovian according to the measures described in Sec. \ref{nonmarkovianity}. In this regard, an equation can lead to Markovian dynamics, even if it is not obtained through a Markov approximation. An example of this is discussed in Sec. \ref{canonical}.

In the reminder of this section, we discuss some of the most relevant situations where a non-Markovian OQS theory that goes beyond the Markov approximation becomes necessary.

\subsection{Non-Markovian effects in different scenarios}

Non-Markovian effects are present in many different contexts, ranging from solid state physics to hybrid systems, quantum biology and quantum optics, as discussed further. 

\subsubsection{Solid state and quantum information: Superconducting flux-qubits and quantum control}

{\bf Solid state physics} is a broad arena where OQSs exhibiting non-Markovian effects may appear \cite{weissbook}. 

As derived by \textcite{feynman1963a}, when the system is weakly coupled to its environment, the coupling can be considered to be linear and the environment described by a set of harmonic oscillators. In this context, one of the best-known models is the one developed by \textcite{caldeira1983,weissbook}, which describes a harmonic oscillator linearly coupled through its displacement coordinate $q$ to a fluctuating dynamical reservoir, which may represent, for instance, the phonons of a lattice. This model will be analyzed in more detail in Sec. \ref{ch1sec3}. 

A Brownian motion type of system exhibiting non-Markovian effects may also arise in the dynamics of a Bose-Einstein condensate (BEC) in a trap, which is coupled to a final atomic state outside of the trap. 
The dynamics of the occupation number of the BEC exhibits oscillations that can be interpreted as a quantum interference effect, and clearly displays non-Markovian behavior and strong departures from the golden rule that predicts exponential decay \cite{hope1997,hope2000,breuer2001b}. This behavior can also be found when a quantum dot is coupled to a superfluid reservoir via laser transitions \cite{recati2005,jaksch2005}, when a quantum dot is coupled to a BEC in a double-well potential \cite{sokolovski20091}, or when atoms trapped in an optical lattice are coherently coupled to an untrapped level, giving rise to a highly non-Markovian dissipation \cite{devega2008,navarrete2010}. 

A similar system, recently proposed and experimentally realized by \textcite{reichel2001,treutlein2007,hunger2010} consists in  a nanomechanical oscillator interacting with a BEC in a double-well potential. The atoms of the condensate are confined in a double well and can tunnel from one side of the potential to another, depending on the position of the oscillator. As shown by \textcite{brouard20111,alonso20141}, if one considers the condensate as an environment for the oscillator, highly non-Markovian effects appear that can be observed in the nonexponential decay of the oscillator coherences. 

\begin{figure}[t]
\centerline{\includegraphics[width=0.45\textwidth]{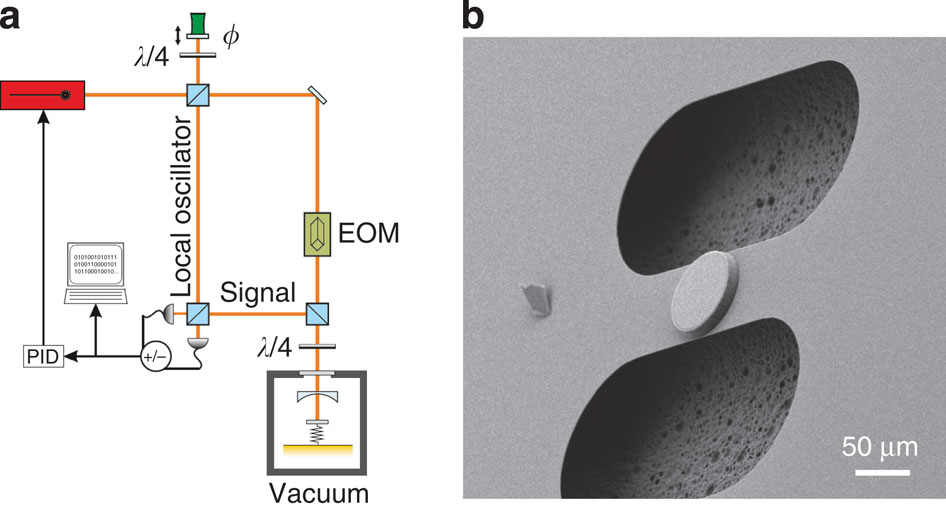}}
\caption{(a) The experimental set-up consists of a laser, which is split into a signal beam and a local oscillator (LO). The signal beam acquires a phase from the motion of the mechanical resonator, which is detected on two photodiodes after a previous beating of the signal with a strong LO. 
(b) Scanning electron microscope picture of the tested device. From \textcite{groblacher2015}. \label{eisert}}
\end{figure}

Quantum Brownian motion can also be observed in an opto-mechanical resonator coupled to a heat bath. A recent experiment by \textcite{groblacher2015} showed that the spectral density of such an environment is highly non-ohmic, which produces non-Markovian dynamics in the resonator. The spectral density is characterized by monitoring the mechanical motion of the resonator with a high degree of sensitivity, which is achieved by weakly coupling the mechanics to an optical cavity field whose phase response encodes the mechanical motion (see Fig. \ref{eisert}).

A different system where a Markov approximation may not be suitable is an OQS coupled to a fermionic oscillator environment. An example of this is a noninteracting fermion coupled to a fermionic bath, analyzed by \textcite{schoen1990}. A more complex situation is the one described by the Anderson impurity model \cite{anderson1961}. It describes clusters of interacting electron impurities, coupled to a continuous conduction band of noninteractingelectrons. The Anderson model is the basis for dynamical mean-field theory \cite{metzner1989,georges1992,georges1996},  
which is the most widely used numerical method to describe strongly correlated systems 
in higher than one dimensions \cite{kotliar2006,maier2005} and is popular also in quantum chemistry \cite{zgid2010}.
This model can also be used to describe electron transport in quantum dots interacting with electron leads. It has been analyzed using different approaches within the theory of OQSs, including rate equations \cite{gurvitz1996}, master and Fokker-Planck equations \cite{li2005,pedersen2005,harbola2006,timm2008,ghosh2012,buesser2014}, stochastic Schr{\"o}dinger equations \cite{zhao2012}, or path integrals \cite{tu2008} [see also \cite{brandes2005} for a review]. 
A similar situation is the one described by the Hubbard-Holstein model \cite{holstein1959,hubbard1964}, which describes the electron-phonon interaction.  This model can be conveniently described with the theory of OQSs, when considering that the phonons are characterized by a narrow energy band in comparison to the electronic band. This justifies treating the electrons as an environment, which evolve in a much faster time scale than the phonons. 

In other condensed-matter systems the dynamics of a quantum system can be dominated by its coupling with surrounding defects, impurity spins or nuclear spins, that effectively lead to a spin environment \cite{prokofev1993,stamp1994,prokofev2000,saykin2002}. The coupling of a system to a spin environment does not scale with the number of environment particles as $1/\sqrt{N}$, as occurs with oscillator environments, but rather is independent of $N$. One of the most well-known examples of these systems is the central spin problem \cite{prokofev2000,breuer2004,breuerbook}, where the OQS itself is considered a spin particle. An example of the central spin model is an electron in a quantum dot coupled due to a hyperfine interaction with the surrounding nuclear spins \cite{khaetskii2002,merkulov2002,schliemann2003,coish2004,schuetz2012,kessler2012}. Another example is an electron spin of a single nitrogen-vacancy (NV) center coupled to the spin environment of substitutional nitrogen defects known as P1 centers \cite{hanson2008}. The central spin model also appears in the context of quantum computation, when analyzing the decoherence of a qubit, such as a superconducting qubit, produced by the coupling with other qubits. 
Other important decoherence sources for each different type of superconducting qubits (charge, flux and phase) were recently discussed by \textcite{xiang2013}. An excellent review on OQS in mesoscopic systems and devices can be found in \cite{rotter2015}.

The dynamics of an OQS can be represented as a quantum channel mapping an initial state to a final state. This representation facilitates the use of quantum information theory to analyze these systems and to explore the effects and possible advantages of non-Markovian dynamics in the quantum channel capacity \cite{maniscalco2007,bylicka2014}, and in preserving quantum memory \cite{man2015,rosario2015,man2015b,hinarejos2016} [see a complete review of this subject by \textcite{caruso2014}]. In addition, it was  pointed out by \textcite{alicki2002} that fault-tolerant quantum computation theory may not be applicable when the environment of the quantum computer has a long correlation time. Nevertheless, a threshold analysis for some non-Markovian error models was performed by \textcite{terhal2005}. Similarly, the quantum optimal control theory, which provides a framework for variationally calculating the optimal choice of shape and parameters of a succession of pulses to control a quantum system, has been extended to deal with systems that are additionally coupled to a non-Markovian environment \cite{rebentrost2009c,hwang2012}. This is of great importance when it comes to controlling decoherence during the operation of a set of gates performing quantum computational tasks. An excellent review on the subject can be found in \cite{koch2016}.

\subsubsection{Quantum biology and chemical physics}

In photosynthetic complexes, the transport of energy between pigments is affected by a phononic environment produced by surrounding vibrating proteins \cite{blankenship2011}. Recent experiments showed the existence of long-lasting interexciton coherences in several types of photosynthetic complexes even at physiological temperatures \cite{engel2007,collini2010,panitcha2010}. Because of these relatively long-lasting coherences, pigments involved in this energy transport should be considered in principle as quantum systems \cite{plenio2008,rebentrost2009,caruso2009,mohseni2013} coupled to the surrounding phononic environment. In addition, in a typical situation, the relaxation time of this environment can be comparable to or even slower than the electronic energy transfer dynamics within the pigment complex, meaning that a Markov approximation is therefore no longer accurate \cite{ishizaki2009b,chin2010}. The dynamics of these systems has been studied beyond the Markov approximation, by considering the full system dynamics (see Sec. \ref{unitary}), or by calculating the reduced density operator with a hierarchy approach (see Sec. \ref{hierarchy}). See also \cite{lambert2013} for an excellent review on quantum biology. 

Although not covered in this review, note that in the context of molecular physics the problem of a quantum system interacting with an environment has a long tradition and there has been an intense research in dealing with non-Markovian effects. Early examples can be found for instance in \cite{mukamel1979non,breton19841,breton19842}.

\subsubsection{Quantum optics: Photonic band gap materials}

Atomic emission is affected by the photonic density of states (DOS) of the radiation field, a quantity that depends critically on whether the field is in free space or within a quantum cavity, waveguide, or nanostructured material like photonic crystals (PC) or metamaterials. 
The importance of the medium in the atomic emission was first pointed out by \textcite{purcell1946}. According to this result, the spontaneous emission rate of an atom in a quantum cavity is enhanced by a factor of $Q$ with respect to that of the vacuum if the atomic transition is in resonance with the cavity. In the same way, if atomic transitions are far from any cavity resonance, the spontaneous emission process will be inhibited. The same kind of inhibition of spontaneous emission occurs if atoms are located in a waveguide and their transition frequency is below the waveguide€™'s fundamental frequency \cite{kleppner1981,barut1987}. 

\begin{figure}[t]
\centerline{\includegraphics[width=0.45\textwidth]{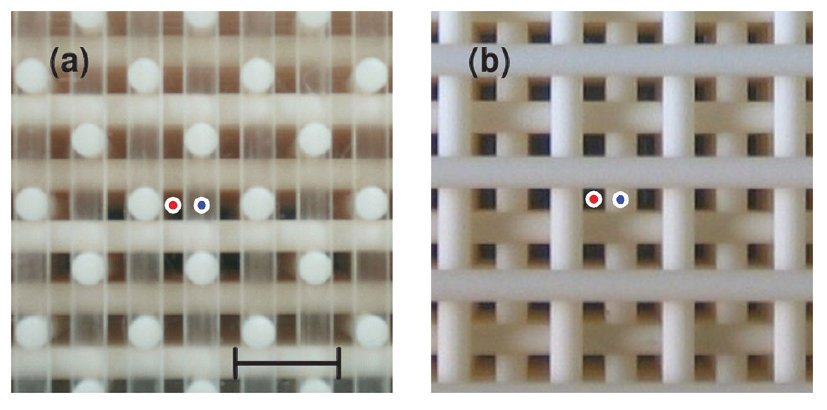}}
\caption{Side view (a) and top-view (b) photographs of the photonic crystal (scale bar $10$nm). The emission dynamics of an emitter was measured considering two different positions:  position 1 close to a dielectric rod (left, red dot), and position 2, at a maximum distance from any dielectric rod (right, blue dot). From \textcite{hoeppe2012}. \label{dens}}
\end{figure}

Atoms or impurities coupled to the modified radiation field within a photonic crystal also exhibit strong deviations from their behavior in the vacuum. Photonic crystals, which were first envisioned by \textcite{yablonovitch1987,john1987}, are periodic optical nanostructures that strongly modify the properties of the electromagnetic field (EM), affecting the photons in a similar way as ionic lattices affect the motion of electrons in solids. The radiation field in this material presents a gap or frequency range where the photonic DOS vanishes, and no propagating photons are allowed. Atoms or impurities coupled to such a modified radiation field exhibit strong non-Markovian effects, like nonexponential decay, or the formation of a photon-atom bound state when the atomic frequency is within the gap \cite{john1994,florescu2001}. In addition, the superradiant emission of a collection of atoms in PC is strongly modified with respect to such emission in the vacuum \cite{john1995}. Non-Markovian effects are also present in impurities coupled to PC-nanocavities exhibiting an ultrahigh quality factor \cite{tanaka2007}, or to waveguides. The non-Markovian character of the emission of a ferromagnetic sphere in a static magnetic field in a PC, which behaves like a single atomic emitter, was recently experimentally observed in \cite{hoeppe2012} (see Fig. \ref{dens}). Further, experimental progress in the control of spontaneous emission by manipulating optical cavity modes and quantum dots within photonic crystals has demonstrated that the spontaneous emission from light emitters embedded in photonic crystals is not only suppressed within the gap, but also enhanced in the direction where optical modes exist \cite{englund2005,noda2007,thompson2013}. Also, recent proposals \cite{hung2013,goban2013,tudela2014} explore the atom-atom interactions that may be produced in these materials mediated by a strong light-matter interaction. 

The analysis of such phenomena requires the use of tools that go well beyond the Markov approximation, in order to capture the relevant aspects of the processes. Among these are the weak coupling Heisenberg equations in Sec. \ref{perturbativeH} \cite{florescu2001}, or the SSE in Sec. \ref{SSE} \cite{devega2005}. In addition, the exact spontaneous emission of an atom within a photonic crystal was studied by \textcite{john1994,bay1997} following variants of the exact method in Sec. \ref{Exact1}, \textit{\textit{i.e.}} within the single photon sector. This method was extended in \cite{nikolopoulos1999,nikolopoulos2000} for two photons. Reviews of these and related results can be found in \cite{lambropoulos2000,woldeyohannes2003}.

\section{Characterization of the problem}
\label{characterization}

At low energies, OQSs can often be described using a few \textit{canonical models}, where a simple central system (the OQS) is linearly coupled to an environment which belongs to one of two different universal classes: spin environments, composed of a set of independent spins, or harmonic oscillator environments composed of a set of either fermionic or bosonic independent harmonic oscillators. See also the discussion by \textcite{prokofev2000}. 

In this review we focus mainly on OQSs coupled to a harmonic oscillator environment.  In general, a complex environment can be mapped into an effective harmonic environment following the linear response theory. This approximation is often deemed to correctly capture qualitative behavior in many relevant situations. Particularly, it describes light-matter interaction at low energies (see Sec. \ref{lightmatter}), and as argued by \textcite{feynman1963a}, describes several models in condensed-matter physics corresponding to a central system weakly coupled to its environment \cite{wallsbook}. In general, it is understood to be valid when aiming at extracting the dynamics of the open quantum systems only, and provided that the environment remains with an approximately Gaussian behavior \cite{makri1999a,makri1999b}. We note also that many of the techniques presented here can be (and in a few cases have been) extended to deal with spin environments. 

In the following section we present the most general Hamiltonian describing a linear interaction between the OQS and its environment. Following this, we introduce the Caldeira-Leggett model, after which we show that the light-matter interaction Hamiltonian leads to a similar model. In addition, we provide a qualitative analysis of the characteristics of the system-environment interaction, offering physical insight into the time scales involved in the problem as well as an overview of the different approximations and strategies available to tackle it. In doing so, we introduce the concept of non-Markovianity from a phenomenological point of view; a more quantitative analysis is provided in the next section and also reviewed by \textcite{breuer2012,rivas2014,breuer2015}.

\subsection{General interaction Hamiltonian}
\label{generalization}
The Hamiltonian of an OQS coupled to an environment can always be written as the composition of two terms:
\begin{eqnarray}
H_{\ttot}=H_0+H_I  
\label{chapuno1}
\end{eqnarray}
where $H_0= H_S +H_B$ is the free Hamiltonian, consisting of a sum of the system and the environment Hamiltonians, and $H_I$ is the interaction Hamiltonian that describes the coupling between the OQS and the environment. A general coupling Hamiltonian $H_I$ can be written as a sum of many couplings between a set of $M$ environment $\{B_\eta \}$ and system operators $\{S_\eta \}$ \cite{gaspard1999b}\footnote{Except for cases where a more precise notation is needed for  clarity, we denote the external product of system $S$ and environment operators $B$, $S\otimes B$, simply as $SB$.},
\begin{eqnarray}
H_I =\sum^{2M}_{\eta=1}B_\eta S_\eta,
\label{chapuno12}
\end{eqnarray}
with $B_\eta=B^\dagger_\eta$, $S_\eta=S^\dagger_\eta$. The Hamiltonian (\ref{chapuno12}) is in fact the most general form of interaction Hamiltonian. Nevertheless, our analysis is restricted to the case where $B_\eta$ are linear combinations of the creation and annihilation operators, which is why we refer to the interaction Hamiltonian (\ref{chapuno12}) as linear. In addition, any Hamiltonian of the type $H_I=\sum^M_{\eta=1}(X^\dagger_\eta Y_\eta+X_\eta Y_\eta^\dagger)$, with $X_\eta$ and $Y_\eta$ system and environment operators respectively, can be written as Eq. (\ref{chapuno12}) and vice-versa. This is done by just considering that any operator can be decomposed as $X_\eta=X_\eta^{(a)}+iX_\eta^{(b)}$, in terms of the Hermitian operators $X_\eta^{(a)}=(X^\dagger_\eta+X_\eta)/2$, and $X_\eta^{(b)}=i(X_\eta-X^\dagger_\eta)/2$. Considering a similar decomposition for $Y_\eta$, we find that $H_I=\sum^M_{\eta=1}(X^\dagger_\eta Y_\eta+X_\eta Y_\eta^\dagger)$ is equal to Eq. (\ref{chapuno12}) with $S_\eta=2X^{(a)}_\eta$, and $B_\eta=2Y^{(a)}_\eta$ for $\eta=1,\cdots, M$, and $S_\eta=2X^{(b)}_{\eta-M}$, and $B_\eta=2Y^{(b)}_{\eta-M}$ for $\eta=M+1,\cdots, 2M$ \cite{rivas2011a}. For instance, if we replace in Eq.(\ref{chapuno12}), 
\begin{equation}
S_1=L+L^{\dagger} \,;\, S_2=i(L-L^{\dagger}),
\label{chapuno30}
\end{equation}
with $L$ as a linear operator of the OQS, and 
\bea
B_1=\frac{1}{2}\sum_\lambda g_\lambda (a_\lambda +a_\lambda^{\dagger}) \,;\ B_2=\frac{i}{2}\sum_\lambda g_\lambda (a_\lambda-a_\lambda^{\dagger}),
\label{chapuno31}
\eea
we arrive at a form for the Hamiltonian in terms of the operators $L$ and $L^\dagger$,
\begin{eqnarray}
H_{tot}=H_S +H_B +\sum_\lambda g_\lambda (L a_\lambda^{\dagger}+L^{\dagger}a_\lambda).
\label{chapuno32}
\end{eqnarray}
As we seen in the following sections, for a harmonic environment, $H_B$ is quadratic in the environment modes, which makes $H_S$ and $L$ crucial in determining whether the dynamics is exactly solvable or not. For instance, if one assumes that the OQS is a harmonic oscillator with annihilation operator $b$, and $L=b$, the full Hamiltonian (\ref{chapuno32}) is quadratic and the system is exactly solvable as a Brownian particle (see Sec. \ref{brownian}). If $H_S$ is not harmonic, containing, for instance, an interaction term of the form $\sim U n^2$, with $n=b^\dagger b$, then the problem is in general no longer exactly solvable, and the only way to tackle it is either by assuming some approximations, or by numerically solving the whole system and environment dynamics (see Sec. \ref{unitary}). 

\subsection{Caldeira-Leggett model}
\label{ch1sec3}
We consider the general Hamiltonian of a system with $1$ or a few degrees of freedom coupled to an environment of harmonic oscillators as described by \textcite{caldeira1983,caldeira1983b,leggett1987,weissbook}. The system Hamiltonian is written as Eq. (\ref{chapuno1}) with the full Hamiltonian having the form\footnote{In this review, except in the path integral method section, we settle natural units with $\hbar=1$.}
\begin{eqnarray}
H_S =\frac{p^2 }{2M}+V(q),
\label{chapuno2}
\end{eqnarray}
where $q$ and $p$ are, respectively, the system position and momentum coordinates of the particle ($[q,p]=i$), and $M$ is its mass. The Hamiltonian of the environment is
\begin{eqnarray}
H_B =\sum_{\lambda} \frac{1}{2} \left[\frac{p^2 _{\lambda}}{m_{\lambda}}+m_{\lambda}\omega^2_{\lambda}x^2_{\lambda}\right],
\label{chapuno3}
\end{eqnarray}
where $p_\lambda$ and $x_\lambda$ are the momentum and position coordinates operators of the $\lambda$ harmonic oscillator. The interaction of the system with each mode of the reservoir is inversely proportional to the volume of the reservoir, so that for a spatially large environment this coupling is small. Therefore, it is a good approximation for macroscopic environments to consider that the system-reservoir coupling is a linear function of the environment coordinates, giving the interaction Hamiltonian the form
\begin{eqnarray}
H_I =-\sum_\lambda S_{\lambda}(q)x_{\lambda}+\Delta V(q).
\label{chapuno4}
\end{eqnarray}
Here, a counter term has been added to renormalize the potential $V(q)$. Indeed, in the presence of the interaction, the minima of the potential for a given $q$ are displaced by a certain quantity, in such a way that the effective potential in Eq. (\ref{chapuno2}) should be written as $V_{\eeff}(q)=V(q)-\Delta' V(q)$. The renormalization consists of choosing in Eq. (\ref{chapuno4}) $\Delta V(q)=\Delta' V(q)$, so that the minima of the potential are placed at zero. 
For the special case of a separable interaction \cite{weissbook},
\begin{equation}
S_\lambda (q)=C_\lambda S(q).
\label{chapuno5}
\end{equation}
In the simplest case in which $S(q)=q$, the total Hamiltonian can be written as
\begin{eqnarray}
H_{tot}=H_S+
\frac{1}{2}\sum_\lambda \left[\frac{p^2_\lambda}{m_\lambda}+m_\lambda \omega^2_\lambda \left(x_\lambda -\frac{qC_\lambda }{m_\lambda \omega^2_\lambda }\right)^2 \right],
\label{chapuno6}
\end{eqnarray}
where the renormalization factor is identified as 
\begin{eqnarray}
\Delta V(q)=\sum_\lambda \frac{C^2 _\lambda}{m_\lambda \omega^2_\lambda}q^2.
\label{chapuno7}
\end{eqnarray}
Replacing Eq. (\ref{chapuno5}) in Eq. (\ref{chapuno4}), the interaction term of the Hamiltonian (\ref{chapuno6}), without the renormalization term, is a simplified version of the general Hamiltonian (\ref{chapuno12})
\begin{eqnarray}
H_I =BS,
\label{chapuno7b}
\end{eqnarray}
with $B=-\sum_\lambda C_{\lambda} x_{\lambda}$ and $S=S(q)$.
The Hamiltonian (\ref{chapuno6}) has been widely used to describe dissipation in OQSs and is often referred to in the literature as the Caldeira-Leggett model \cite{weissbook,leggett1987}. 

\subsection{The spin-boson model}
\label{SB}
In many physical and chemical systems, the generalized coordinate $q$ is associated with an effective potential with two separate minima placed at the same energy. Since only these two states are available, the Hilbert space of the system is reduced to a two-dimensional space. This situation, described with the well-known spin-boson model, occurs, for instance, in the motion of light particles in metals, in certain chemical reactions involving electron transfer processes (see for instance the review by \textcite{leggett1987}), or in a  superconducting qubit, which can be coupled to propagating photons within an open transmission line as described by \textcite{peropadre2013}. In addition, OQSs such as vibrating molecules can be represented as an anharmonic oscillator, having an energy spectrum which is no longer infinite and evenly spaced as in the harmonic case. Then, provided that the interaction strength producing the anharmonicity, $U$, is sufficiently large, such spectrum can be truncated at the lowest few energy levels. If the truncation is at the first two levels, then the resulting OQS is also a truncated two-level system. The systems described above are often referred to as \textit{truncated} two-state systems \cite{leggett1987}, as opposed to \textit{intrinsic} two-state systems, such as a nucleus of spin $1/2$, or a photon with two polarization states. The spin-boson model can also be considered to describe dissipative energy transfer in a pair of two-level systems (each of them representing a molecule, for instance) within the one-excitation sector \cite{gilmore2005,nazir2009}. 

The spin-boson Hamiltonian can be written as (\ref{chapuno1}), with $H_0=\frac{1}{2} \omega_{12} \sigma_z-\frac{1}{2}\Delta_0 \sigma_x+\sum_{\lambda}\omega_\lambda a^{\dagger}_\lambda a_\lambda$,
where $\sigma_\alpha$ ($\alpha=x,y,z$) are the standard Pauli matrices for a two-level system, $\omega_{12}$ is the energy separation between the two states, and $\Delta_0$ is the coupling energy, representing the tunneling between them. Also, the interaction term has the form (\ref{chapuno7b}), $H_I=\sigma_z\sum_{\lambda}C_\lambda x_\lambda$, such that 
\begin{eqnarray}
H_{\ttot}&=&\frac{1}{2} \omega_{12} \sigma_z-\frac{1}{2}\Delta_0 \sigma_x +\sum_{\lambda}  \omega_\lambda a^{\dagger}_\lambda a_\lambda\cr
&+&\sum^N_{\lambda=1} g_\lambda (a_\lambda +a^\dagger_\lambda)\sigma_z,
\label{chapuno11}
\end{eqnarray}
where we have explicitly written the environment operators in terms of creation and annihilation operators, 
\bea 
x_\lambda =\sqrt{\frac{1}{2m_\lambda \omega_\lambda}}(a_\lambda+a^{\dagger}_\lambda),\,\, p_\lambda =-i\sqrt{\frac{ m_\lambda \omega_\lambda}{2}}(a_\lambda -a^{\dagger}_\lambda), \nonumber
\eea
and the coupling parameter is 
\bea
g_\lambda =\sqrt{\frac{1}{2 m_\lambda \omega_\lambda}}C_\lambda.\nonumber
\eea 
Note that alternatively the spin-boson model can be expressed as 
\bea
H_{\ttot}=\frac{1}{2} \omega_{12} \sigma_x-\frac{1}{2}\Delta_0 \sigma_z +H_B +\sigma_x\sum_{\lambda}C_\lambda x_\lambda,\nonumber
\eea
by simply performing a unitary rotation of the previous Hamiltonian. 

In general, the dynamics and ground state properties of the spin-boson model are both extremely rich, and have been a continuous object of study during the past decades. 
Regarding the dynamics, the main topic of this review, for a weak coupling between the system and the environment, the evolution of the system can be computed with the master equation discussed in Sec. \ref{perturbativeM}, the SSE covered in Sec. \ref{expansion}, and the Heisenberg approach explained in Sec. \ref{perturbativeH}. In addition, as discussed in Sec. \ref{polaron}, the Hamiltonian (\ref{chapuno11}) can be unitarily transformed with a polaron transformation that allows using the perturbative methods of Sec. \ref{perturbativeM} to derive a master equation that is valid also for strong coupling. In addition, in close connection to this idea is the noninteracting blip approximation discussed in Sec. \ref{path}, which is also valid for strong coupling. These polaron based approaches are particularly accurate when $\Delta_0$ is small with respect to all other energy scales of the problem.
In situations where none of these methods are suitable, one may consider alternatives such as the path integral or Monte Carlo approaches briefly discussed in Sec. \ref{path}, that include in Sec. \ref{hierarchy} the hierarchy expansions (useful when the environment spectral function is a Lorentzian or a sum of Lorentzians), or the chain mapping approaches discussed in Sec. \ref{unitary}.

\subsection{Light-matter interaction Hamiltonian}
\label{lightmatter}
The light-matter interaction Hamiltonian, which describes the radiation field and an electron wave field, is written as 
\cite{wallsbook}
\begin{eqnarray}
\tilde{H}_{tot}=\frac{1}{2m}\big[{\bf p}-e{\bf A}({\bf r})\big]^2 +eV({\bf r})+H_B,
\label{chapuno13}
\end{eqnarray}
when discarding the spin of the electron. Here, $e$ and $m$ are the electronic charge and mass, respectively, ${\bf p}=-i\nabla$ is the momentum of the electron, 
\bea
{\bf A}({\bf r})=\sum_\lambda \sqrt{\frac{1}{2\omega_\lambda \epsilon_0 }}\left[a_\lambda {\bf A}_{\lambda}({\bf r})+a^{\dagger}_\lambda {\bf A}^*_{\lambda}(\bf r)\right] \nonumber
\eea
is the vector potential of the electromagnetic field, and $H_B=\sum_\lambda  \omega_\lambda a^{\dagger}_\lambda a_\lambda$ is the Hamiltonian of the free radiation field. Both quantities are written in terms of the field annihilation (creation) operators $a_\lambda$ ($a^\dagger_\lambda$). Also, $\epsilon_0$ is the vacuum permittivity, while $\lambda$ denotes the polarization $\sigma$ and the wave vector ${\bf k}$. In addition, ${\bf A}_{\lambda}({\bf r})$ are the mode functions of the electromagnetic field, which in free space may be expanded as ${\bf A}_{{\bf k}\sigma}({\bf r})=\upsilon^{-1/2}e^{i{\bf k}\cdot{\bf r}}\hat{\bf e}_{{\bf k}\sigma}$,
with $\hat{\bf e}_{{\bf k}\sigma}$ as the unit polarization vector, and $\upsilon$ as the quantization volume. 

We now consider the electric dipole approximation, which replaces ${\bf A}_{{\bf k}\sigma}({\bf r})$ by its value in the position of the atomic nucleus ${\bf r}_0$.
Considering this, the Hamiltonian (\ref{chapuno13}) can be written as\footnote{Here, we neglected the term $\sim q^2 {\bf A}^2_{\lambda}({\bf r})/2m$, compared to $\sim e{\bf p}\cdot{\bf A}_{\lambda}({\bf r})/2m$. To estimate their magnitude \cite{schulten2001}, one should consider that the magnitude of the vector potential is $||{\bf A}_{\lambda}||\sim\sqrt{{\mathcal N}_{\lambda}/(\omega_\lambda\upsilon)}$, with ${\mathcal N}_{\lambda}$ the number of photons of frequency $\omega_\lambda$ in the field, and that for a hydrogen atom $||p^2/(2m)||\sim e^2/a_0$, with $a_0$ the Born radius. In general, neglecting $\sim q^2 {\bf A}^2_{\lambda}({\bf r})/2m$ is valid as long as the photon density in the radiation field ${\mathcal N}_{\lambda}/\upsilon$ is small, particularly for the resonant frequencies of the field, $\omega_\lambda=\omega_{jl}$.}
\begin{eqnarray}
H_{\ttot}&=&H_\eel+H_B+\sum_{\lambda,j,l}g_{\lambda,j,l} b^{\dagger}_j b_{l}\left(a_\lambda +a^{\dagger}_\lambda \right),
\label{chapuno25}
\end{eqnarray}
where $H_\eel=\sum_j E_j b^{\dagger}_j b_{j}$ corresponds to the electron field, and we have considered fermionic annihilation (creation) operators $b_j$ ($b^\dagger_j$). The last term in Eq. (\ref{chapuno25}) corresponds to the interaction between them, $H_I$, with the coupling constants defined as
\begin{eqnarray}
g_{\lambda,j,l}=-i \sqrt{\frac{1}{2 \omega_\lambda \epsilon_0}}\omega_{jl}{\bf A}_{{\bf k},\sigma}({\bf r}_0)\cdot{\bf d}_{jl},
\label{chapuno23}
\end{eqnarray} 
such that $g_{\lambda,j,l}=g^*_{\lambda,l,j}$. Here, $\epsilon_0$ is the electric permittivity, $\omega_{jl}=E_j -E_l\equiv \omega_j -\omega_l$ ($\hbar=1$), and ${\bf d}_{jl}=e\int d^3 r\phi^*_j ({\bf r}){\bf \hat{x}}\phi_{l} ({\bf r})$ is the atomic dipolar moment. Here, we consider that ${\bf \hat{p}}=(im/\hbar)[H_{el},{\bf \hat{x}}]$, and also the fact that $\{\phi_j \}$ are eigenfunctions of $H_{el}$ with eigenvalues $E_j$.
For a two-level atom coupled to the radiation field, we locate the atom at the origin of coordinates ${\bf r}_0=0$, such that $g_{\lambda,2,1}=g_{\lambda,1,2}$, and therefore 
\begin{eqnarray}
H_I=  \sum_{\lambda}g_{\lambda,1,2}\left( a_\lambda + a^{\dagger}_\lambda \right)\left[ b^{\dagger}_1 b_{2}+b^{\dagger}_2 b_{1}\right],
\label{chapuno26}
\end{eqnarray}
which is of the form of Eq. (\ref{chapuno12}).

\subsection{The rotating wave approximation for the interaction Hamiltonian}
\label{RWA}

The interaction Hamiltonian can often be simplified by considering the rotating wave approximation (RWA), which allows us to neglect processes that do not conserve energy, \textit{\textit{i.e.}} those that correspond to the simultaneous creation (annihilation) of a quanta in both the open system and its environment. 
Let us consider for instance the Hamiltonian (\ref{chapuno25}), and reexpress it in the interaction picture, $e^{iH_0t}H_Ie^{-iH_0t}$. Then, it is found that the terms that conserve the energy, also known as resonant terms, oscillate with frequencies $\omega_\lambda -\omega_{jl}$, whereas those that do not conserve energy oscillate with frequencies $\omega_\lambda +\omega_{jl}$. Performing the RWA, which consists of neglecting such energy nonconserving terms, is particularly suitable to describe light-matter interaction with transitions at optical frequencies (above $600$ THz), since in this regime the nonresonant terms oscillate so fast that they cancel out along the evolution\footnote{Indeed, considering the dominant frequency of the field as the resonant frequency, $\omega_{\bf k_0}=\omega_{jl}$ (such that ${\bf k}_0$ is the resonant wave-vector), the dominant rotating phase of the energy non-conserving terms is $\omega_{{\bf k}_0} +\omega_{jl}=2\omega_{jl}$.}.
Thus, with the RWA the interaction Hamiltonian in (\ref{chapuno25}) is expressed as 
\begin{eqnarray}
H_I = \sum_{\gamma\lambda} g_{\lambda,\gamma}\left(L_\gamma a^{\dagger}_\lambda +a_\lambda L^{\dagger}_\gamma \right),
\label{chapuno27}
\end{eqnarray}
where we defined the coupling operators as $L_\gamma = b^{\dagger}_j b_{l}$, with $\gamma\equiv j,l$, and $j>l$.
For a two-level atom interacting to a bosonic field with Hamiltonian (\ref{chapuno26}), we have $L_\gamma=L=\sigma^-$ (similarly $L^\dagger_\gamma=L^\dagger=\sigma^+$), where we expressed the electron operators in terms of the spin ladder operators, $\sigma^{+}=b^{\dagger}_2 b_{1}$ and $\sigma^{-}=b^{\dagger}_1 b_{2}$. Also, $g_{\lambda}\equiv g_{\lambda,2,1}=g_{\lambda,1,2}$.

Note that the RWA is closely related to the secular approximation discussed in Sec. \ref{thermalization}. However, while in the secular approximation the fast rotating terms are eliminated after tracing out the environment degrees of freedom, the RWA discussed here is introduced before the trace, \textit{\textit{i.e.}} at the level of the Hamiltonian. As discussed by  \textcite{fleming2010} the RWA made before the trace is more problematic than the secular approximation, and may lead to incorrect values for the environmentally induced shifts to system frequencies (see also \cite{eastham2015}). Also, it is known that the RWA tends to fail in the ultrastrong coupling regime, since although the terms with phases $2\omega_{jl}$ rotate very fast, they might still represent a significant contribution \cite{prior2012}. Such ultrastrong coupling limit can be achieved in superconducting circuits \cite{niemczyk2010}, superconducting qubits in open transmission lines \cite{peropadre2013}, coupled-cavity polaritons \cite{gunter2009}, or plasmon polaritons in semiconductor quantum wells \cite{geiser2012}.

With respect to the connection with non-Markovianity, according to \textcite{makela2013} the rapidly oscillating terms are responsible for the majority of the non-Markovianity in a two-level system interacting with a bosonic environment at zero temperature. In this regard, in the limit of weak coupling and considering the RWA, the non-Markovianity appears to be relevant only at short times that are smaller than or of the order of the environment correlation time, a quantity that will be further discussed in the next section. However, without the RWA the fast-rotating terms contribute significantly to non-Markovianity during the whole evolution and not only at short times. This and other effects that relate the RWA with non-Markovianity were recently analyzed by \textcite{wang2008,zeng2012,makela2013}.

Note that a Hamiltonian of the form (\ref{chapuno27}) can be obtained in some cases without using the RWA, particularly when the bosonic field is a massive particle field, and the particle number shall be conserved. 
For many-body OQSs, the most general linear interaction Hamiltonian under the RWA is
\begin{eqnarray}
H_{I}= \sum_{\lambda,j,k} \left( g_{\lambda}({\bf r}_j) L_j a^\dagger_\lambda + g^*_{\lambda}({\bf r}_j) a_\lambda L^\dagger_j \right),
\label{ch1collective}
\end{eqnarray}
where now $L_j$ represents the coupling operator of the particle $j$ with the environment. In the case of atoms coupled to the radiation field, $L_j=\sigma_j^-$ and $g_{\lambda}({\bf r}_j)=g_\lambda e^{-i{\bf k}\cdot {\bf r}_j}$, where ${\bf k}$ and ${\bf r}_j$ represent respectively the wave vector and the position of the particle $j$. 

\subsection{Relevant scales of the problem}
\label{ch1sec7}

For environments described as a set of independent harmonic oscillators, the interaction is characterized by a spectral density, 
\begin{eqnarray}
J(\omega)=  \sum_\lambda g^2_\lambda \delta(\omega-\omega_\lambda)
\label{chapuno33}
\end{eqnarray}
where $g_\lambda$ are the coupling strengths defined previously. This function fully characterizes the action of the environment on the OQS dynamics. Such action can also be encoded in the environmental correlation function.
For most of the applications here, we are dealing with an environment that has an infinite number of degrees of freedom and that, at least initially, is in a thermal equilibrium state. Such thermal reservoirs or baths are characterized by the universality of their fluctuation-dissipation relation, and the existence of detailed balance conditions and a Kubo-Martin-Schwinger relation \cite{kubo1957,martin1959}. This relation is further discussed in Sec. \ref{perturbativeM}.
For such thermal environments, considering also the case where $L=L^\dagger$ (see also Sec. \ref{perturbativeM}) the correlation function can be written as 
\begin{eqnarray}
\alpha_T(t)=\int_0^{\infty}d\omega J(\omega)\left[\coth{\bigg(\frac{\omega \beta}{2}\bigg)}\cos{(\omega t)-i\sin{(\omega t)}}\right],\cr
\label{chapuno33bb}
\end{eqnarray}
where $\beta=(\kappa_B T)^{-1}$, with $\kappa_B$ the Boltzmann constant and $T$ the environmental temperature. This function can also be defined as 
\bea
\alpha_T(t)=\frac{1}{\pi}\int_0^\infty d\omega J(\omega) \frac{\cosh[\omega(\beta/2-it)]}{\sinh{(\beta\omega/2)}}.\nonumber
\eea
For $T=0$ the correlation function becomes a partial Fourier transform of the function $J(\omega)$
\begin{eqnarray}
\alpha(t)=\int_0^{\infty}d\omega J(\omega)e^{-i\omega t}.
\label{chapuno33b}
\end{eqnarray}
Note that in the limit of a large number of oscillators and the frequency representation $g_\lambda \rightarrow g(\omega)$, so that Eq. (\ref{chapuno33}) in the case of a dispersion relation with a single branch may be written in the continuum as
\begin{eqnarray}
J(\omega)= g^2(h(\omega))D_{\ddos}(\omega),
\label{ch134}
\end{eqnarray}
where $h(\omega)$ is the inverse of the dispersion relation (such that $h(\omega)=k$), and $D_{\ddos}(\omega)=|\partial\omega(k)/\partial k|^{-1}$ is the density of states of the field.

Thus, the behavior of the open system crucially depends on the correlation function $\alpha(t)$, which is determined by the shape of $J(\omega)$. The correlation function is an essential ingredient in every system dynamical equation, acting as the kernel of integral terms over past times. Roughly speaking, the time-scale of its decaying defines the \textit{environmental correlation} or \textit{relaxation time}, $\tau_c$, which corresponds to the time that the environment takes to return to its initial (usually equilibrium) state. If $\tau_c$ is much smaller than the evolution time of the system, $T_S$, then a Markovian approximation shall be considered to derive the OQS evolution equations. Hence, one can approximate $\Rre[\alpha(t-\tau)]\sim \Gamma \delta(t-\tau)$, with $\Gamma$ the decay rate, which cancels out the dependency over the past in the system evolution equations.

If the frequencies of the oscillators form a finite discrete spectrum, the associated correlation function is periodic or quasiperiodic for commensurate and incommensurate frequencies, respectively. Then, there is another time scale associated with the presence of a \textit{finite recurrence time} $\tau_R$ on the correlation function. In general, the denser the spectrum, the longer the recurrence time. 
When reaching the recurrence time, the OQS suffers a revival in its dynamics, regaining at least partially its lost energy and coherence. 

Here, we denote as \textit{environment} the larger subsystem with which the OQS is coupled  having either a continuous or a discrete spectrum, and as \textit{reservoir} the environment with a continuum or quasicontinuum spectrum. In general, here we will deal with environments with a spectrum that is sufficiently dense so as to assure that $\tau_c\ll T_S\ll \tau_R$. In accordance with \cite{breuerbook}, the expression \textit{bath} is to be reserved for those reservoirs in a thermal equilibrium state. 

\subsubsection{Derivations of the spectral density}
\label{ch1sec8}

There are two different possibilities to determine the spectral density $J(\omega)$, depending on the particular situation. As noted previously, when the environment is a reservoir of harmonic oscillators and the OQS can be described by a single extended coordinate $q$, the spectral density can be determined \textit{phenomenologically}, particularly from the knowledge of the coefficients of the classical equation of motion \cite{weissbook,leggett1987}. With complex environments such as those found in photosynthetic complexes, the spectral density may also have to be built based on experimental evidence. For other systems, a \textit{microscopic} knowledge of the interaction may be available, in particular of the coupling constants $g_\lambda$ and the dispersion relation $\omega(k)$. Therefore, these quantities can be used to derive $J(\omega)$ through the sum (\ref{chapuno33}). 
In the following we present an example of a microscopic and a phenomenological derivation of $J(\omega)$.

In general, a \textit{microscopic derivation} is possible for atoms interacting with electromagnetic fields, since the coupling constants are given by Eq. (\ref{chapuno23}) for a dipolar coupling between levels $|j\rangle$ and $|l\rangle$. Let us now take Eq.  (\ref{chapuno33b}) in its discrete version,
\begin{eqnarray}
\alpha(t)=\sum_\lambda |g_\lambda|^2 e^{-i\omega_\lambda t}.
\label{micro2}
\end{eqnarray}
In the continuum limit, we then have
\begin{eqnarray}
\sum_\lambda \equiv \sum_\sigma \sum_{{\bf k}}\rightarrow 2\frac{\upsilon}{(2\pi)^3}\int_0^{2\pi}d\phi \int_0^{\pi}d\theta \sin\theta \int_0^\infty dk k^2,\nonumber
\label{micro3}
\end{eqnarray}
where the factor of $2$ in the last expression comes from a sum in the two polarization modes, $\sigma$. Considering Eq. (\ref{chapuno23}) for a two-level system, 
\begin{eqnarray}
g^*_\lambda g_\lambda =\frac{1}{2 \omega_\lambda \epsilon_0 \upsilon}\omega_{12}^2 d_{12}^2 \cos^2 \theta,
\label{micro4}
\end{eqnarray}
where $|{\bf \hat{e}}_{k,\sigma}\cdot {\bf \hat{d}}_{12}|^2 =\cos^2 \theta$, and $\theta$ is the angle between the atomic dipole moment ${\bf \hat{d}}_{12}$ and the electric field polarization vector ${\bf \hat{e}}_{k\sigma}$. 
Solving the angular integrals, and considering the dispersion of the electromagnetic field in the vacuum, $k=\omega/c$, the correlation function can be rewritten as the result of an integral in frequencies as (\ref{chapuno33b}), with 
$D_{\ddos}(\omega)=\upsilon\omega^2/(2\pi)^2 c^3$, and the function $g^2 (\omega)=\omega^2_{12}d^2_{12}/6\upsilon  \omega\epsilon_0$. By virtue of Eq. (\ref{ch134}) these quantities allow us to recover the environmental spectral density. 

In many cases, the behavior of a system can be described by considering a phenomenological modeling of the spectral density at low frequencies. In this regard, one of the most well-known models is the one by \textcite{leggett1987} [see also \cite{caldeira1983,weissbook}], 
\begin{eqnarray}
J(\omega)=\eta_s \omega^s \omega^{1-s}_c e^{-\omega/\omega_c},
\label{chapuno41}
\end{eqnarray}
for all $s>0$, where $\eta_s$ has the dimensions of a viscosity and describes the coupling strength of the system and the environment \cite{weissbook}. 

The spectral density (\ref{chapuno41}) constitutes a very general model to describe many different types of reservoir, depending on the choice of the parameter $s$. The exponential factor in this model, modulated by the frequency $\omega_c$, is generally added \textit{ad-hoc} to provide a smooth regularization for the spectral density. A hard cut-off can also be considered at the characteristic frequency $\omega_c$, $J(\omega)=\eta_s \omega^s \omega^{1-s}_c\theta(\omega_c-\omega)$. The cut-off frequency should be conveniently chosen in accordance with other scales and parameters of the problem, and if it is sufficiently large, the OQS dynamics does not depend on $\omega_c$ for a vast range of parameters. The environments with $0<s<1$ are called \textit{sub-ohmic}, while those corresponding to $s=1$ and $s>1$ are known as \textit{ohmic} and \textit{super-ohmic}, respectively. 

The case of Ohmic dissipation is important for charged interstitials in metals. Also, an Ohmic model with a Lorenz-Drude regularization instead of an exponential one as in Eq. (\ref{chapuno41}), describes quantum dissipation in chemical and biophysical systems (see Sec. \ref{hierarchy}). 
A sub-Ohmic spectral density describes the dominant noise sources in solid state devices at low temperatures, such as superconducting qubits \cite{shnirman2002}, nanomechanical oscillators \cite{seoanez2007} and quantum dots \cite{tong2006}. It also appears in the context of glass dynamics \cite{rosenberg2003} or quantum impurity systems \cite{si2001}. In addition, spectral densities with $s=1/2$ and $s=3/2$ describe the radiation field in isotropic and anisotropic photonic crystals, respectively \cite{florescu2001,devega2005}. Other nonintegral values of $s$ may be relevant for fractal environments. Also, a phonon environment in $p$ spatial dimensions corresponds to the case $s=p$ or $s=p+2$, depending on the symmetry properties of the field. 

Interestingly, in the Ohmic and sub-Ohmic regimes the ground state of the spin-boson model (\ref{chapuno11}) displays a quantum phase transition when tuning the coupling strength between the system and the environment \cite{anders2007,frithjof2007,florens2011,chin2011} [see also the excellent review by \textcite{lehur2008}]. In detail, the magnetization parameter given by $\langle\sigma_z\rangle$ displays a transition between a delocalized (with $\langle\sigma_z\rangle=0$) and a localized (with $\langle\sigma_z\rangle\neq 0$) phase. This phase transition was studied by \textcite{vojta2005} with a density matrix renormalization group, and in \cite{chin2011} with a variational model and a chain mapping. Both approaches are further discussed in Sec. \ref{unitary}. 

The Ohmic dissipation is sometimes referred to as Markovian, which refers to the fact that an Ohmic spectral density leads to a constant friction kernel in the corresponding Langevin equation. However, as discussed by \textcite{rivas2010b}, an Ohmic dissipation may lead to non-Markovian effects, when the coupling strength is higher than a certain value. 

\subsubsection{The weak coupling approximation}
\label{weakcoupling}

One of the most important approximations used to describe the dynamics of an OQS is to consider that in the general form (\ref{chapuno1}), the magnitude of the coupling term $H_I$ (often described with a parameter $g$) is much smaller than the magnitude of the relevant energy transitions of the system\footnote{For instance, for a two-level system with energy transition $\hbar\omega_s\sim 3eV$, the term that describes its coupling to the light field is of the order of $||q{\bf p}\cdot{{\bf A}_{\lambda_s}}/m||\sim 10^{-3}eV$, where $\lambda_s$ is the wavelength corresponding to $\omega_s$, and we considered as a basis the dimensional analysis of Sec. \ref{lightmatter} [see also \cite{schulten2001}].}. 
The validity of such a weak coupling limit \cite{vanhove1954} is also conditioned to the environment correlation time $\tau_c$ [see also Sec. \ref{perturbativeM}]. 
Indeed, as discussed by \textcite{rivas2011a}, a \textit{necessary condition} for the existence of a weak coupling limit is that the environment has to have a well-defined correlation time. This means that it should have infinite degrees of freedom, so that there are no recurrences or periodicities in the environment correlation function. When $\tau_c$ is not even defined (e.g. because the recurrence time is smaller than the correlation time), the coupling between system and environment has to be zero in order to justify the use of a perturbative expansion. 
A \textit{sufficient condition} for the existence of a well-defined weak coupling limit was derived by \textcite{davies1976,davies1974} [see also \cite{rivas2011a}]. It states that such a well-defined limit exists if there is an $\epsilon>0$, such that $\int_0^\infty dt |\alpha(t)|(1+t)^\epsilon<\infty$, where as seen later, $\alpha(t)$ is of the order of $g^2$.

\section{Concepts of the theory of OQS}
\label{concepts}
In this section we analyze several concepts of OQSs that are independent from the tools to describe their dynamics, treated in Sec. \ref{ME}, \ref{SSE}, \ref{path}, and \ref{HR}. We start discussing the relevance of the system-environment initial state to determine the nature and properties of the resulting OQS dynamics. 
Following this, in Sec. \ref{nonmarkovianity} we introduce the different non-Markovianity measures that have been proposed during the past few years. In Sec. \ref{correlationsNM} we discuss the effect of having initial system-environment correlations in the back-flow of information from the environment into the system, and also explore the relationship between the non-Markovianity and the system-environment correlations that are built up during the evolution. Thereafter, we discuss the effects of temperature on the non-Markovianity of a process. To end this part, we analyze in Sec. \ref{asym} the influence of non-Markovianity to reach a particular steady state.

\subsection{Initially correlated and uncorrelated states between the system and the environment}
\label{initialstate}
The structure of the system-environment initial state is fundamental to determine the nature of the evolution for the reduced density matrix of the OQS, defined as $\rho_s(t)=\Ttr_B\{\rho(t)\}$, where $\rho_{{\ttot}}(t)$ is the density operator of the full system.
In this regard, the initial state is often considered an uncorrelated state of the form
\bea
\rho_{{\ttot}}(0)=\rho_s(0)\otimes \rho_B, 
\label{uncorrelated}
\eea
where $\rho_B$ is the environment density operator having a spectral decomposition $\rho_B=\sum_q \lambda_q |E_q\rangle\langle E_q|$, in terms of its eigenvectors $|E_q\rangle$, and with $\lambda_q\ge 0$. The reduced density matrix at a time $t$, $\rho_s(t)=\Ttr_B\{{\mathcal U}(t)\rho_s(0)\otimes \rho_B {\mathcal U}^{\dagger}(t)\}$, with ${\mathcal U}(t)=e^{-iHt}$, can then be written in terms of a Kraus decomposition,
\bea
\rho_s(t)=
\sum_l E_l(t)\rho_s(0)E^\dagger_l(t)=\Lambda(t)[\rho_s(0)],
\label{map1}
\eea
where $E_l=\sqrt{\lambda_q}\langle E_{q'} |{\mathcal U}(t)|E_q\rangle$ ($l\equiv \{q,q'\}$) are Kraus operators fulfilling the  property $\sum_l E^\dagger_l E_l=\unit_S$. Eq. (\ref{map1}) shows that when the reduced density operator can be written in terms of a Kraus decomposition, it can also be written in terms of a map $\Lambda(t)$ acting on its initial state $\rho_s(0)$. Such a map, often called universal dynamical map, can be shown to preserve complete positivity (CP). Following the discussion by  \textcite{rivas2011a}, CP can be explained as follows. Imagine that apart from the system $S$ and the environment $B$, there is another component $W$ that interacts neither with $S$ nor with $B$. The partial dynamics of the subsystem $SW$ can be written as 
\bea
\rho_{SW}(t)&=&\sum_l E_l(t)\otimes {\mathcal U}_W(t)\rho_{SW}(0)E^\dagger_l(t)\otimes {\mathcal U}^\dagger_W(t)\cr
&=&\Lambda(t)\otimes {{\mathcal U}}_W(t)[\rho_{SW}(0)]
\eea
with ${{\mathcal U}}_W(t)[A]={\mathcal U}_W(t)A {\mathcal U}^\dagger_W(t)$, where ${\mathcal U}_W(t)$ is the unitary evolution operator on $W$. Then, we decompose
\bea
\Lambda(t)\otimes {{\mathcal U}}_W(t)=[\Lambda(t)\otimes\unit_W][\unit_s\otimes{{\mathcal U}}_W(t)].
\eea
Here, $\Lambda(t)$ is a universal dynamical map, and so $\Lambda(t)\otimes {{\mathcal U}}_W(t)$ is a universal dynamical map too, and is therefore positive-preserving. The quantity $\unit_s\otimes{{\mathcal U}}_W(t)$ is a unitary operator, which means that $\Lambda(t)\otimes\unit_W$ should be positive-preserving. Linear maps $\Lambda(t)$ fulfilling this property are CP maps. 

When the system and the environment are initially correlated, the situation is more complicated, and it is still an open problem to understand the relationship between the structure of the initial system-bath states and the nature of the resulting dynamics, including whether or not the dynamics are CP. The requirement of complete positivity for dynamical maps was first questioned by \textcite{pechukas1994}, who pointed out that initially correlated states might not lead to CP dynamics. This idea was subject of an intense debate with \textcite{alicki1995}, who argued that all physically meaningful initial states lead to CP dynamics. To show this, he considered the initially correlated state that is experimentally obtained as a result of projective measurements on an OQS in equilibrium with its environment, 
\bea
\rho_{{\ttot}}(0)=\sum_n \lambda_nP_n\otimes \frac{\Ttr_S\{\rho_{{\ttot}}^{\eeq}(0)P_n\}}{\Ttr_{\ttot}\{\rho_{{\ttot}}^{\eeq}(0)P_n\}}
\label{projective}
\eea
with $P_n=|\psi_n\rangle\langle \psi_n|$ representing a projection on a system orthogonal state $|\psi_n\rangle$ and $\rho_{{\ttot}}^{\eeq}(0)$ is a thermal equilibrium state for the full system, and probed that it leads to CP dynamics. 

In general, initial states of the total system can be expressed in terms of the so-called assignment maps, which map a certain system state $\rho_s(0)$ to a (possibly correlated) state in the system-environment space $SB$. As it is now known, depending on their structure assignment maps may preserve or not certain properties in the reduced dynamics, such as linearity, consistency or complete positivity  \cite{modi2012a,rosario2010}. In this regard, it was shown by \textcite{jordan2004,carteret2008} that assignment maps producing certain entangled initial system-bath states may indeed lead to non-CP dynamics. 
In addition, was proven by \textcite{rosario2008} that initial states with zero discord \cite{ollivier2001,modi2012}, such as (\ref{projective}), \textit{\textit{i.e.}} fulfilling the property $[\rho_s(0)\otimes\unit_B,\rho_{\ttot}(0)]=0$, lead to completely positive reduced dynamics. \textcite{shabani2009} proposed that quantum discord is not only sufficient, but also necessary for CP dynamics. Later, \textcite{brodutch2013,buscemi2014} showed that complete positivity may also be fulfilled for some particular states with nonvanishing quantum discord, \textit{\textit{i.e.}} including quantum correlations [see also \cite{erratumshabani}].
A complete discussion on the subject can be found in \cite{dominy2013}.


\subsection{Non-Markovianity measures}
\label{nonmarkovianity}

Many non-Markovianity measures have been proposed, each of them having different strengths, \textit{\textit{i.e.}} a different ability to capture the non-Markovian nature of a process. Based on this, and on their conceptual basis, these non-Markovianity measures can be classified in different ways [see for instance  \cite{rivas2014} and \cite{hall2014}]. An exhaustive discussion of the different non-Markovianity measures is out of the scope of this review, and can be found in the reviews by \textcite{rivas2014} and \textcite{breuer2015}. Nevertheless, in the following we give an account of the most important proposals which, to our knowledge, have been presented up to date, and organize them in an approximately chronological (rather than conceptual) order. 

According to \textcite{wolf2008} a map is Markovian if it is a trace-preserving CP map and satisfies the {\bf semigroup} property,
\beqa
\Lambda(t_1+t_2)=\Lambda(t_1)\Lambda(t_2).
\label{semigroup}
\eeqa
In that case, $\Lambda(t)=e^{{\mathcal L}t}$, and it leads to a Markovian equation in Lindblad form (\ref{Linb}), also written as 
\bea
\frac{d\rho_s(t)}{dt}={\mathcal L}\rho_s(t).
\label{semigroup}
\eea
Here, ${\mathcal L}$ is a Liouville operator or superoperator (since it acts on the system density matrix flattened as a vector), which generates a dynamical semigroup. The definition (\ref{semigroup}) leads to a computable measure, which quantifies how Markovian a snapshot of a quantum evolution is, thus revealing the nature of the intermediate continuous time evolution. This approach is particularly useful to understand experimental results where input-output relations are measured via quantum process tomography. 

A less restrictive definition was proposed by \textcite{rivas2010a} [see also \cite{rivas2014}], where a map is defined as Markovian when it is a trace-preserving {\bf divisible} map, so that\footnote{Here, and when necessary for clarity, we make explicit the initial time dependence of the map.}
\beqa
\Lambda(t_1+t_2,0)=\Lambda(t_1+t_2,t_2)\Lambda(t_2,0),
\label{semigroup_2}
\eeqa
where $\Lambda(t_1+t_2,0)$ is completely positive for any $t_1,t_2>0$. Let us consider the maximally entangled state between two copies of the OQS, the system (S) and the ancilla (A), $|\Phi\rangle=(1/\sqrt{d})\sum_{n=0}^{d-1}|n\rangle_S|n\rangle_A$, where $d$ is the dimension of the OQS basis $\{|n\rangle\}$. Then, a map is CP if and only if $(\Lambda(t+\epsilon,t)\otimes\unit_d)(|\Phi\rangle\langle\Phi|)\ge 0$. Hence, since the map is trace preserving, $||(\Lambda(t+\epsilon,t)\otimes\unit_d)(|\Phi\rangle\langle\Phi|)||_1=1$ if and only if it is also CP, and higher than $1$ otherwise. Here $||\cdot||_1$ denotes the trace norm and $\unit_d$ denotes an identity map. With this idea at hand, we can define a function 
\bea
g(t)=\lim_{\epsilon\to 0+}\frac{||(\Lambda(t+\epsilon,t)\otimes\unit_d)(|\Phi\rangle\langle\Phi|)||_1-1}{\epsilon}
\label{rivas}
\eea
Then, a system is non-Markovian, \textit{\textit{i.e.}} indivisible, when $g(t)>0$ for certain interval $t\in I$, so that the total amount of non-Markovianity can be quantified by the so-called RHP (the initials standing from the author's names) measure as
\bea
{\mathcal N}(\Lambda):=\int_I g(t)dt.
\label{rivasM}
\eea 
Divisibility is crucial for the derivation of the quantum regression theorem, studied in Sec. \ref{QRT}, and for the determination of the properties of the steady state of the system, as analyzed in the following section.
As discussed by \textcite{rivas2010a,rivas2011a} and Sec. \ref{canonical}, any divisible, invertible, and differentiable completely positive map in the Hilbert space of a $d$-level system leads to a master equation that is local in time, having the form $d\rho_s(t)/dt={\mathcal L}(t)\rho_s(t)$, where ${\mathcal L}(t)$ is a Liouvillian superoperator related to the generators as ${\mathcal L}(t)=\dot{\Lambda}(t)\Lambda(t)^{-1}$. 
When a Markov process is homogeneous, \textit{\textit{i.e.}} with a time-independent generator ${\mathcal L}(t)={\mathcal L}$ as in (\ref{semigroup}), then its map is such that $\Lambda(t+\tau,t)=\Lambda(\tau)$, and the divisibility and semigroup properties are equivalent. 


Similarly, \textcite{breuer2009} developed an alternative derivation which considers as non-Markovian those systems in which there is a back-flow of information from the environment to the system during the dynamics. This back-flow of information is characterized by an increase in the distinguishability of pairs of evolving quantum states. In detail, a system is non-Markovian if there is a pair of system initial states $\rho_s^1(0)$ and $\rho_s^2(0)$, such that for certain times $t>0$ their distinguishability increases, 
\bea
\sigma(\rho_s^1(0),\rho_s^2(0);t)=\frac{d}{dt}{\mathcal D}[\rho_s^1(t),\rho_s^2(t)]>0.
\eea
Here ${\mathcal D}(\rho_s^1,\rho_s^2)=\frac{1}{2}||\rho_s^1(t)-\rho_s^2(t)||_1$ is the distinguishability of $\rho_s^1$ and $\rho_s^2$, 
and $\rho_s^j(t)=\Lambda(t,0)\rho_s^j(0)$. In this criterion, the amount of non-Markovianity of a quantum process can be quantified with the BLP measure (the initials standing from the author's names)
\bea
{\mathcal N}(\Lambda):=\mmax_{\rho_{1,2}(0)}\int_{\sigma>0}dt\sigma(\rho_s^1(0),\rho_s^2(0),t),
\label{nonBreuer}
\eea
which reflects the maximum amount of information that can flow back to the system for a given process. As proven by \textcite{wissmann2012}, for all finite dimensional quantum systems the evaluation of Eq. (\ref{nonBreuer}) can be optimized by considering initial states $\rho_s^1(0)$ and $\rho_s^2(0)$ that are orthogonal and lie at the boundary of the subset of physical states. An analogous definition of the BLP measure based on the Bures metric was studied by \textcite{vasile2011}.

A relationship between the two non-Markovianity measures RHP and BLP can be derived from the fact that all divisible maps continuously reduce the distinguishability of quantum states. Therefore, if a map is Markovian according to the RHP measure, it is Markovian according to the BLP measure, while the converse is in general not true \cite{rivas2011a,haikka2011}. \textcite{zeng2011} performed a further comparison between these two non-Markovianity measures, demonstrating that both are equivalent to each other when they are applied to open two-level systems coupled to environments via the Jaynes-Cummings or dephasing models. 

\begin{figure}[ht]
\centerline{\includegraphics[width=0.4\textwidth]{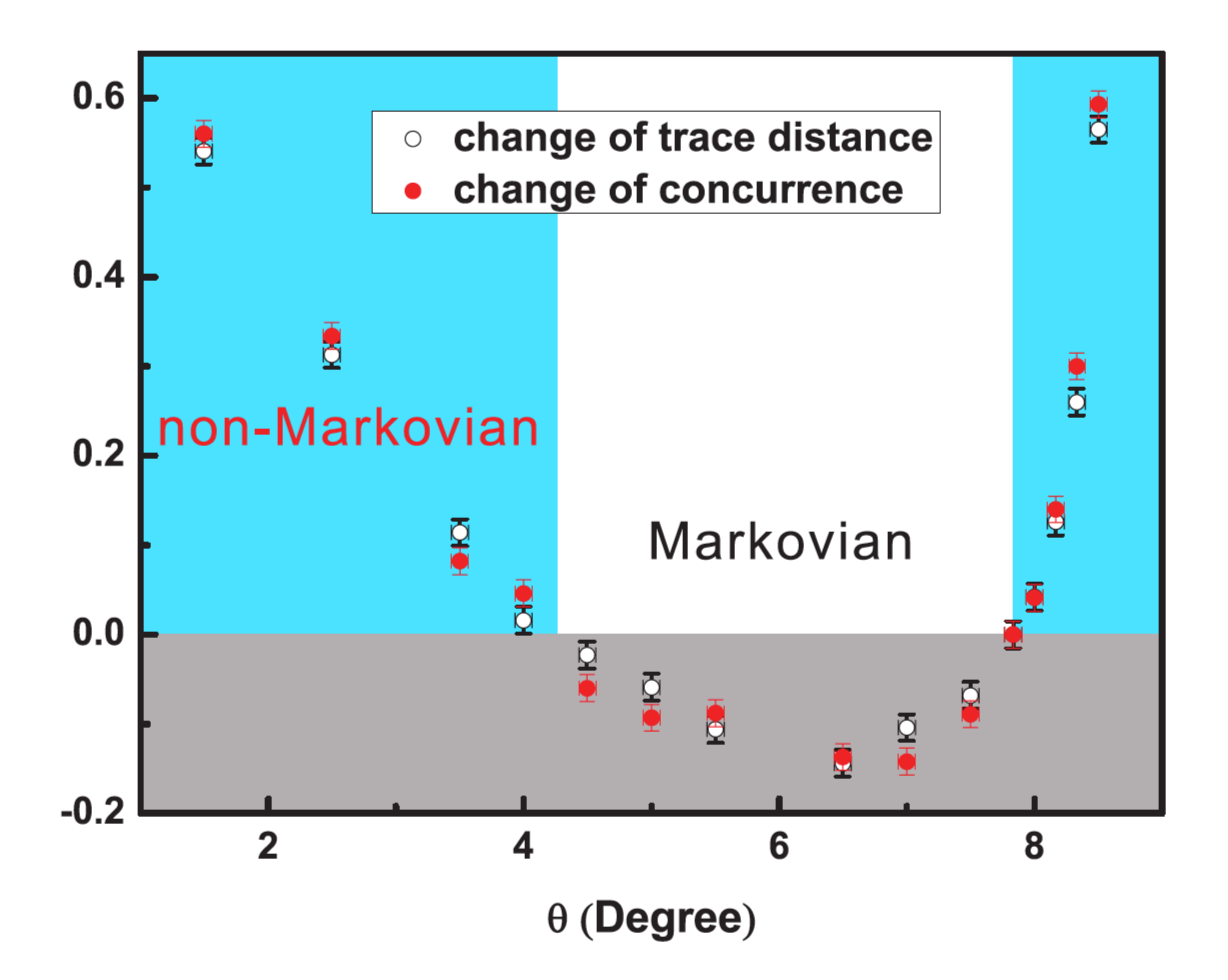}}
\caption{Changes in the trace distance and the concurrence as functions of the tilting angle $\theta$. Such an angle determines the structure of the frequency spectrum and, thus, the environmental initial state $|\xi\rangle$. The positive values in the blue regions give directly the non-Markovianity measure ${\mathcal N}(\Lambda)$ of the process, while the negative values correspond to ${\mathcal N}(\Lambda)=0$, \textit{\textit{i.e.}} Markovian dynamics. From \textcite{liu2011}. \label{figexp0}}
\end{figure}

Recently, the transition from Markovian to non-Markovian dynamics was experimentally observed by \textcite{liu2011} (see Fig. \ref{figexp0}). In this proposal, the OQS is the polarization degree of freedom of photons, described with the horizontal $|H\rangle$ and vertical $|V\rangle$ polarization states. The environment is represented by the photonic frequency degree of freedom, and it is prepared initially in a one-photon state $|\xi\rangle=\int d\omega f(\omega)|\omega\rangle$, where the frequency distribution $f(\omega)$ is normalized as $\int d\omega |f(\omega)|^2=1$. Both degrees of freedom are coupled with each other, and $f(\omega)$ can be experimentally controlled to produce different initial states for the environment $|\xi\rangle=\int d\omega f(\omega)|\omega\rangle$. In this way, initial states of the form $|\psi_{+,-}\rangle=\frac{1}{\sqrt{2}}(|H\rangle+|V\rangle)\otimes |\xi\rangle$ can be generated with different $|\xi\rangle$. The non-Markovianity of the process is then quantified through the non-Markovianity measure (\ref{nonBreuer}).

Another interesting application of non-Markovianity mesaures is the one by \textcite{znidaric2011} which analyzes the non-Markovianity of a qubit strongly coupled to an environment, considering the RHP and BLP measures. To this end, everything but the coupling operator is neglected in $H_{\ttot}$, which is chosen to be such that the statistical properties of its eigenvectors can be described by a random unitary matrix. This is a very good approximation for quantum chaotic systems \cite{haakebook}.  By analytically computing the quantum channel acting on the qubit, it is shown that a non-Markovian behavior always occur in such a strong coupling limit, independently of the environment dimension. 

Other non-Markovianity measures are based on the quantum Fisher information flow \cite{lu2010,zhong2013} and Bures distance \cite{liu2013}, 
a quantification of the deviation from divisibility in terms of the negative values of transition maps \cite{rajagopal2010}, or the non-monotonicity of the decay of the mutual correlations between the OQS and an ancilla \cite{luo2012}. Also, a non-Markovianity measure has been derived by \textcite{lorenzo2013} based on the idea that the volume of physical states accessible to a system decreases monotonically for Markov evolutions, while non-Markovian evolutions may present some time intervals where it increases. In addition, \textcite{churumode2014} propose a non-Markovianity measure based on a formal analogy with the entanglement theory, such that a Markov evolution corresponds to a separable state, while a non-Markovian evolution is characterized by the Schmidt number of an entangled state. A non-Markovianity measure based on the concept of temporal€ steering and its quantification similar to that of the original spatial EPR-steering has been developed \cite{chen2015}. Also, the canonical form of time-local master equations (see Sec. \ref{canonical}), is the basis of the non-Markovianity mesaure presented in \cite{hall2014}, which also discuss the relative strength of this measure and the previously proposed measures. More recently  \textcite{liu2015} quantifies the non-Markovianity of a chromophore-qubit pair in a super-Ohmicbath, using the distance between an evolved state and the steady state. 

\subsection{System-environment correlations and non-Markovianity}
\label{correlationsNM}

It is apparent that during the decoherence process there is an exchange of information between the system and the environment. An initial flow of information from the environment to the system has been found by \textcite{laine2011b} to be linked to the presence of initial correlations between the system and the environment. In this analysis, they have considered two initial states of the full system, $\rho_{{\ttot}}^1(0)$ and $\rho_{{\ttot}}^2(0)$, concluding that an initial increase of the trace distance of the reduced states implies that there are initial system-environment correlations or that the initial environmental states are different. Such increase in trace distance is found to be upper-bounded as
\bea
&&{\mathcal D}(\rho_s^1(t),\rho_s^2(t))-{\mathcal D}(\rho_s^1(0),\rho_s^2(0))\le\cr
&&\sum_{j=1,2}{\mathcal D}(\rho_{{\ttot}}^j(0),\rho_s^j(0)\otimes\rho_B^j(0))+{\mathcal D}(\rho^1_B(0),\rho^2_B(0)),\cr
\eea
where $\rho_s^j(t)=\Ttr_B\{\rho_{{\ttot}}^j(t)\}$ ($j=1,2$).
A more complete discussion can also be found in \cite{breuer2015}.
More recently, \textcite{mazzola2012,smirne2013} have linked the Non-Markovianity with the occurrence of system-environment correlations created during the interaction. The analysis is based on the consideration that the total system-environment density matrix can always be written as 
\begin{equation}
\rho_{{\ttot}}(t)=\rho_s (t)\otimes \rho_B(t)+\chi_{SB}(t),
\label{totalrho}
\end{equation} 
where $\chi_{SB}(t)$ describes the correlation between the system and the environment at time $t$. Then, the difference between density matrices of the total system at time $t$ that have departed from different initial states $\rho^{1}_{{\ttot}}(0)$ and $\rho^{2}_{{\ttot}}(0)$ can be decomposed as 
\bea
&&\rho_{{\ttot}}^1(t)-\rho_{{\ttot}}^2(t)=(\rho_s^1(t)-\rho_s^2(t))\otimes\rho_B^1(t)\cr
&+&\rho_s^2(t)\otimes(\rho^1_B(t)-\rho_B^2(t))+(\chi^1_{{\ttot}}(t)-\chi^2_{{\ttot}}(t)),
\label{expandrho12}
\eea
as a function of the system and environment operators corresponding to the two initial conditions (denoted by super-indexes $1$ and $2$). Computing the difference of the trace distance between $\rho_{{\ttot}}^1$ and $\rho_{{\ttot}}^2$ at times $t$ and $t+t'$, $\Delta D(t',t,\rho_{{\ttot}}^{1,2})=D(t+t',\rho_{{\ttot}}^{1,2})-D(t,\rho_{{\ttot}}^{1,2})$, it is found that a sufficient condition for non-Markovianity, \textit{\textit{i.e.}} $\Delta D(t',t,\rho_{{\ttot}}^{1,2})>0$, is that 
\bea
B(t',t,\rho_{{\ttot}}^{1,2})>D(t,\rho_{{\ttot}}^{1,2})+F(t',t,\rho_{{\ttot}}^{1,2}).
\label{correlationsth}
\eea
Here, $B(t',t,\rho_{{\ttot}}^{1,2})$ keeps track of the effects of system-environment correlations (\textit{\textit{i.e.}} it is originated by the last two terms in Eq. (\ref{expandrho12})) at times $t+t'$. Also, $F(t',t,\rho_{{\ttot}}^{1,2})$ expresses how the distinguishability between reduced states would be at $t+t'$ if the two total states at time $t$ were product states, and it is thus originated by the first term in the rhs of (\ref{expandrho12}). Hence, system-environment correlations, given by $B$ in (\ref{correlationsth}), must exceed a threshold in order to produce an increase in the distinguishability, and thus lead to a non-Markovian evolution. 

A different question is whether the information exchanged is of quantum or classical nature. In particular, there are circumstances where the system decoheres without becoming entangled with the environment at any time \cite{eisert2002,pernice2011,pernice2012}. 

Naturally, the former analysis refers to system-environment correlations existing in the total density matrix, which is obtained as a sum of the results of many different experimental runs starting from the same initial configuration. Hence, even if it is found that $\chi_{SB}=0$, system-environment correlations could be (and in fact are) present at each experimental run. In fact, system-environment correlations are the basis of indirect measurement techniques, in which, for instance, information about the atomic state is obtained from scattered photons. As discussed in Sec. \ref{SSE}, indirect measures may be a basis for deriving stochastic Schr\"odinger equations. 

\subsection{Environment-environment correlations and non-Markovianity}
\label{BBcorrelationsNM}
As proposed by \textcite{laine2011}, non-Markovian effects can emerge in a composite open system (for instance a bi-partite open system with reduced state $\rho_{s}$), when each OQS's component interacts locally with a subsystem of a composite environment. Then, provided that the subsystems of the composite environment are initially correlated, non-Markovianity can be observed in the reduced dynamics of the composite OQS state $\rho_{s}$, while the local dynamics of the reduced density operator of each member of the composite open system ($\rho^1_s=\Ttr_2\{\rho_s\}$ and $\rho^2_s=\Ttr_1\{\rho_s\}$) remains Markovian. Such nonlocal memory effects have been shown to be a resource for quantum information tasks, such as quantum communication \cite{laine2014,hengliu2013} and efficient superdense coding in the presence of dephasing noise \cite{liu2015}. 

Environment-environment correlations can be either experimentally prepared as in \cite{hengliu2013,liu2015}, or emerge dynamically. As discussed by \textcite{chan2014}, a key aspect for the appearance of correlations among multiple baths is the presence of non-Markovianity in the interaction between the subsystems and their environments. In more detail, only when such interaction is non-Markovian quantum interference between the baths emerge, as opposed to the Markovian limit where the action of the different baths is additive. An OQS coupled to multiple reservoirs can be found in different situations, as in cavity quantum electrodynamics \cite{gea-banacloche2005}, opto-mechanical cavities \cite{safavi-Naeini2014}, traveling-wave photon-phonon transduction in nanophotonic waveguides \cite{shin2015}, photo-active molecules coupled to a vibrating environment such as photosynthetic complexes \cite{blankenship2002}, or the dynamical Casimir effect \cite{impens2014} just to name a few. 

\subsection{Temperature and non-Markovianity}

An insightful case study is to consider an initial uncorrelated state of the form (\ref{uncorrelated}) with the environment in a thermal equilibrium, 
\bea
\rho_B^{\eeq}=\frac{e^{-\beta H_B}}{\Ttr_B\{e^{-\beta H_B}\}}.
\label{thermal_B}
\eea
In general, it is well known that temperature enhances the decay and therefore tends to decrease the relaxation time of the system \cite{affleck1981,weiss1983,grabert1984}. A different question is how temperature affects the non-Markovianity of the evolution. It is generally believed that non-Markovian effects are more important at low temperatures \cite{weissbook}. In this regard, for a two-level system in a spin bath, \textcite{huzheng2014} showed that the non-Markovianity decreases close to the critical point of the system, and that this decrease is indeed higher at higher temperatures. In addition, \textcite{haikka2013} analyze a two-level system subject to a dephasing bath with spectral density (\ref{chapuno41}), observing that there is a critical value of $s$ for the onset of non-Markovianity. This critical value is higher for high temperatures. Also, \textcite{liu2015} concluded that the non-Markovianity of a chromophore qubit in a super-Ohmic bath, and thus the backflow of information from the environment, is reduced when the temperature increases.

\begin{figure}[ht]
\centerline{\includegraphics[width=0.4\textwidth]{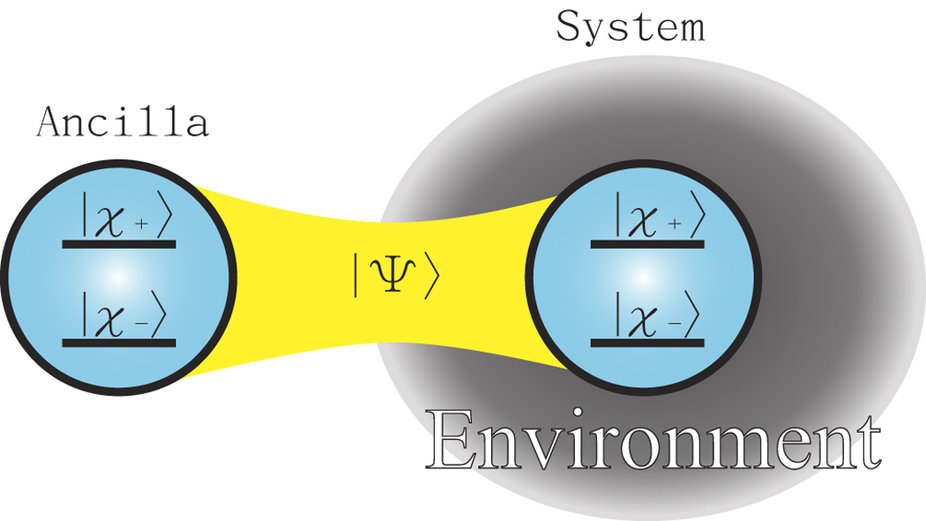}}
\caption{Schematic illustration of the entanglement measure, which is closely associated with the non-Markovianity measure proposed by \textcite{rivas2010a}. A system and an environment isolated copy of it, acting as an ancilla, is considered. Initially, they form a maximally-entangled state. When the system starts to be coupled with its environment (denoted by the gray shadow) it will evolve and the system-ancilla entanglement will be sensitive to the environment coupling. From \textcite{hogbin2015}. \label{nori}}
\end{figure}

However, as shown recently by \textcite{ruggero2014} for a bosonic noninteracting system, all quantities environment size, temperature, proximity of a cutoff frequency $\omega_c$ in the spectra, spectral density shape (sub-Ohmic, Ohmic, super-Ohmic), and strength of the coupling to the system are crucial factors in determining the non-Markovianity of an evolution. Interestingly, \textcite{ruggero2014} determines that for certain parameter values, the non-Markovianity increases with the temperature. 
Along the same line, \textcite{hogbin2015} also showed that non-Markovianity can increase with temperature and with the coupling to the environment. In this proposal, both entanglement and non-Markovianity measures are used to reveal whether second-order weak coupling non-Markovian master equations (Sec. \ref{ME}) underestimate or overestimate memory effects. This is done by comparing the approximated equations to the numerically exact hierarchical equations of motion (HEOM) discussed in Sec. \ref{hierarchy}. The entanglement measure considered is detailed in Fig. \ref{nori}.

\subsection{Asymptotic and equilibrium states in Markovian versus non-Markovian dynamics}
\label{asym}
In order to characterize an OQS in the long time limit, several concepts come into play. In the long time limit the OQS may relax to a steady state, characterized by a time-independent density matrix, $\lim_{t\to\infty}\frac{d\rho_{S}(t)}{dt}=0$. 
Moreover, while relaxation describes the convergence of the reduced density matrix of a system to a fixed but arbitrary state, thermalization corresponds to a relaxation or stabilization of the system to its thermal or Gibbs state \footnote{In the context of isolated many-body systems, a generalization of the Gibbs state was proposed \cite{rigol2007}, which is also valid for systems obeying conservation laws.},
\bea
\rho^{\eeq}_S=\frac{e^{-\beta H_S}}{Z_S(\beta)}.
\label{boltzman}
\eea
In the limit of vanishing coupling strength, a system coupled to a thermal reservoir relaxes to such thermal state \cite{vanhove1954,davies1976,laird1991},
\bea
\lim_{g\to 0}\lim_{t\to\infty} \rho_s(t)=\rho^{\eeq}_S,
\label{boltzman2}
\eea
irrespective of the initial state of the system, but only if certain conditions are fulfilled \cite{romerorochin1989,geva2000}.  However, this may not be the case in the strong coupling limit, or for specific spectral densities of the environment. A detailed discussion of this is provided in Sec. \ref{thermalization}.

In addition, as pointed out by \textcite{dario2010}, there is a crucial difference between the long time limit states resulting from Markovian and non-Markovian evolutions (Markovian evolutions are here understood as those described by a divisible map). In order to appreciate this difference, it is necessary to recall the following definitions: first, for a Markov evolution a steady state $\rho_{ss}$ is defined as
\beqa
\lim_{t\to\infty}\Lambda(t,0) \rho_0=\rho_{ss},
\eeqa
for any arbitrary state $\rho_0$, where the solution of the general Markovian master equation is written as $\rho_s(t)=\Lambda(t,0)\rho_0$; second, because a Markov evolution possesses the divisibility property $\Lambda(t+t_2,0)=\Lambda(t,t_2)\Lambda(t,0)$, we find
\beqa
\lim_{t_2\to\infty}\Lambda(t+t_2,0) \rho_0=\rho_{ss}.
\eeqa
But also $\lim_{t_2\to\infty}\Lambda(t+t_2,0) \rho_0=\Lambda(t,0)\rho_{ss}$. Hence, $\Lambda(t,0)\rho_{ss}=\rho_{ss}$, \textit{\textit{i.e.}} for a Markovian evolution the steady state $\rho_{ss}$ is always invariant, wherein a state $\rho_0$ is said to be invariant if $\Lambda(t) \rho_0=\rho_0$ for any $t\ge 0$. 
Because non-Markovian evolutions do not fulfill the semigroup property, a non-Markovian evolution may lead to a steady state that is not invariant. A non-Markovianity measure based on this idea was proposed by \textcite{dario2010}. 

A related question is whether an OQS relaxes to steady state (either thermal or not) which is independent of the initial condition. 
For the case of a Markov semigroup map, as the one corresponding to the Lindblad equation, the steady state is unique as long as the semigroup is relaxing, in which case the equation ${\mathcal L}\rho_{ss}=0$ admits only one solution. A semigroup is relaxing when the zero eigenvalue of the generator ${\mathcal L}$ is non-degenerate, and the rest of the eigenvalues have negative real part. Otherwise, the final state of the system might depend on the initial state. 

The dependency of the steady state on the spectral density structure was extensively studied for the case of the OQS being a harmonic oscillator (see also Sec. \ref{brownian}). In this case, the system annihilation operator can be expressed as $a(t)=A(t)a(0)+\sum_\lambda u_\lambda(t)b_\lambda(0)$ \cite{cai2014,louisell}, and a nonvanishing asymptotic value of $A(t)$ (which is solved through an integro-differential equation that depends on the spectral density) is clearly identified with a non-thermal relaxation. In this regard, \textcite{zhang2012,xiong2013} concluded that when the spectral density has band gaps or a finite band [so that $J(\omega)$ vanishes in a certain region], localized modes exist in the environment that give rise to dissipationless dynamics (and hence nonthermal relaxation) in the OQS. Remarkably, nonthermal relaxation also occurs for non-gapped spectral densities, provided that the coupling strength exceeds a certain threshold \cite{xiong2013,cai2014}. Moreover, such nonthermal relaxation was also explored by \textcite{iles-smith2014} by considering a two-level system coupled to an environment with a Drude spectral density (see also Secs. \ref{hierarchy} and \ref{unitary} for details on how to deal exactly with this case).

The long time limit of an evolution is often difficult to obtain, either because of inherent limitations of the approximations used, or because of the difficulty of performing numerical calculations at long times. Nevertheless, as shown by \textcite{cerrillo20141}, the initial evolution of an OQS up to $\tau_c$, already contains all the relevant information of the multiple-time correlations of the OQS observables. This information can be extracted to determine a set of non-Markovian transfer tensors, which can be used to propagate the system state to arbitrary long times.

\subsubsection{Quantum correlations and entanglement in the steady state}

The coupling of a multipartite open system with an environment does not always produce decoherence and decay of its quantum correlations. 
In fact, entanglement may be preserved and even generated within the system due to a combined action of environment noise with either a driving source \cite{huelga2007,li2009,galve2010} a nonequilibrium situation \cite{cai2010,lambert2007}, or a continuous monitoring of the decay dynamics \cite{plenio1999}, for instance. 

Entanglement generation has also been analyzed in situations where the systems involved are coupled to common or independent reservoirs (see discussion in Sec. \ref{cooperative} regarding the conditions for these two limits). In more detail, entanglement may be generated by considering that the systems involved are coupled to a common reservoir, Markovian in the case of \cite{benatti2003}, and non-Markovian in \cite{braun2002} where the dynamics of $H_S$ is neglected. 
But even when considering the systems coupled to independent Markovian reservoirs, a careful design of the system-environment couplings can lead to an entangled state as the unique stationary state of a dissipative process \cite{kraus2008,verstraete2008}.

A different situation is analyzed by \textcite{huelga2012} who considered two spins with nearest-neighbor interactions and locally coupled to two damped harmonic oscillators, showing that non-Markovianity is a resource to support the formation of steady state entanglement in situations where purely Markovian dynamics leads to separable steady states. In a subsequent analysis of the many-body generalization of this model, \textcite{cormick2013} showed that long time limit entangled states can also be achieved in the Markov case, and that the role of non-Markovianity is just to allow for a faster convergence to such steady state. 

Regarding the dynamics from an initially entangled state, the entanglement between a pair of two-level systems has been shown to vanish at short times compared to the usual spontaneous lifetime \cite{yu2004b}.  However, if the reservoirs are non-Markovian, the dynamics of a pair of two-level systems in an initial Bell-like state \cite{bellomo2007}, or Werner-like state \cite{bellomo2008a}, or the dynamics of two oscillators \cite{paz2008,paz20091}, may show the presence of entanglement oscillations and revivals after a finite period of time of its complete disappearance. Other studies of continuous variable systems coupled to non-Markovian environments have suggested its relevance in the preservation of two-mode \cite{maniscalco2007,horhammer20081,liu20071,cormik20101,correa20121,estrada20151} and three-mode \cite{valido20131,valido20132,valido20141,hsiang20151,valido20151} entanglement. 
Revivals of quantum correlations may also occur when the environment is classical \cite{zhou2010,franco2012,franco2012b,bordone2012,xu2013} and thus it does not have a back-action into the system.

\section{Master equations} 
\label{ME}

One of the most important approaches describing the dynamics of an OQS is to compute the master equation evolving the reduced density matrix of the system $\rho_s(t)$. Some of the most relevant master equations available are discussed in this section.

\subsection{Brief historical review: Rate equations and Markov master equations}
\label{historicalME}
The theory for describing the dynamics of an OQS is well developed
under the Markov hypothesis, assuming that the relaxation time of the environment is much smaller than any relevant time scale of the system. One of the first evolution equations was derived by \textcite{einstein1917}, and describes the atomic population dynamics of an atom emitting and absorbing light in a thermal field. The generalization of this equation, made by \textcite{pauli1928} [see also \cite{quantumnoise}], reads as follows:
\begin{eqnarray}
\frac{d P_n (t)}{dt}=\sum_{m>n}(A^n_m +B^n_m D)P_m (t)+\sum_{m<n} B^n_m D P_m (t)\nonumber\\
-\sum_{m<n} (A^m_n +B^m_n D)P_n (t)-\sum_{m>n}B^m_n D P_n (t),\nonumber
\label{rateeq}
\end{eqnarray}
where $P_n (t)$ are the occupation probabilities of the energy levels. The coefficients $A^m_{n} $ and $B^m_n $ represent the transition rates from the atomic state $n$ to the atomic state $m$ due to spontaneous and stimulated emission, respectively. In this equation, $D\equiv D(\omega_{mn})$ is the energy density of the electromagnetic field at the emitting frequency, $\omega_{mn}=(E_m -E_n )$, where $E_n$ is the energy of the level $n$. The energy density is given by Plank's radiation law $D(\omega_{mn})=\alpha \omega_{mn}^3 \exp(-\omega_{mn}\beta)$ \footnote{Let us recall here that we have set $\hbar=1$, so that $\omega_{mn}=(E_m -E_n )/\hbar \equiv E_m -E_n$ and $D(\omega_{mn})=\alpha \omega_{mn}^3 \exp(-\hbar\omega_{mn}\beta)\equiv D(\omega_{mn})=\alpha \omega_{mn}^3 \exp(-\omega_{mn}\beta)$. }.

The positive terms represent the gain of probability from transitions into the state $n$, and the negative terms represent the loss of probability by transitions from the state $n$. The transition rates between populations $\left\{A^m_{n},B^m_n \right\}$ are given by the Fermi golden rule within the weak coupling approximation \cite{quantummechanicsII}. When the Hamiltonian of the system is unknown, transition rates can be calculated from experimental data or chosen by a phenomenological ansatz. The use of a quantum theory that only has to deal with probabilities was justified by Pauli with the \textit{repeated random phase assumption}, which consists of assuming that the phase relations between wave functions are always (repeatedly) randomized, so that only the square of the wave function (\textit{\textit{i.e.}} the probabilities) are relevant. Nevertheless, this assumption is not valid whenever the quantum coherences remain finite during the system evolution time scale.

In the second half of the last century, the {\bf density operator} $\rho(t)$ was introduced by \textcite{landau1927,luders1951,neumann1955} [see also \textcite{landau,diosi19901,quantummechanicsII,quantumnoise,breuerbook}]. Such an object is more convenient for describing systems where the repeated random phase assumption cannot be applied. A good example of such systems is lasers, which, as highly coherent fields, cannot be described with the Pauli equation (nor can systems interacting with them). 

The best-known master equation, which is obtained under the Born-Markov approximation, is the Lindblad equation\cite{gorini1976,lindblad1976}, which corresponds a dynamical semigroup as discussed in Sec. (\ref{nonmarkovianity}). For a Hamiltonian of the form (\ref{chapuno27}) and considering an OQS having a $d$-dimensional Hilbert space (see Sec. (\ref{canonical}) for a derivation)
\begin{eqnarray}
\frac{d{\rho}_s(t)}{dt}&=&-i[H_s,\rho_s(t)]+\sum^{d^2-1}_{k=1} \Delta_k \bigg[2C_k\rho_s(t)C^\dagger_k\cr
&-&\{C^\dagger_k C_k,\rho_s(t)\}\bigg],
\label{Linb}
\end{eqnarray}
where $C_k$ are system operators in the Lindblad form, and $\Delta_k$ is a constant and positive parameter. 
This master equation represents one of the key elements of the theory of OQSs, and is particularly suited to quantum optics and quantum thermodynamic scenarios. In the latter case it allows us to address the thermodynamic processes taking place at a finite time \cite{spohnbook,campisi2009,esposito2012,strasberg2013,kosloff2013,gelbwaser20131,correa20131,szczygielski20131,correa20141,correa20142,correa2015internal,palao2016efficiency}. The intimate connection between quantum thermodynamics and the theory of OQSs, and hence the thermodynamic consistency of the latter, beyond the weak coupling condition remains the subject of ongoing developments [see \cite{esposito2009a} for a review].

A critical analysis of the validity of the Markov approximation for a single oscillator and two interacting harmonic oscillators coupled to a harmonic oscillator environment, which is exactly solvable problem (see Sec. \ref{brownian}), was performed by \textcite{rivas2010b}. 

\subsection{Non-Markovian master equations}

\label{canonical}
Even without the Markov approximation, the dynamics of the reduced density operator of an OQS obey a time-local master equation, as long as its map, given by Eq. (\ref{map1}), is invertible and differentiable. To show this, we compute the time derivative of this equation to get 
\bea
\frac{d\rho_s(t)}{dt}
=\sum_l\bigg(\frac{dE_l(t)}{dt}\rho_s(0)E^\dagger_l(t)+E_l(t)\rho_s(0)\frac{dE^\dagger_l(t)}{dt}\bigg).\cr
\label{map2}
\eea
If the map $\Lambda(t)$ describing the evolution is invertible, then we can always reexpress $\rho_s(0)=\sum_{m}F_m(t)\rho_s(t)Q_m(t)$, where $F_m$ and $Q_m$ are system operators. Conditions for invertibility of a map have been discussed for instance by  \textcite{maldonado2012}, and consequences on the complete positivity of the resulting equation are further discussed by  \textcite{breuer2015}. Inserting this expression in (\ref{map2}), this equation can be reformulated as 
\bea
\frac{d\rho_s(t)}{dt}=
\sum_{k} A_k(t)\rho_s(t)B^\dagger_k(t),
\label{map3}
\eea
where the label $k=\{\eta,l,m\}$, with $\eta=1,2$, such that $A_{1,l,m}(t)=(dE_l(t)/dt) F_m(t)$,
$A_{2,l,m}(t)=E_l(t)F_m(t)$, $B^\dagger_{1,l,m}(t)=Q_m(t)E_l^\dagger(t)$, $B^\dagger_{2,l,m}(t)=Q_m(t)(dE^\dagger_l(t)/dt)$.
Here we defined 
\bea
V_{t-t_0} X ={\mathcal U}^{\dagger}_{0}(t,t_0)X{\mathcal U}_{0}(t,t_0), 
\label{defVs}
\eea
for any system or environment operator X, and also $\rho^{I}_{\ttot}={\mathcal U}_{0}(t,t_0)\rho_{\ttot}(t){\mathcal U}^{\dagger}_{0}(t,t_0)$ with the free evolution operator ${\mathcal U}_{0}(t,t_0)=e^{-iH_{0}(t-t_0)}$. 
%
Following the techniques developed by \textcite{gorini1976} for deriving the Lindblad equation, and further discussed by \textcite{hall2014} for the non-Markovian case, we rewrite the system operators $A_k(t)$ and $B_k(t)$ in terms of the complete set of $N=d^2$ basis operators $\{G_i\,;\,i=0,\cdots,N-1\}$, with the properties $G_0=\unit_S/\sqrt{d}$, $G_i=G_i^\dagger$, $\Ttr\{G_i\}=\delta_{i0}$, and $\Ttr\{G_{i}G_j\}=\delta_{ij}$. Note that for a two-level system, these are just the unit matrix $\unit_S$ and the Pauli matrices, $\sigma_i$ with $i=x,y$ and $z$. Then the expansion takes the form
\bea
A_k(t)=\sum_i a_{ik}(t)G_i\,,\qquad B_k(t)=\sum_j b_{jk}(t)G_j,
\eea
with $a_{ik}(t)=\Ttr\{A_k(t)G_i\}$, and $b_{ik}(t)=\Ttr\{B_k(t)G_i\}$. In these terms, the general master equation (\ref{map3}) becomes $\dot{\rho}_s(t)=\sum^{N-1}_{ij=0}c_{ij}G_i\rho_s(t)G_j$, with $c_{ij}=\sum_k a_{ik}(t)b^*_{jk}(t)$ being the elements of an $N\times N$ matrix, which because of the Hermiticity of $\rho_s(t)$ shall be Hermitian too, such that $c_{ij}=c^*_{ji}$.

Separating out the terms $i=0$ and $j=0$, the master equation can formally be written as 
\bea
\frac{d\rho_s(t)}{dt}&=&-i[\hat{H}_S(t),\rho_s(t)]+C\rho_s(t)+\rho_s(t) C^\dagger\cr
&+&\sum^{N-1}_{ij=1}c_{ij}G_i\rho_s(t)G_j,
\label{map4}
\eea
where we defined 
\bea
C=\frac{\unit_S}{2d}c_{00}+\sum_{i=1}^{N-1}\frac{c_{i0}}{\sqrt{d}}G_i. \nonumber
\eea
Trace preservation implies that $C+C^\dagger=-\sum^{N-1}_{ij=1}c_{ij}G_jG_i$. Rewriting Eq. (\ref{map4}) in terms of combinations of $C-C^\dagger$ and $C+C^\dagger$, we find that 
\begin{eqnarray}
\frac{d{\rho}_s(t)}{dt}&=&-i[\hat{H}_S(t),\rho_s(t)]+\sum^{d^2-1}_{ij=1} d_{ij}(t) \bigg(G_i\rho_s(t)G_j\cr
&-&\frac{1}{2}\{G_j G_i,\rho_s(t)\}\bigg),
\label{NMLinb0}
\end{eqnarray}
where $\hat{H}_S(t)=(i/2)(C-C^\dagger)$, and $d_{ij}=c_{ij}$ for $i,j>0$.
Then, considering that the decoherence matrix ${\bf d}$ is Hermitian, it can be rewritten in its diagonal form $d_{ij}(t)=\sum_{k}U_{i k}(t)\Delta_k(t)U^*_{j k}$, where $\Delta_k(t)$ and $U_{j k}(t)$ are, respectively, its eigenvalues and unitary eigenvectors. Defining the time-dependent operators
\bea
C_k(t)=\sum_{i=1}^{N-1}U_{ik}(t)G_i,
\eea
we can rewrite (\ref{NMLinb0}) in the canonical form
\begin{eqnarray}
\frac{d{\rho}_s(t)}{dt}&=&-i[\hat{H}_S(t),\rho_s(t)]+\sum^{d^2-1}_{k=1} \Delta_{k}(t) \bigg(2C_k(t)\rho_s(t)C^\dagger_k(t)\cr
&-&\{C^\dagger_k (t)C_k(t),\rho_s(t)\}\bigg),
\label{nmbreuer}
\end{eqnarray}
which is the non-Markovian generalization of the Lindblad equation (\ref{Linb}).
Note that this equation corresponds to the general time-local master equation previously defined in \cite{breuer2004a}. In Eq. (\ref{nmbreuer}), complete positivity can be ensured only when $\Delta_k(t)\ge 0$ throughout the whole evolution. If this condition is not fulfilled, nothing can be assured, and CP may or may not be preserved depending on the case. Moreover, according to \cite{hall2014,breuer2015}, a time-local master equation is Markovian if an only if the canonical decoherence rates are positive during the whole evolution. Non-Markovianity can then be defined as a sum of all intervals where the decaying rates $\Delta_k(t)$ are negative. This measure is shown to be equivalent in strength to the one defined by \textcite{rivas2010a}. Finally, in the Markov case originally considered by \textcite{gorini1976}, the coefficients are time independent, $\Delta_k(t)=\Delta_k$, and CP can be assured provided that they are all positive. 

Eq. (\ref{NMLinb0}), or its canonical version (\ref{nmbreuer}), formally describes the evolution of the reduced density matrix of an OQS. Its coefficients can only be computed exactly in the specific systems discussed in Sec. \ref{Exact}, namely for a quantum Brownian particle, in the dephasing case $L\sim H_s$, or when the full problem can be solved within the one excitation sector.  Nevertheless, in a recent derivation \textcite{ferialdi2016a} has provided the most general form of non-Markovian map. Based on this, \textcite{ferialdi2016b} has shown that the coefficients of the master equation for the spin-boson and Jaynes-Cummings models (the last one consisting on several spins coupled to a common bosonic field) come in terms of an infinite series of mutually dependent terms. The convergence and properties of this series for different spectral densities and particle configurations (in the many-body case) remains to be studied.

In the following sections, we analyze several approximations to tackle the dynamics of a general OQS. In this regard, the first non-Markovian master equation was derived by \textcite{redfield1957,redfield1965} within the context of nuclear magnetic resonance. A more accurate non-Markovian master equation, which allows us to recapture the Redfield equation itself in a limit, was later derived by considering a weak coupling approximation between the system and the environment. This equation can be obtained by means of two different methods, which are explained in the following sections: the first is based on assumptions made on the evolution time scales and on the Born-Markov approximation, and the second is based on an expansion in the coupling parameter between the system and the environment. For both methods, a total Hamiltonian of the form $H_{tot}=H_0+gH_I$ is considered, where $g$ is a small parameter that, for simplicity, is absorbed here into $H_I$, so that terms proportional to $H_I^n$ are at least of the order $g^n$. 

\subsubsection{The Born-Markov approximation}
\label{bornmarkov}
The von-Neumann equation for the density operator of the total system in the interaction picture, $\rho^{I}_{tot}(t)$, reads as follows:
\begin{equation}
\frac{d\rho^{I}_{\ttot}(t)}{dt}=\frac{1}{i}[V_t H_I,\rho^{I}_{\ttot}(t)],
\label{total1}
\end{equation}
where we considered the definition (\ref{defVs}).
To simplify the notation, we set $\rho^{I}_{\ttot}(t)=\rho(t)$. We can integrate Eq. (\ref{total1}) between $t_0$ and $t$. After two iterations and a trace over the environmental degrees of freedom, this leads to the following equation,
\begin{eqnarray}
&&\Delta \rho_s (t)=\frac{1}{i}\int_{t_0}^{t}d\tau {\textmd{Tr}_B}\{[V_\tau H_I,\rho(t_0)]\}+ {\left( \frac{1}{i} \right)}^2 \cr
&\times&\int_{t_0}^{t}d\tau\int_{t_0}^{\tau} d\tau' {\textmd{Tr}_B}\{[V_\tau {H}_I,[V_{\tau'}{H}_I,\rho(\tau')]]\},
\label{total2}
\end{eqnarray}
where $\rho_s (t)={\textmd{Tr}_B}\{\rho(t)\}$ is the system reduced density operator and
\begin{equation}
\Delta \rho_s (t)=\rho_s (t)-\rho_s (t_0).
\end{equation}
Equation (\ref{total2}) is exact, but some assumptions have to be made in order to express it as a closed equation for $\rho_s(t)$. For an initially uncorrelated state of Eq. (\ref{uncorrelated}), $\rho(t_0)=\rho_s (t_0)\otimes \rho_B$, and considering the case where 
\bea
\Ttr_B\{{V_{t_0} H}_I \rho_B^{\eeq}\}=0,
\label{odd} 
\eea
so that the first term in Eq. (\ref{total2}) can be eliminated. Note that this occurs for instance when the environment is initially in thermal equilibrium $\rho_B=\rho_B^{\eeq}$ given by (\ref{thermal_B}).

After the change of variable $T=\tau$ and $s=\tau-\tau'$, the Eq. (\ref{total2}) becomes
\begin{eqnarray}
&&\rho_s(t)=\rho_s (t_0)-\int_{t_0}^{t}dT\int_{0}^{T-t_0} ds {\textmd{Tr}_B}\,\{[V_T {H}_I ,\cr
&\times&[V_{T-s} {H}_I ,\rho(T-s)]]\}.
\label{total3chap3}
\end{eqnarray}
The evolution equation for the reduced density operator can be obtained by taking the derivative of Eq. (\ref{total3chap3}) with respect to $t$, 
\begin{equation}
\frac{d\rho_s (t)}{dt}= - \int_{0}^{t-t_0}d\tau {\textmd{Tr}_B}\bigg\{[V_t {H}_I,[V_{t-\tau}{H}_I ,\rho(t-\tau)]]\bigg\},
\label{total4}
\end{equation}
with initial condition $\rho_s (t_0)$. 
The density operator appearing in the rhs of Eq. (\ref{total4}) has the general form (\ref{totalrho}). However, the integral in Eq. (\ref{total4}) contains a kernel, the correlation function, that decays with $\tau_c$. In addition, the term $\chi_{SB}(t)$, which describes the correlations between the system and the environment at time $t$, persists only during a time approximately equal to $\tau_c$. Hence, such correlations can be neglected with the assumption that $\tau_c\ll T_S$. This is the Born approximation, which is valid only up to the order $g^2$ in the perturbation parameter \cite{atomphotoninteractions,breuer1999}. 
In order to transform the resulting equation into a time-local form, we further replace $\rho_s(t-\tau)=\rho_s(t)$ within the integral term. This approximation is valid provided that the system evolution time $T_S$ is much slower than the correlation time of the environment, which settles the scale in which the integrand decays to a certain value. This is sometimes known in the literature as the \textit{first Markov approximation}. 

Choosing $t_0 =0$, the evolution equation (\ref{total4}) then becomes, after a trivial change of variable $t-\tau\rightarrow \tau$
\begin{equation}
\frac{d\rho_s (t)}{dt}= - \int_{0}^{t}d\tau {\textmd{Tr}_B}\{[V_t {H}_I ,[V_\tau{H}_I ,\rho_{B}(t)\otimes\rho_s (t)]]\},
\label{total5}
\end{equation}
where $\rho_B (t)=Tr_{S}\{\rho(t)\}$, and the initial condition is $\rho_s(0)$. 
As discussed later, a further approximation consists of assuming that the integral limits can be extended to $\infty$, which is often known in the literature as the \textit{second Markov approximation}.

\subsubsection{Perturbative approximation in the coupling constant}
\label{perturbativeM}
The equivalence between approximations on time scales and the Born approximation and the weak coupling assumption can easily be seen by returning to Eq. (\ref{total1}) and performing a perturbative integration of $\rho(t)$. After tracing out the environment's degrees of freedom, we get an expression similar to Eq. (\ref{total3chap3}), 
\begin{eqnarray}
&&\rho_s(t)=\rho_s(t_0)- \int_{t_0}^{t}dT\int_{t_0}^{T} d\tau {\textmd{Tr}_B}\{[V_T {H}_I ,\cr
&\times&[V_{\tau} {H}_I,\rho(t_0)]]\},
\label{total6}
\end{eqnarray}
but now with $\rho(t_0)$ instead of $\rho(\tau)$ inside the integral term. Taking the derivative of Eq. (\ref{total6}) with respect to $t$, Eq. (\ref{total5}) is again obtained, where it has been used that in the term of the order $g^2$ of Eq. (\ref{total6}), $\rho_s(t_0)$ can be replaced by $\rho_s(t)$, so that the discarded terms are of a higher order than $g^2$. In summary, the assumptions over the time-scale hierarchy ($\tau_{c}\ll T_S$) are related to the weak coupling limit ($g\ll 1$)\footnote{Indeed, when $g\rightarrow 0$, $T_S\rightarrow\infty$ and the condition $\tau_{c}\ll T_S$ is more easily fulfilled.}.
In order to obtain Eq. (\ref{total5}) we considered an initial condition of the form (\ref{uncorrelated}), with $\rho_B$ fulfilling the property (\ref{odd}). More general initial conditions are studied in Sec. \ref{initcorr}. 

A more specific form for the master equation can be obtained by replacing in Eq. (\ref{total5}) the general $H_I$ given by Eq. 
(\ref{chapuno12}),
so that $V_t H_I =\sum_\eta V_t \left\{S_\eta B_\eta\right\}=\sum_\eta V_t S_\eta V_tB_\eta$, with $V_t$ specified in Eq. (\ref{defVs}). 
In that way, we get
\begin{eqnarray}
&&\frac{d\rho_s (t)}{dt}=
-\sum_{\gamma,\eta}\int^t_0 d\tau C_{\gamma\eta}(t-\tau)[V_t S_\gamma ,V_{\tau} S_\eta \rho_s (t)]\cr
&-&\sum_{\gamma,\eta}\int^t_0 d\tau C^*_{\gamma\eta}(t-\tau)[\rho_s (t) V_{\tau} S_\eta, V_t S_\gamma],
\label{mborn}
\end{eqnarray}
where we set $t_0=0$ and defined 
\begin{eqnarray}
C_{\gamma \eta}(\tau)&=&\Ttr_B\{V_t B_\gamma V_{t-\tau} B_\eta \rho_B \}\nonumber\\
C_{\gamma \eta}(-\tau)&=&C^*_{\gamma \eta}(\tau)=\Ttr_B\{V_{t-\tau} B_\eta V_t B_\gamma \rho_B\},
\label{correla37}
\end{eqnarray}
using the cyclic property of the trace, and considering $C^*$ as the complex conjugate of $C$.

For the choice (\ref{chapuno30}) and (\ref{chapuno31}) of coupling operators, and considering that the environment is in a thermal state $\rho_B=\rho_B^{\eeq}$ given by Eq. (\ref{thermal_B}), we find that the correlation functions in Eq. (\ref{mborn}) combine as $\alpha^{\pm}(t)=2\left[C_{11}(t)\pm iC_{21}(t)\right]$ \cite{devega2005a}, with 
\begin{eqnarray}
\alpha^- (t-\tau)=\sum_\lambda g^2_\lambda 
[n(\omega_\lambda)+1]e^{-i\omega_\lambda (t-\tau)},
\label{icc19b}
\end{eqnarray}
and
\begin{eqnarray}
\alpha^+ (t-\tau)=\sum_\lambda g^2_\lambda n(\omega_\lambda)e^{i\omega_\lambda (t-\tau)},
\label{icc13}
\end{eqnarray}
The function $n(\omega_\lambda)=[\exp(\omega_\lambda\beta)\mp 1]^{-1}$ is the average thermal number of quanta in the mode $\omega$ corresponding to bosonic ($-$) and fermionic ($+$) reservoirs.
In terms of these, Eq. (\ref{mborn}) can be expressed as \cite{yu1999}
\begin{eqnarray}
\frac{d\rho_s (t)}{dt}&=&
\int_0^t d\tau \alpha^{+}(t-\tau) [V_{\tau} L^\dagger \rho_s (t) ,V_{t}L]\nonumber\\
&+&\int_0^t d\tau\alpha^{-}(t-\tau)[V_{\tau}L\rho_s (t) ,V_{t}L^{\dagger}]+\Hhc\nonumber\\
\label{master}
\end{eqnarray}
Note that for zero temperature, $n(\omega_\lambda)=0$, so that $\alpha^+(t-\tau)=0$ and $\alpha^- (t-\tau)$ becomes equal to Eq. (\ref{micro2}), and the master Eq. (\ref{master}) is further simplified.
Yet a further simplification can be obtained when $L=L^\dagger$, so that the terms in Eq. (\ref{master}) combine in such a way that the resulting equation depends only on the correlation function 
$\alpha_T (t-\tau)$ defined in Eq. (\ref{chapuno33bb}).

Although the former master equation is valid only up to $g^2$, its form already suggests the result of the thermofield approach proposed by \textcite{bargmann1961,araki1963,takahashi1975} [see \cite{blasonebook} for a review], \textit{\textit{i.e.}} that a thermal environment can be expressed as two different environments at zero temperature. This can be formally described by introducing an auxiliary environment with operators $c_\lambda$ and $c_\lambda^\dagger$, so that the total Hamiltonian can be rewritten as $H=H_{{\ttot}}-\sum_\lambda \omega_{\lambda} c^\dagger_\lambda c_\lambda$, with $H_{{\ttot}}$ given by Eq. (\ref{chapuno32}), and considering as initial state $|\Omega_0\rangle\propto e^{-\beta H_B/2}|I\rangle$, with $|I\rangle=\sum_n|n\rangle_a |n\rangle_c$. This is the maximally entangled state between the real and the auxiliary environments, defined in terms of their energy eigenstates, $|n\rangle_a$, $|n\rangle_c$, and it is such that the reduced state of each environment (physical $B$, and auxiliary $C$) is a thermal state, $\Ttr_{C}\{|\Omega_0\rangle\langle\Omega_0|\}=\Ttr_B\{|\Omega_0\rangle\langle\Omega_0|\}=\rho_B^{\eeq}$. Then, a thermal Bogoliubov transformation is considered 
\bea
&&a_{1\lambda}=e^{-iG}b_\lambda e^{iG}=\cosh(\theta_\lambda )a_\lambda-\sinh(\theta_\lambda )c^\dagger_\lambda\cr
&&a_{2\lambda}=e^{-iG}c_\lambda e^{iG}=\cosh(\theta_\lambda )c_\lambda-\sinh(\theta_\lambda ) a^\dagger_\lambda,
\eea
where $G=i\sum_\lambda\theta_\lambda (a_\lambda^\dagger c^\dagger_\lambda-c_\lambda a_\lambda)$, with $\theta_\lambda $ a function of the temperature such that $\cosh(\theta_\lambda )=\sqrt{1+n(\omega_\lambda)}$. The transformed Hamiltonian has the form \cite{diosi1998,yu2004,devega2015}
\begin{eqnarray}
&&\tilde{H}_{{\ttot}}=H_S +\sum_{\lambda} \omega_{\lambda} \large(a_{1\lambda}^\dagger a_{1\lambda}-a^\dagger_{2\lambda}a_{2\lambda}\large)\cr
&+&\sum_{\lambda}g_{1\lambda} (L^\dagger a_{1\lambda}+a^\dagger_{1\lambda}L)+\sum_{\lambda} g_{2\lambda}(L a_{2\lambda}+a^\dagger_{2\lambda}L^\dagger)\quad
\label{h3}
\end{eqnarray}
where $g_{1\lambda}=g_\lambda \cosh(\theta_{k})$ and $g_{2\lambda}=g_\lambda \sinh(\theta_{\lambda})$.
The key point is that the transformed initial state, known as the thermal vacuum, $|\Omega\rangle=e^{-iG}|\Omega_0\rangle$, is the vacuum for the transformed modes, $a_{1\lambda}|\Omega\rangle=a_{2\lambda}|\Omega\rangle=0$. 
Hence, solving the dynamics of the initial problem, given by the Hamiltonian (\ref{chapuno32}) with an initial condition $\rho_0^{\ttot}=\rho^S_0\otimes\rho_B$, is equivalent to solving the dynamics with (\ref{h3}), but with an initial condition $\rho_0^{\ttot}=\rho^S_0\otimes|\Omega\rangle\langle \Omega|$.
Thus, the thermofield approach permits one to treat a thermal state of the environment as a vacuum state (\textit{i.e.} the thermal vacuum) of two transformed environments, which therefore does not contain any initial excitation. This enables the use of the SSEs derived for an environment in the vacuum state to describe thermal environments \cite{diosi1998,yu2004}, and gives rise to a potentially better scaling of the basis dimension needed for exact numerical calculations such as matrix product states (MPS) \cite{devega2015}.


\subsubsection{Markov limit and secular approximation of the weak coupling master equation}
\label{thermalization}
When $V_{t}L$ evolves very slowly in time as compared to the environment correlation time, the integration limits in Eq. (\ref{mborn}) [and also in (\ref{master})] can be extended to infinity, leading to a Markovian master equation also referred as the Redfield master equation \cite{redfield1957,redfield1965}. Nevertheless, in general this equation does not generate a dynamical semigroup and therefore does not guarantee CP \cite{davies1974,dumcke1979}. 

To get this property, and thus obtain an equation in the Lindblad form, the secular approximation has to be considered. Following the discussion in \cite{breuerbook} [see also \cite{rivas2011a}], the interaction Hamiltonian (\ref{chapuno12}), written as an interaction picture, can be expanded as
\bea
V_tH_I=\sum_{\eta,\omega} e^{-i\omega t} S_\eta(\omega)B_\eta(t),
\label{spectral}
\eea
where we considered the spectral decomposition of the system operators 
\bea
S_\eta(\omega)=\sum\limits_{\substack{n,n' \\ \epsilon_{n}-\epsilon_{n'}=\omega}}\Pi(\epsilon_n)S_\eta\Pi(\epsilon_{n'}), \nonumber
\eea
where $\Pi(\epsilon_n)$ represents a projection onto the eigenspace belonging to the eigenvalue $\epsilon_n$ of $H_S$, which is assumed to have a discrete spectrum. Also, $S_\eta^+(\omega)=S_\eta(-\omega)$. Previously considering a change of variable $t-\tau\rightarrow \tau$, Eq. (\ref{mborn}) can be rewritten in terms of these quantities as
\bea
&&\frac{d\rho_s(t)}{dt}=\sum_{\eta\gamma}\sum_{\omega\omega'}e^{i(\omega'-\omega)t}\Gamma_{\gamma\eta}(\omega)\bigg(S_{\eta}(\omega)\rho_s(t) S^+_\gamma(\omega')\cr
&-&S_\gamma^+(\omega')S_{\eta}(\omega)\rho_s(t)\bigg)+\Hhc,
\label{master1}
\eea 
where we defined $\Gamma_{\gamma\eta}(\omega)=\int_0^\infty d\tau e^{i\omega \tau}C_{\gamma\eta}(\tau)$. If the spectrum of the system Hamiltonian, $H_S=\sum_n \epsilon_n| n\rangle\langle n|$ is nondegenerate, and the typical value for $|\omega-\omega'|^{-1}$ defines a time scale that is much smaller than the dissipation time scale, the terms in Eq. (\ref{master1}) with $\omega\neq\omega'$ lead to a vanishing contribution in the equation, and can be discarded following the secular approximation. As discussed earlier, this approximation is similar to the rotating wave approximation in quantum optics. The resulting equation is in the Lindblad form, with corrected system Hamiltonian $\hat{H}_S=H_S+\sum_\omega \Delta_{\gamma\eta}(\omega)S^\dagger_\eta S_\gamma$, and a dissipative term ${\mathcal L}\rho_s=\sum_{\eta\gamma}\sum_{\omega}\tilde{\gamma}_{\gamma\eta}(\omega)[S_{\eta}(\omega)\rho_s S^+_\gamma(\omega)-(1/2)\{S_\gamma^+(\omega)S_{\eta}(\omega),\rho_s\}]$, with $\Delta_{\gamma\eta}(\omega)=\Iim\{\Gamma_{\gamma\eta}(\omega)\}$ the Lamb shift, and $\tilde{\gamma}_{\gamma\eta}(\omega)=\Rre\{\Gamma_{\gamma\eta}(\omega)\}$. 

In addition, since $\rho_B$ is a thermal equilibrium state, the correlations (\ref{correla37}) follow the Kubo-Martin-Schwinger condition, and therefore can be written as $C_{\gamma\eta}(t)=C_{\eta\gamma}(-t-i\beta)$. This emerges from the property $n(\omega)+1=e^{\beta\omega}n(\omega)$, and leads to $\tilde{\gamma}_{\gamma\eta}(-\omega)=\int_{-\infty}^\infty d\tau e^{-i\omega\tau}C_{\gamma\eta}(\tau)=\tilde{\gamma}_{\eta\gamma}(\omega)e^{-\beta\omega}$. This, together with the properties $\rho^{\eeq}_S S_\eta(\omega)=e^{\beta\omega}S_\eta(\omega)\rho^{\eeq}_S$ and $\rho^{\eeq}_S S^+_\eta(\omega)=e^{-\beta\omega}S^+_\eta(\omega)\rho^{\eeq}_S$, can be used  to prove that the thermal state $\rho^{\eeq}_S$, given in Eq. (\ref{boltzman}) cancels the rhs of the Markovian master Eq. (\ref{master1}) obtained after the secular approximation, and therefore is a steady state of this equation [see for instance \cite{breuerbook}]. Note that as discussed in Sec. \ref{asym}, the uniqueness of such steady state, and thus its independence of the initial state, depends on whether the corresponding map is relaxing or not.

Equation (\ref{master1}), together with the secular approximation, 
gives rise to a closed equation of motion for the populations $P(n,t)=\langle n|\rho_s(t)|n\rangle$ with a similar form as the rate equation (\ref{rateeq}), $\frac{dP(n,t)}{dt}=\sum_m[W(n|m)P(m,t)-W(m|n)P(n,t)]$ \cite{breuerbook}. This equation is now governed by two types of rates, $W(n|m)=\sum_{\gamma,\eta}\tilde{\gamma}_{\gamma\eta}(\epsilon_m-\epsilon_n)\langle m|S_\gamma|n\rangle\langle n|S_\eta|m\rangle$, and $W(m|n)$, defined similarly. From the Kubo-Martin-Schwinger condition discussed previously, the detailed balance condition follows $W(m|n)e^{-\beta \epsilon_n}=W(n|m)e^{-\beta\epsilon_m}$, which leads to the conclusion that the equilibrium populations $P^{\sst}(n)$ follow the Boltzmann distribution $P^{\sst}(n)\sim e^{-\beta\epsilon_n}$. 

This rough picture of spontaneous emission is equivalent to the one that follows from the Fermi golden rule\cite{atomphotoninteractions,quantumoptics,woldeyohannes2003}. This rule determines that the spontaneous emission rate corresponding to a process driving the system from an initial state to a final state with an energy difference $\omega$ is just given by $\Rre[\Gamma_{\gamma\eta}(\omega)]$. 

\begin{figure}[ht]
\centerline{\includegraphics[width=0.35\textwidth]{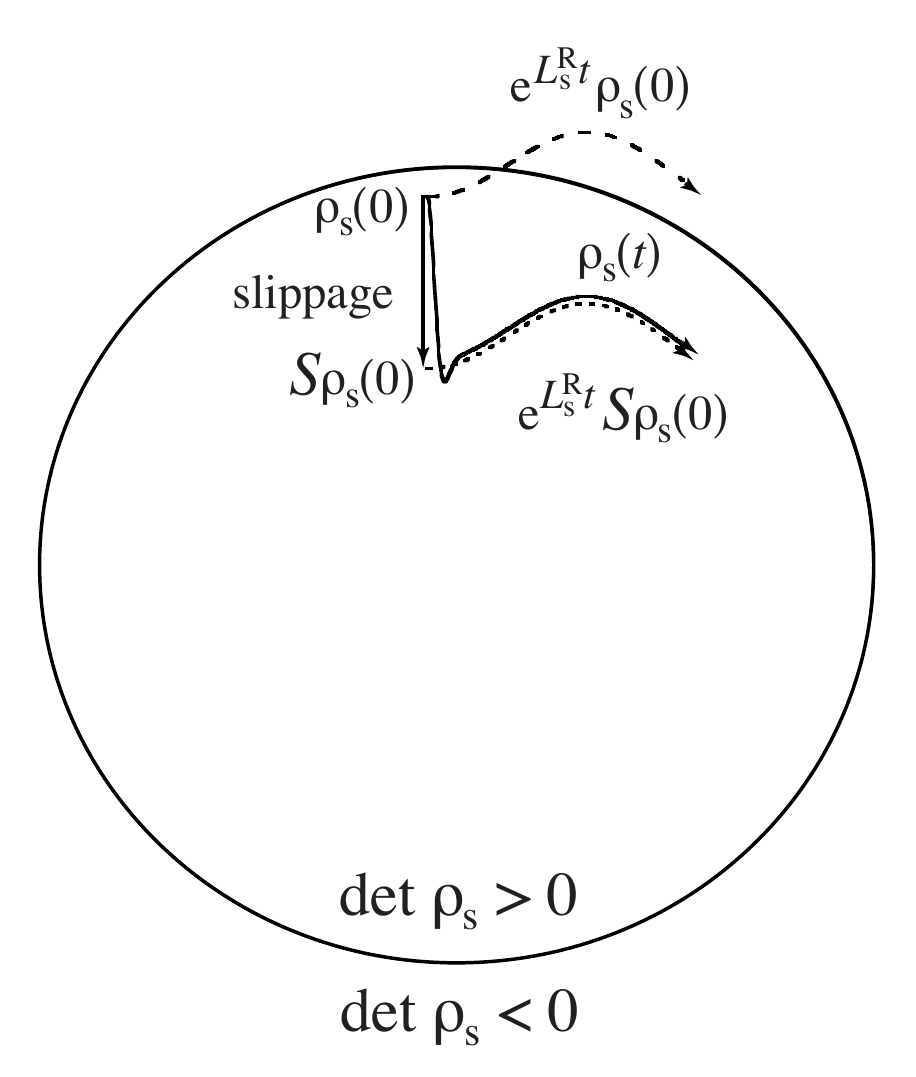}}
\caption{The space of the reduced matrices $\rho_s$ for a two-level system is divided into the set of admissible density matrices for which $\ddet(\rho_s)\ge 0$ so that all the eigenvalues of $\rho_s$ define non-negative probabilities, and the set of non-admissible density matrices for which $\ddet(\rho_s)<0$. From \textcite{gaspard1999a}. \label{slippage}}
\end{figure}


As discussed previously, without the secular approximation, the Redfield equation cannot in general be written in the Lindblad form and thus does not preserve positivity. For the case of a two-level system, the breaking of positivity is related to having initial conditions near the border of the space of physically admissible density matrices, \textit{i.e.} when $\ddet(\rho_s (0))\geq 0$, but very close to $0$\footnote{A density matrix should obey three properties: (i) $\Ttr\{\rho\}=1$, (ii) $\rho=\rho^\dagger$, and (iii) $\langle u|\rho|u\rangle\ge 0$ for any state $|u\rangle$. From the decomposition $\rho_s=\sum_n\lambda_n|\phi_n\rangle\langle\phi_n|$ in terms of eigen-states $|\phi_n\rangle$, these can be rewritten as: (i) $\sum_n\lambda_n=1$, (ii) $\lambda_n$ real, and (iii) $\lambda_n\ge 0$, which implies that $\det(\rho_s)=\prod_n\lambda_n\ge 0$. For a two-level system, a trace-preserving and Hermitian density matrix with $\det\rho_s\ge 0$ also fulfills (iii), since both eigenvalues cannot be negative, and thus they should both be positive.}. This issue occurs because the non-Markovian effects that happen at the initial stage of the evolution are not being taken into account when the integral limits of equation (\ref{master}) are extended to infinity. The application of a slippage (\textit{i.e.} a displacement) of initial conditions, first suggested by \textcite{suarez1992} for the case of a spin-boson model and then extended by \textcite{gaspard1999a} for general systems, appears to solve this problem, at least within the domain of the weak coupling approximation (see Fig. \ref{slippage}).

In simple cases, a relationship can be established explicitly between the correlation time $\tau_c$, the weak coupling parameter $g$, and the maximum time $t_m$ up to which the evolution calculated with the second-order weak coupling approximation gives rise to a positive density matrix. This is calculated by formally solving the evolution equation of the populations up to the second-order, and calculating the maximum time $t_m$ up to which they are still positive. For a two-level system with $H_S=\omega_{12}\sigma_z$ coupled to the zero-temperature reservoir, this relation can be simply written as $1/g^2=2\int^{t_{m}}_{0}dl\int^l_{0} d\tau \Re[\hat{\alpha}(l-\tau)]$, with $\hat{\alpha}(t-s)=e^{-i\omega_{12}(t-s)}\alpha(t-s)$. Considering a simple exponentially decaying correlation $\alpha(t)=\exp(-\Gamma t)$, the limiting condition is just $t_m =1/\tau_c g^2$. 

\subsubsection{Coarse-graining approach to weak coupling master equations}
An interesting alternative to derive a second-order master equation is the coarse-graining approach discussed in \cite{alicki1989,schaller2008,benatti2009}. Indeed, the formal solution of Eq. (\ref{total1}) can be written as $\rho(t+\tau)={\mathcal W}(t+\tau,t)\rho(t){\mathcal W}^{\dagger}(t+\tau,t)$, with ${\mathcal W}(t+\tau,t)={\mathcal T}\exp(-i\int_t^{t+\tau} dt_1V_{t_1}H_I)$, and ${\mathcal T}$ a time-ordering operator. 
Following \textcite{schaller2008}, one can perform the second-order perturbative expansion of ${\mathcal W}(t+\tau,t)$, and replace the result back in the above definition of $\rho(t+\tau)$. Then, truncating at second-order and considering the Born approximation, we find
\begin{eqnarray}
&&\rho_s(t+\tau)=\rho_s(t)-\frac{1}{2}\sum_{\gamma\eta}\int_t^{t+\tau}dt_1\int_t^{t+\tau}dt_2\cr
&\times&C_{\gamma\eta}(t_2-t_1)\ssign(t_2-t_1)[V_{t_2}S_\gamma V_{t_1}S_\eta,\rho_s(t)]\cr
&+&\sum_{\gamma\eta}\int_t^{t+\tau}dt_1\int_t^{t+\tau}dt_2C_{\gamma\eta}(t_2-t_1)\cr
&\times&\bigg(V_{t_1}S_\eta\rho_s(t) V_{t_2}S_\gamma-\frac{1}{2}\{V_{t_2}S_\gamma V_{t_1}S_\eta,\rho_s(t)\}\bigg)\cr
&\equiv&\rho_s(t)+\tau{\mathcal L}_c^\tau(t)\rho_s(t),
\label{total6_benatti}
\end{eqnarray}
where ${\mathcal L}_c^\tau(t)$ represents the Liouville super operator specified above. Then, provided that $\tau g^2$ is small with respect to the time scale where $\rho_s(t)$ varies, we can replace $\frac{\rho_s(t+\tau)-\rho_s(t)}{\tau}\approx \partial_t\rho_s(t)$. Therefore, Eq. (\ref{total6_benatti}) can be rewritten as $\partial_t\rho_s(t)={\mathcal L}_c^\tau(t)\rho_s(t)$. As discussed in the previous references, the coefficients of this equation are positive for any $\tau\ge 0$. Without the need to invoke the secular approximation, the resulting equation has the Lindblad form and thus preserves complete positivity. Also, the Lindblad equation which is obtained after applying the secular approximation in Eq. (\ref{master1}) is automatically recovered in the limit $\tau\rightarrow\infty$.

\subsubsection{Weak-coupling master equations for time-dependent system Hamiltonians}

The characterization of OQSs additionally subject to a time-dependent perturbation such that the system Hamiltonian is time-dependent, is a long-standing topic \cite{davies1978,alicki1979}. Indeed, a rigorous derivation of Eq. (\ref{master}) [and hence (\ref{mborn})], using either projection methods or a perturbative expansion, leads one to conclude that it is also valid for time-dependent $H_S$ \cite{breuer2004a,sarandy2005,amin2008,amin2009,devega2010}. Back in Schr\"odinger picture, Eq. (\ref{master}) can be written, for a time dependent system Hamiltonian $H_S$ as
\begin{eqnarray}
\frac{d\rho_s (t)}{dt}&=&-i[H_S(t),\rho_s(t)]\cr
&+&\int_0^t d\tau \alpha^{+}(t-\tau) [V_{\tau-t} L^\dagger \rho_s (t),L]\nonumber\\
&+&\int_0^t d\tau\alpha^{-}(t-\tau)[V_{\tau-t}L\rho_s (t) ,L^{\dagger}]+\Hhc\nonumber
\label{master_tdep}
\end{eqnarray}
However, a practical use of this equation requires the ability to rewrite it on a system basis. 
In detail, the master equation depends on system operators with the form $V_{-\tau}L^\dagger={\mathcal U}^\dagger_s(-\tau)L^\dagger{\mathcal U}_s(-\tau)$, with
\bea
{\mathcal U}_s(t)={\mathcal T} e^{-i \int_0^t H_S(\tau) d\tau}, 
\label{evoltime}
\eea
and ${\mathcal T}$ the usual time-ordering operator. Such evolution operators should be expressed in terms of the time-dependent eigen-states of the system, $|n(t)\rangle$, corresponding to the set of eigen-values $E_n(t)$ that diagonalize instantaneously $H_S(t)$.

A way to avoid this is to eliminate the explicit time dependence of $H_S(t)$. This can be done for example when the system is subject to a time-dependent perturbation that is periodic in time, such as an atom in a laser field, and the latter is considered in the semiclassical limit. In this limit, the free part of the Hamiltonian can be written as $H_0=H_B+H_S+\hbar\epsilon(\sigma^+e^{-i(\omega_Lt+\phi_T)}+\sigma^-e^{i(\omega_Lt+\phi_T)})$, where $\omega_L$ and $\epsilon=d_{21}{\mathcal E}$ are the laser frequency and the Rabi frequency respectively, ${\mathcal E}$ is the laser electrical field magnitude, and the factor $\phi_T=\phi_L-\pi/2$ groups all phase contributions. Hence, a unitary can be considered to transform the system into a rotating frame of reference with respect to the laser, which effectively leads to a time-independent Hamiltonian (see for instance  \cite{law1991,quantumoptics,florescu2001}). Following this, it is often convenient to diagonalize the system part of such Hamiltonian, thus reexpressing it in the well-known dressed basis. 
 
Nevertheless, in general a time-independent form of the Hamiltonian cannot be obtained via a unitary transformation, and thus it is unavoidable to express (\ref{evoltime}) in terms of the system instantaneous eigen-states. This can be done when the system is subject to a general periodic time dependent perturbation, $H_s(t)=H_s^0+H_L(t)$, with $H_L(t+T)=H_L(t)$, and $T=2\pi/\omega_L$ where $\omega_L$ is the driving frequency. The instantaneous basis is then the Floquet basis, obtained from the Floquet eigenvalue problem $(H_S(t)-i\frac{d}{dt})|n(t)\rangle=\epsilon_n|n(t)\rangle$, and obeying periodic boundary conditions in time $|n(t)\rangle=|n(t+T)\rangle$ \cite{breuerbook}. In this basis, we can rewrite (\ref{evoltime}) as
\begin{eqnarray}
{\mathcal U}^{\pper}_s(t,t')=\sum_n e^{-i\epsilon_n(t-t')}|n(t)\rangle\langle n(t')|,
\label{periodicU}
\end{eqnarray}

A second tractable situation is when the system undergoes an exact adiabatic evolution, which allows us to express (\ref{evoltime}) as \cite{mostafazadeh1997,albash2012}
\begin{eqnarray}
{\mathcal U}^{\aad}_s(t,t')=\sum_n e^{-i\mu_n(t,t')}|n(t)\rangle\langle n(t')|,
\label{adiabaticU}
\end{eqnarray}
where $\mu_n(t,t')=\Delta_n(t,t')-\gamma_n(t,t')$, with $\Delta_n(t,t')=\int_{t'}^t dsE_n(s)$, and $\gamma_n(t,t')=i\int_{t'}^t ds\langle n(s)|\frac{d}{ds}|n(s)\rangle$. 
Equation (\ref{adiabaticU}) is valid as long as the adiabatic condition $h\ll \Delta^2 t_f$ is fulfilled. Here $t=t_f$ is the maximum evolution time and $\Delta=\mmin_{t\in[0,t_f]}\left(E_1(t)-E_0(t)\right)$ is the minimum ground state energy gap, with $E_0$ and $E_1$ as the ground and the first excited eigen-energies of $H_S(t)$, and $h=\mmax_{s\in[0,1];n,m}\left|\langle n(s)|\partial_sH_S(s)|m(s)\rangle\right|$, with $s=t/t_f$ as a dimensionless parameter. The adiabatic condition provided earlier can be defined in alternative ways [see for instance \cite{mostafazadeh1997}].

Let us now consider, for instance, a master equation of the form (\ref{mborn}), and reexpress it in terms of system operators in the time-dependent basis for an adiabatic evolution. To this end, following \cite{albash2012}, we need to adiabatically approximate $V_{t-\tau}S_\eta={\mathcal U}^\dagger_s(t-\tau,0) S_\eta {\mathcal U}_s(t-\tau,0)$. Then, we first take into account that ${\mathcal U}_s(t-\tau,0)={\mathcal U}_s(t-\tau,t){\mathcal U}_s(t,0)={\mathcal U}^\dagger_s(t,t-\tau){\mathcal U}_s(t,0)$, and then consider two approximations: first replace ${\mathcal U}_s(t,0)\approx {\mathcal U}^{\aad}_s(t,0)$, and then replace ${\mathcal U}^\dagger_s(t,t-\tau)\approx e^{i\tau H_S(t)}$, which is justified by the fact that this term appears in an integral with a fast-decaying correlation function $C_{\gamma\eta}(\tau)$. With these considerations, we find that ${\mathcal U}_s(t-\tau,0)\approx e^{i\tau H_S(t)}{\mathcal U}^{\aad}_s(t,0)$, such that, for instance, one of the terms of the interaction picture master equation (\ref{mborn}) can be rewritten as
\bea
&&\int_0^t d\tau C_{\gamma\eta}(\tau)V_{t-\tau}S_\eta\rho_s(t)V_tS_\gamma\cr
&\approx & \sum_{nm} \Gamma^{mn}_{\alpha,\eta}(t) e^{-i\mu_{mn}(t,0)}V_tS_{n,m,\eta}\Pi_{nm}(0)\rho_s(t)V_tS_\gamma,\nonumber
\eea
where $\Gamma^{mn}_{\alpha,\eta}(t)=\int_0^t d\tau e^{i\tau E_{mn}(t)}C_{\gamma\eta}(\tau)$, with $E_{mn}(t)=E_{m}(t)-E_n(t)$, $\mu_{mn}(t,t')=\mu_m(t,t')-\mu_n(t,t')$, $\Pi_{nm}(t)=|n(t)\rangle\langle m(t)|$ and $S_{n,m,\eta}(t)=\langle n(t)|S_\eta|m(t)\rangle$. Performing a similar adiabatic approximation to $S_\gamma(t)$, and expressing the other terms of Eq. (\ref{mborn}) in a similar way, we obtain an adiabatic interaction picture master equation of the form
\bea
&&\frac{d\rho_s(t)}{dt}=\sum_{nmpq}e^{-i(\mu_{qp}(t,0)-\mu_{mn}(t,0))}\sum_{\gamma\eta}\Gamma^{mn}_{\gamma\eta}(t)\cr
&\times&V_tS_{p,q,\gamma}V_tS_{n,m,\eta}[\Pi_{nm}(0)\rho_s,\Pi_{pq}(0)]+\Hhc\nonumber
\eea
An excellent review on driven quantum systems, including their dissipation can be found in \cite{grifoni1998}.

\textit{Many-body OQS master equation--}
Following the same procedure as in Sec. \ref{perturbativeM}, it can be concluded that the master equation for a many-body system with Hamiltonian (\ref{ch1collective}) is, up to second-order in the perturbative parameter and back in the Schr\"odinger picture,
\begin{eqnarray}
\frac{d\rho_s (t)}{dt}&=&-i[H_S,\rho_s(t)]\cr
&+&\int_0^t d\tau \sum_{lj} \alpha^{+}_{lj}(t-\tau) [V_{\tau-t } L_j^\dagger \rho_s (t) ,L_l]\nonumber\\
&+&\int_0^t d\tau \sum_{lj} \alpha^{-}_{lj}(t-\tau)[V_{\tau-t}L_j\rho_s (t) ,L_l^{\dagger}]+\Hhc\nonumber\\
\label{icc20}
\end{eqnarray}
where we defined 
\begin{eqnarray}
\alpha_{lj}^- (t)&=&\sum_\lambda g^*_{l\lambda}g_{j\lambda}
(n(\omega_\lambda)+1)e^{-i\omega_\lambda t},\cr
\alpha_{lj}^+ (t)&=&\sum_\lambda g_{l\lambda}g^*_{j\lambda}n(\omega_\lambda)
e^{i\omega_\lambda t}.
\label{alphamas}
\end{eqnarray}
The previous master equation may also represent the evolution of a multilevel OQS with $j=1,\cdots,n$ decaying channels, each of them represented by a coupling operator $L_j$. 

The master equations  (\ref{master}) or (\ref{icc20}) can be written in the canonical form (\ref{nmbreuer}), by just expanding all system coupling operators, $L_j$ in terms of the complete set of basis operators $G_i$, as $V_tL_j=\sum_{\gamma}a_{ij}(t)G_i$. The key is that the time dependency is absorbed into the expansion coefficients $a_{ij}=\Ttr_S\{V_tL_j G_i\}$. 

\subsubsection{Weak coupling master equations for alternative initial conditions and initially correlated states}
\label{initcorr}

In previous sections we analyzed several derivations of master equations for an initially uncorrelated state between the system and the environment and considering that the environment is at thermal equilibrium. However, as discussed in Sec. \ref{initialstate} an experimentally realistic situation is that where the initial state is obtained when the total system is in an equilibrium state [for instance $\rho_{\ttot}(0)\sim e^{-\beta H_{{\ttot}}}$], and a set of projective measurements on the system is performed, resulting in a state to the form (\ref{projective}). This type of initial conditions was discussed by \textcite{grabert1988} for the case of a quantum Brownian particle, when considering the state prepared by measuring a dynamical variable such as the position of the particle $q$. More recently,  \textcite{chaudhry2013} derived a weak coupling master equation from the initial state obtained after a single measure that projects the system to the state $|\psi_0\rangle$. In such a case, the initial state can be written as 
\bea
\rho_{\ttot}(0)=|\psi_0\rangle\langle\psi_0|\otimes \frac{\langle\psi_0|e^{-\beta H_{\ttot}}|\psi_0\rangle}{Z},\nonumber
\eea
where $Z$ is a normalization factor, such that the total trace is preserved. Such initial state may be simplified by considering the Kubo identity, which states that for all operators $A$ and $C$, $e^{\gamma(A+C)}=e^{\gamma A}\large(1+\int_0^\gamma d\lambda e^{-\lambda A}Ce^{\lambda(A+C)}\large)$,
where $\gamma$ is a parameter. Using this expression, it is possible to expand $e^{-\beta H_{\ttot}}$ in different orders of the coupling constant $g$, by considering $\gamma=\beta$, $A=H_0$, and $C=H_I$. At first order, $e^{-\beta H_{{\ttot}}}\approx e^{-\beta H_0}(1-\int_0^\beta d\lambda e^{\lambda H_0}H_Ie^{-\lambda H_0})$, and therefore
\bea
\rho_{\ttot}(0)=|\psi_0\rangle\langle\psi_0|\otimes (\rho_B^{(0)}+\rho_B^{(1)}+\cdots),
\label{rhotot0}
\eea
where $\rho_B^{(0)}=\langle\psi_0|e^{-\beta H_S}|\psi_0\rangle e^{-\beta H_B}$, and $\rho_B^{(1)}=-ge^{-\beta H_B}E(\beta)$, with 
\bea
E(\beta)=\int_0^\beta d\lambda e^{\lambda H_B}B e^{-\lambda H_B}\langle\psi_0|e^{-\beta H_S}e^{\lambda H_S}Se^{-\lambda H_S}|\psi_0\rangle.\nonumber
\eea
Here, the interaction Hamiltonian is considered of the form (\ref{chapuno12}), \textit{i.e.} $H_I=AB$. Inserting (\ref{rhotot0}) in the perturbative expansion (\ref{total6}), and diferentiating, we obtain
\bea
\frac{d\rho_s(t)}{dt}&=&i[\rho_s(t),H_S]-if_{\ccor}(t)[\rho_s(t),S]\cr
&+&\int_0^s ds \large([V_{t-s}S\rho_s(t),S]C(t-s)+\Hhc\large),\nonumber
\eea
where, as usual, $V_{t}S=e^{iH_S t}Se^{-iH_S t}$, and $f_{\ccor}(t)=\Ttr_B\{\rho^{\eeq}_BE(\beta)V_{t}B\}/Z'$, with $Z'=\langle\psi_0|e^{-\beta H_S}|\psi_0\rangle-g\Ttr_B\{\rho^{\eeq}_BE(\beta)\}$. Also, the quantity $C(t-s)$ is defined according to (\ref{correla37}) as $C(t-s)=\Ttr_B\{\rho^{\eeq}_BV_{t}BV_sB\}$. The structure of this master equation preserves the trace and the Hermiticity. However, CP is not ensured. 

In addition, as argued by \textcite{pechukas1994,meier1999,smirne2011,liu2011}, there are situations that are experimentally relevant where the system and the environment are initially correlated. In this regard, \textcite{meier1999} derived a master equation for such correlated initial states, which is based on the Nakajima-Zwanzig projection-operator approach, discussed in the next section, up to second-order in the system-bath interaction. \textcite{chen2016} also extended this analysis to investigate under which conditions the initial factorization approximation of the system-environment state is valid. Another method for tackling this problem is the correlated projection operator, also discussed in the next section. In addition, the reduced hierarchical equations of motion of Sec. \ref{hierarchy} have been extended to deal with correlated initial conditions \cite{tanimura2014}. A stochastic propagation similar to those discussed in Sec. \ref{SSE} can also be considered, based on expressing the initial state in a Bargmann coherent state representation for the environment \cite{devega2006}. Based on this result, a master equation for general initial conditions have been derived by \textcite{halimeh2016}.
 
%

Other initial conditions correspond to the OQS being coupled to an environment that is initially in a squeezed state. A master equation to describe such systems was derived by \textcite{gardiner1986} for the Markovian case. This was recently extended to non-Markovian interactions by \textcite{manirul2010} to study bipartite entanglement dynamics in the presence of dissipation.

\subsubsection{Projection techniques}
\label{projectionM}

In the projection-operator techniques, a projection superoperator ${\mathcal P}$ is defined such that ${\mathcal P}\rho$ captures the relevant part of the total density matrix $\rho\equiv \rho^I_{\ttot}$, which offers an approximate description of the OQS dynamics, while the irrelevant part ${\mathcal Q}\rho$ is defined by the complementary superoperator ${\mathcal Q}=\unit-{\mathcal P}$, with $\unit$ denoting the unit map. The projection superoperator should: (i) be a linear map $\rho\rightarrow{\mathcal P}\rho$ which takes any operator of the total state space ${\mathcal H}$ to an operator ${\mathcal P}\rho$ of ${\mathcal H}$; (ii) have the properties
\begin{eqnarray}
&&{\mathcal P}^2={\mathcal P}={\mathcal P}^{\dagger},\ \ {\mathcal Q}^2={\mathcal Q}={\mathcal Q}^{\dagger} \nonumber\\
&&{\mathcal Q}{\mathcal P}={\mathcal P}{\mathcal Q}=0, \ \ {\mathcal P}+{\mathcal Q}=\unit;
\label{F22}
\end{eqnarray}
and (iii) be such that $\rho_s=\Ttr_B\{\rho\}=\Ttr_B{{\mathcal P}\rho}$. 

In order to obtain a dynamical equation for ${\mathcal P\rho(t)}$, there are basically two different possibilities \cite{breuer2004}. The first is to follow the Nakajima-Zwanzig method \cite{nakajima1958,zwanzig1960}, which leads to an equation for ${\mathcal P}\rho$ that contains a time integration over the past history of the system. This equation reads as 
\bea
&&\frac{d}{dt}{\mathcal P}\rho(t)=\int_0^t ds {\tilde K}(t,s){\mathcal P}\rho(s)
+g{\mathcal P}{\mathcal L}_{\ttot}(t){\mathcal G}(t,t_0)Q\rho(t_0)\cr
&+&g{\mathcal P}{\mathcal L}_{\ttot} (t){\mathcal P}\rho(t)
\label{NZmaster}
\eea
where ${\mathcal L}_{\ttot}$ is the Liouvillian corresponding to the von-Neumann equation for the total density operator $\rho(t)$,$\frac{d\rho(t)}{dt}=-i[V_tH_I,\rho(t)]={\mathcal L}_{\ttot} (t)\rho(t)$. Also, we defined the memory kernel as
\bea
\tilde{K}(t,s)=g^2{\mathcal P}{\mathcal L}_{\ttot} (t){\mathcal G}(t,s){\mathcal Q}{\mathcal L}_{\ttot}(s),
\eea
and
\bea
{\mathcal G}(t,s)=T_{\leftarrow}\exp{\bigg(g\int_0^tds'{\mathcal Q}{\mathcal L}_{\ttot}(s')\bigg)}.
\eea
Here, $T_{\leftarrow}$ denotes the chronological time ordering, which orders any product of superoperators such that the time arguments increase from right to left. Also, this quantity satisfies the evolution equation $\frac{d{\mathcal G}(t,s)}{dt}=gQ{\mathcal L}_{\ttot} (t){\mathcal G}(t,s)$, with ${\mathcal G}(s,s)=1$.
As noted by \textcite{breuer1999,breuerbook}, (\ref{NZmaster}) is an exact equation and therefore its resolution is as difficult as the resolution of the original von-Neumann equation. 
Nevertheless, it provides a good starting point for considering different simplifications and approximations. 

For instance, for a factorizing initial condition, ${\mathcal P}\rho(t_0)=\rho(t_0)$, such that ${\mathcal Q}\rho(t_0)=0$, the second term of Eq. (\ref{NZmaster}) vanishes. Equation (\ref{NZmaster}) can be further simplified by assuming that, in general, any string containing an odd number of ${\mathcal L}_{\ttot}$ between factors of ${\mathcal P}$ vanishes 
\bea
{\mathcal P}{\mathcal L}_{\ttot}(t_1){\mathcal L}_{\ttot}(t_2)\cdots {\mathcal L}_{\ttot}(t_{2n+1}){\mathcal P}=0.
\label{magicmoment}
\eea
This means that the term ${\mathcal P}{\mathcal L}_{\ttot}(t){\mathcal P}=0$ and the last term of (\ref{NZmaster}) vanishes too. The resulting equation, can be rewritten as a time-local equation \cite{hall2014}. In this regard, considering that it describes an evolution process given by a linear map $\rho_s(t)=\Lambda(t)\rho_s(0)$, which is invertible, such that $\Lambda(t)^{-1}\Lambda(t)=\unit$, we find
\bea
\frac{d\rho_s(t)}{dt}=\int_0^t ds {\tilde K}(t,s)\rho_s(s)={\mathcal L}(t)\rho_s(t),
\label{timeconvolutionless}
\eea
where we defined ${\mathcal L}(t)=\int_0^t ds {\tilde K}(t,s)\Lambda(s)\Lambda(t)^{-1}$. 

Finally, the memory kernel ${\tilde K}(t,s)$ can be expanded in terms of the weak coupling parameter between system and environment. For instance, up to second-order in $g$, we can simply consider that $\tilde{K}(t,s)=g^2{\mathcal P}{\mathcal L}_{\ttot} (t){\mathcal Q}{\mathcal L}_{\ttot}(s)+{\mathcal O}(g^3)$. 

A second possibility for solving the dynamical equation of ${\mathcal P}\rho(t)$ is the time-convolutionless projection-operator technique (TCL), which departs from (\ref{NZmaster}) to derive an equation that is local in time \cite{kubo1963,royer1972,chaturvedi1979} and has the general form \cite{breuer2006b,breuerbook}
\bea
\frac{d}{dt}{\mathcal P}\rho(t)={\mathcal K}(t){\mathcal P}\rho(t)+{\mathcal J}(t){\mathcal Q}\rho(t_0).
\label{TCLmaster}
\eea
Here, we defined 
\bea
{\mathcal K}(t)=g{\mathcal P}{\mathcal L}_{\ttot}(t)[1-\Sigma(t)]^{-1}
\label{kern}
\eea
with $\Sigma(t)=g\int_0^t ds {\mathcal G}(t,s){\mathcal Q}{\mathcal L}_{\ttot}(s){\mathcal P}G(t,s)$. We also considered the backward propagator of the total system as 
\bea
G(t,s)=T_{\rightarrow}e^{-g\int_s^t ds'{\mathcal L}_{\ttot}(s')},
\eea
with $T_{\rightarrow}$ as the antichronological time-ordering operator. 
The operator $[1-\Sigma(t)]^{-1}$ written above can be expressed as 
\bea
[1-\Sigma(t)]^{-1}=\sum^\infty_{n=1}\Sigma(t)^n.
\label{expan}
\eea
Inserting Eq. (\ref{expan}) in (\ref{kern}), it is possible to rewrite this term as a perturbative expansion in $g$
\bea
{\mathcal K}(t)=g\sum^\infty_{n=1}{\mathcal P}{\mathcal L}_{\ttot}(t)\Sigma(t)^n= \sum^\infty_{n=1}g^n{\mathcal K}_n(t).
\label{TCLexp}
\eea
Following the cumulant expansion approach by \cite{kubo1963,royer1972,kampen1974a,kampen1974b} of the equation for ${\mathcal P}\rho(t)$, the n-th order coefficient can be defined as \cite{breuer2006b,breuerbook} 
\bea
&&{\mathcal K}_n(t)=\int_0^tdt_1\int_0^{t_1}dt_2\cdots \int_0^{t_{n-2}}dt_{n-1}\cr
&\times&\langle {\mathcal L}_{\ttot}(t){\mathcal L}_{\ttot}(t_1){\mathcal L}_{\ttot}(t_2)\cdots {\mathcal L}_{\ttot}(t_{n-1})\rangle_{oc}.
\label{expansionk}
\eea
Here, $\langle {\mathcal L}_{\ttot}(t){\mathcal L}_{\ttot}(t_1){\mathcal L}_{\ttot}(t_2)\cdots {\mathcal L}_{\ttot}(t_{n-1})\rangle_{oc}\equiv \sum (-1)^q{\mathcal P}{\mathcal L}_{\ttot}(t)\cdots{\mathcal L}_{\ttot}(t_i){\mathcal P}{\mathcal L}_{\ttot}(t_j)\cdots{\mathcal L}_{\ttot}(t_k){\mathcal P}\cdots{\mathcal P}$ are ordered cumulants. These are built by inserting a number $q$ of ${\mathcal P}$s between one or more ${\mathcal L}_{\ttot}$, and then summing over all possible $q$. The first ${\mathcal L}_{\ttot}$ should be evaluated at time $t$, and the others may carry any permutation of time arguments, with the restriction that these shall be chronologically ordered between two successive ${\mathcal P}'s$. Note that because of (\ref{magicmoment}), the odd moments $n$ vanish. 

The expansion (\ref{TCLexp}) can always be assumed, provided that the map is continuous and with a zero initial condition $\Sigma(t_0)=0$. However, a practical use of such an expansion requires that it is truncated at relatively low orders [see \cite{breuerbook} for the explicit expression of the first few terms of the expansion], which may be accurate only at short times and within the weak coupling regime. Also, after truncation, complete positivity is no longer guaranteed. 
Higher orders lead to increasingly complex equations, and to a solution that might be more exact at short times, but still fails at long times \cite{breuer2004}. In this regard, an optimal choice of the projection-operator $P$ is of primary importance, such that the first few terms of the expansion accurately reproduce the OQS dynamics. The choice should therefore be motivated by the specific characteristics of the problem. 
In the following, we discuss the two standard approaches described in the literature to choose the projection operator, namely, the standard approach and the correlated projection-operator approach. 

In the \textit{standard approach} \cite{breuerbook}, the projection superoperator is chosen such that ${\mathcal P}\rho=\rho_s(t)\otimes\rho_B$, where $\rho_s(t)=\Ttr_B\{\rho(t)\}$. This superoperator satisfies the conditions (i)-(iii), and furthermore is suitable for those problems in which system-environment correlations are small both initially and during the evolution, so that they can be treated as small perturbations of the reduced density matrix. With this choice, the equation convoluted equation (\ref{NZmaster}) with factorized initial conditions leads to Eq. (\ref{total5}), but with $\rho_s(s)$ within the integral in the rhs term. In the convolutionless technique, the second-order term of the expansion (\ref{TCLexp}), ${\mathcal K}_2(t)$, leads to the time-local master equation (\ref{total5}). Both convoluted and convolutionless equations are equivalent in this order, since the reduced density matrix is already in a second-order term, and hence we can replace $\rho_s(s)\approx \rho_s(t)+{\mathcal O}(g^2)$. However, the convoluted and the convolutionless equations at the same order lead to completely different solutions that may differ with each others in all orders of the coupling. A comparison between these two perturbative schemes with respect to the exact solution for a two-state system in an environment with $T=0$ (discussed in Sec. \ref{Exact1}) can be found in \cite{vacchini2010}.

An alternative to the standard approach is the \textit{correlated projection superoperator technique} formalized   \textcite{breuer2007,breuer2006b}, which considers the relevant part of the dynamics as a correlated system-environment state, rather than a tensor product state $\rho_s(t)\otimes\rho_B$. This second approach is naturally adapted to those situations in which system and environment states are non-negligibly correlated initially and/or during the dynamics. The relevant part of the dynamics is expressed in terms of a positively correlated projection superoperator ${\mathcal P}=\unit_S \otimes\Lambda$, where $\Lambda$ maps operators in ${\mathcal H}_B$ to operators in ${\mathcal H}_B$, and can be represented in terms of environment operators $A_i$ and $B_i$, such that $\Ttr_B\{A_jB_i\}=\delta_{ij}$ \cite{breuer2007}. These operators should fulfil certain properties so that $\Lambda$ is a trace-preserving and completely positive map. In this representation, 
\bea
{\mathcal P}\rho(t)=\sum_i\Ttr_B\{A_i\rho(t)\}\otimes B_i. 
\eea
An example of a projection superoperator is obtained with the choice $A_i=\Pi_i$ and $B_i=\frac{\Pi_i\rho_0\Pi_i}{Z_i}$, where $i=1,\cdots,n$ ($n$ being the total number of operators in the expansion), and $Z_i=\Ttr_B\{\Pi_i\rho_0\}$, and $\Pi_i$ are projection operators on ${\mathcal H}_B$ such that $\Pi_i\Pi_j=\delta_{ij}\Pi_i$, and $\sum_i\Pi_i=\unit_B$, 
\bea
{\mathcal P}\rho(t)=\sum_i \Ttr_B\{\Pi_i\rho\}\otimes\frac{\Pi_i\rho_0\Pi_i}{Z_i},
\label{projectcorr}
\eea
where $\rho_0$ is any fixed environmental density matrix.
The reduced density matrix is described as a sum of a set of unnormalized states $\rho_i(t)$,
\bea
\rho_s(t)=\Ttr_B\{\mathcal P\rho(t)\}=\sum_i\rho_{i}(t),
\eea
that should nevertheless be such that $\Ttr_S\rho_s(t)=1$. The states $\rho_i(t)=\Ttr_B\{\Pi_i\rho(t)\}$ belong to a subspace of the total space ${\mathcal H}$, and reflect correlations between the system and the environment. 
Considering an initial condition of the form $\rho(0)=\sum_i\rho_i(0)\otimes B_i$ and using the TCL technique, a system of equations for each $\rho_i$ is obtained, each with the general form 
\bea
\frac{d}{dt}\rho_i={\mathcal K}_i(t)(\rho_1,\cdots,\rho_n),
\label{generalproj}
\eea
where the time-dependent generators ${\mathcal K}_i(t)$ can be approximated as time-independent ones ${\mathcal K}_i$ following a Markov approximation. Note that while in the standard approach this is linked to the Born approximation at second order, implying zero system-environment correlations, this is not the case in this derivation. 
After this approximation, a generalized Lindblad equation can be obtained \cite{breuer2007,budini2006}, 
\bea
&&\frac{d}{dt}\rho_i=
-i[H^i,\rho_i]+\sum_{j\lambda}\bigg(R_\lambda^{ij}\rho_jR_\lambda^{ij\dagger}-\frac{1}{2}\large\{R_\lambda^{ji\dagger}R_\lambda^{ji},\rho_i\}\bigg),\cr
&&
\label{linbproj}
\eea
which ensures complete positivity. Here $H_i$ and $R_\lambda^{ij}$ are system Hermitian operators. 

This derivation formalizes (and generalizes) the previous derivations by \textcite{esposito2003} and \textcite{budini2005,budini2006} to derive master equations up to second-order in perturbation theory. 
The proposal by \textcite{esposito2003} is based on choosing the projectors in (\ref{projectcorr}) as projectors to environment subspaces corresponding to a given energy, \textit{i.e.} $\Pi_\epsilon$. 
Following this choice, an evolution equation was derived for the quantity $\rho_\epsilon(t)$, based on a weak coupling expansion up to second-order in the coupling parameter between the system and the environment. Despite it is also based on a weak coupling expansion, considering an environment described by Gaussian random matrices this approach gives was shown to give more accurate results than the usual Born approximation ${\mathcal P}\rho=\rho_s(t)\otimes\rho_B$. Here, the projection (\ref{projectcorr}) is made into a large region of the total Hilbert space corresponding to states where the environment has a given energy, and considers the fluctuations in the environment energy states as a non-relevant part of the density matrix, so that they can be neglected. According to this derivation, the reduced density matrix of the system is computed as a sum of all possible environment states, considered as a quasi-continuum, $\rho_s(t)=\int d\epsilon \rho_{\epsilon}(t)$. 

The interesting aspect of the resulting equation is that it takes into account the principle of conservation of the total system energy. Following this principle, when the OQS gains a quantum of energy, this should be lost in the environment and vice-versa. This is in contrast with the Lindblad equation (\ref{Linb}), which is derived under the assumption that despite the coupling with the system, the environment remains in the same energy state. This contradiction with the energy conservation principle is acceptable provided that the environment is sufficiently large compared to the system. In that case, it is possible to assume that the environment quantities do not vary significantly on energy scales of the order of the system energy. Hence, if the environment is initially in a microcanonical state of energy $\epsilon$, it will remain in such an energy state without being much affected by the energy exchange with the system. Any situation beyond this case is more accurately described with the equation proposed in \cite{esposito2003}.

The proposal by \textcite{budini2006} considers a projection of the form (\ref{projectcorr}), with $\rho_0$ being the  stationary state of the bath, and uses the notation $\Pi_R$ to refer to the projections to each subspace (hence $i\equiv R$). The projectors $\Pi_R=\sum_{\{\epsilon_R\}}|\epsilon_R\rangle\langle\epsilon_R|$ decompose the Hilbert space of the environment into different sub-reservoirs, each spanned by the base of eigenvectors $|\epsilon_R\rangle$. Hence, this projection corresponds to splitting the environment into a set of sub-reservoirs, such that the interaction Hamiltonian can be written as a direct sum of Hamiltonians $H_I=\sum_{R,R'}H_{I_{RR'}}$, with 
$H_{I_{RR'}}=\Pi_RH_I\Pi_{R'}$. This choice gives rise, in the long time limit, to the same general equation (\ref{linbproj}), which connects each $\rho_R$ to the other $\rho_{R'}$ ($R'\neq R$). 
A simpler situation is discussed in a previous paper \cite{budini2005} by considering the case in which $H_{I_{RR'}}=0$ for $R\neq R'$. In this case, the interaction Hamiltonian can be written as a direct sum of sub-Hamiltonians for each subspace $H_I=H_{I_1}\oplus H_{I_2}\cdots \oplus H_{I_R}\oplus H_{I_{R+1}}\cdots$, and each $\rho_R(t)$ follows a Lindblad type of evolution equation of the form (\ref{Linb}) induced by the coupling with the corresponding sub-reservoir, and independently of other $\rho_{R'}$ ($R'\neq R$). Each $\rho_R$ evolves with a rate $\gamma_R(t)$, and the reduced density operator of the system is obtained as 
\bea
\rho_s(t)=\Ttr_B\large[{\mathcal P}\rho(t)\large]=\sum_RP_R\rho_R(t),
\label{reducedbudini}
\eea
where the weight is given as $P_R=\Ttr_B\large[\rho_{RB}\large]=\sum_{\{\epsilon_R\}}\langle\epsilon_R|\rho_B|\epsilon_R\rangle$, and therefore $\sum_RP_R=1$. The fact that each $\rho_R$ follows a Markovian evolution does not mean that $\rho_s$ will also do so. Indeed, the evolution of $\rho_s$ has the form of a convoluted master equation as long as the weights $P_R$ are different. In the effective approximation \cite{budini2005}, the equation can be written as 
\bea
\frac{d}{dt}{\mathcal P}\rho(t)={\mathcal L}_S\rho_s+\int_0^t ds {\tilde K}(t,s)e^{(t-s){\mathcal L}_S}{\mathcal L}\rho(s).
\label{NZmaster2}
\eea
Here ${\mathcal L}_S$ and ${\mathcal L}$ are the free evolution and Lindblad superoperators, respectively, and ${\tilde K}(t,s)$ is a superoperator that depends on the rates $\gamma_R$ and the probabilities $P_R$. Its Laplace transform is given by 
$k(p)=f(p)/g(p)$, in terms of the Laplace transform of the waiting time distribution and survival probabilities, $f(p)$ and $g(p)$ respectively, which in this case take the form $f(p)=\langle\frac{\gamma_R}{p+\gamma_R}\rangle$, and $g(p)=\langle\frac{1}{p+\gamma_R}\rangle$, with $\langle\cdots\rangle=\sum_R P_R\cdots$ denoting an average over all subenvironments. As described previously, the rates $\gamma_R$ are obtained by applying the Fermi golden rule to each reservoir, which provides a connection between the waiting time distribution and the spectral density of the environment. Thus, the choice of the different $P_R$ and $\gamma_R$ depends on the specific structure of the environment. 

Similarly, \textcite{harbola2006} derived a master equation to analyze electron transport through quantum dots and single molecules weakly coupled to two metal leads. To this end, they define projection operators ${\mathcal P}_n$ onto the Fock state, with $n$ electrons in the quantum dots. The total density matrix can then be expanded as Eq. (\ref{projectcorr}), but now with a sum that extends over all $n$ states, and with $\rho_R(t)\equiv \rho^n(t)$ being the many-body density matrix of the quantum system with $n$ electrons. Also, consistent with the weak coupling assumption, the leads are assumed to remain in thermal equilibrium so that $\rho_{RB}=\rho_B$. Under these conditions, a set of equations for $\rho^n(t)$ are obtained, and found to be coupled in a hierarchy to $\rho^{n-1}$ and $\rho^{n+1}$.
A similar hierarchy of quantum master equations was originally derived by \textcite{gurvitz1996}, which keeps track of the number of electrons transferred from the source-lead to the collector-lead.

The projection superoperator techniques have also been applied to scenarios in which all the parts of the system have similar sizes and characteristic dynamical times, so that there is no clear distinction between system and environment. In particular, the recently-developed self-consistent Mori projector (c-MoP) technique \cite{degenfeld2014}, in which the Nakajima-Swanzig equations for the reduced state of all parts of the system are solved in parallel under different approximations, was applied to many-body scenarios \cite{degenfeld2014}, as well as to few-body bosonic quantum-optical problems \cite{degenfeld2015a,degenfeld2015b}. In the latter case, further Gaussian approximations render this approach a very efficient way of dealing with problems in which one needs to check consistently whether non-Markovian and back-action effects can be neglected between the different parts of the bosonic system.

The TCL technique has been applied to many different problems, ranging from spin relaxation \cite{chang1993,blanga1996} to the spin-boson model \cite{breuer2001a}, the spin star model \cite{breuer2004,barnes2012} for the standard case and \cite{fischer2007} for the correlated one, and to atomic lasers \cite{breuer2001b}. 

\subsubsection{Master equations derived from dynamical maps and a measurement approach}
\label{dynamicM}
In order to ensure that complete positivity is preserved, another possibility is to derive master equations from dynamical maps that are known to preserve this property. A recent derivation in this direction is the one by \textcite{vacchini2013}, which defines a time-evolved state of the reduced density matrix according to the following dynamical map:
\bea
\Lambda(t)\rho_s&=&p_0(t){\mathcal F}(t)\rho_s+\int_0^t dt_n \cdots \int_0^{t_2} dt_1 p_n(t;t_n,\cdots t_1)\cr
&\times&{\mathcal F}(t-t_n){\mathcal E}\cdots{\mathcal E} {\mathcal F}(t_1)\rho_s.
\eea
Also, the quantities ${\mathcal F}(t)$ and ${\mathcal E}$ are time-dependent and time-independent completely positive maps. In addition, $p_n(t;t_n,\cdots,t_1)$ is the exclusive probability density for the realization of $n$ events up to time $t$, at given times $t_1,\cdots,t_n$ with no events in between. This probability density relates to the waiting time distribution $f(t)$ as 
\bea
p_n(t;t_n,\cdots,t_1)=f(t-t_n)\cdots f(t_2-t_1)g(t_1),
\eea
where $g(t)=1-\int_0^t dsf(s)$ is its associated survival probability. 
The evolution equation associated with the map $\Lambda(t)$ can be written as 
\bea
\frac{d\rho_s(t)}{dt}=\int_0^tds {\textit K}(t-s){\mathcal E} \rho_s(s)+{\mathcal I}(t)\rho_s(0),
\label{vacc}
\eea
where the integral kernel is $K(t-s)=\frac{d}{dt}[f(t){\mathcal F}(t)]+f(0)\delta(t)$ and the inhomogeneous term ${\mathcal I}(t)=\frac{d}{dt}[g(t){\mathcal F}(t)]$.

If we now consider ${\mathcal F}(t)=e^{t{\mathcal L}}$, with ${\mathcal L}$ as a Lindblad generator, then the following equation is obtained:
\bea
\frac{d\rho_s(t)}{dt}={\mathcal L}\rho_s(t)+\int_0^tds {\textit k}(t-s)e^{(t-s){\mathcal L}}({\mathcal E}-\unit) \rho_s(s),
\cr
\label{exponvac}
\eea
where the memory kernel ${\textit k}(t)$ is related to the waiting time and survival probabilities through its Laplace transform as in Eq. (\ref{NZmaster2}), or alternatively $f(\tau)=\int_0^\tau dt {\textit k}(\tau-t)g(t)$.
With the choice ${\mathcal L}={\mathcal L}_S=-i[H_S,\rho_s]$ for the first term of Eq. (\ref{exponvac}), and the Lindblad generator $(\epsilon-1)={\mathcal L}$ for the second term, the master equation (\ref{NZmaster2}) is regained. 
Considering now that ${\mathcal L}=0$ in Eq. (\ref{exponvac}), a quantum semi-Markov equation is obtained,
\bea
\frac{d\rho_s(t)}{dt}=\int_0^tds {\textit k}(t-s)({\mathcal E}-\unit) \rho_s(s),
\label{semimarkov}
\eea
This type of equation was introduced by \textcite{breuer2008}, and its non-Markovian character was further analyzed by \textcite{vacchini2011}. The interesting thing about the quantum semi-Markov process is that the solution of the corresponding equation has a relatively simple form 
\bea
\rho_s(t)=\Lambda(t,0)\rho_s(0)=\sum_{n=0}^\infty p_n(t){\mathcal E}^n\rho_s(0),
\eea
which represents that the reduced density operator at time $t$ is the result of the repeated action of the map ${\mathcal E}$, where $p_n(t)=\int_0^t d\tau f(t-\tau)p_{n-1}(\tau)$ is the probability that at time $t$ there has been $n$ of such projections, with a given waiting time distribution $f(t)$. Finally, the case where we consider in (\ref{exponvac}) that ${\mathcal E}=\unit$ allows us to recover the Lindblad equation. 

Another master equation that preserves complete positivity, and at the same time includes environment memory effects is the so-called post-Markovian master equation derived by \textcite{shabani2005} from a measurement approach.
\subsubsection{Collisional models}
\label{collisionM}

\begin{figure}[ht]
\centerline{\includegraphics[width=0.35\textwidth]{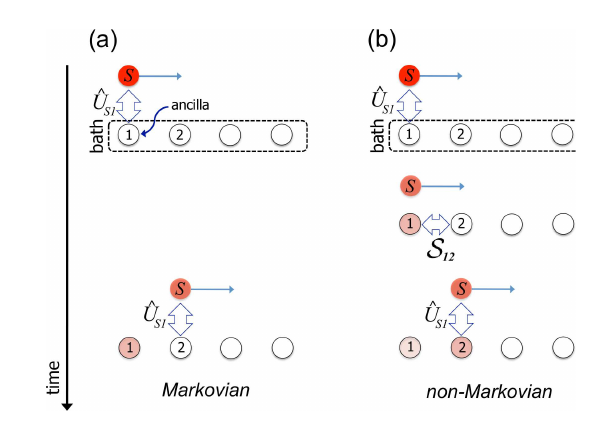}}
\caption{Schema showing the first steps of the collisional model in the (a) Markovian case, and in the (b) non-Markovian case. In the second case, ancilla-ancilla collisions occur in between system-ancilla collisions. From \textcite{ciccarello2013}. \label{collisional}}
\end{figure}

Collisional models give rise to a visual and intuitive way of deriving master equations \cite{rau1963}. In these models, it is assumed that the environment is a collection of $M$ harmonic oscillators or ancillas organized in a chain. Then, it is assumed that the system $S$ interacts, or \textit{collides}, at each time step with each ancilla, such that at $t_1$ there is a collision $S\leftrightarrow 1$, at $t_2$ there is a collision $S\leftrightarrow 2$, and so on. It was shown by \textcite{scarani2002,ziman2002,ziman2010} that when no initial correlation is assumed between the ancillas and no correlations are created between them along the process, a Lindblad master equation can be derived. 
More recently, it was realized in \cite{rybar2012} that introducing correlations in the initial state of the ancillas allows one to recover the dynamics of any indivisible and therefore non-Markovian channel. An alternative to introducing a non-Markovian evolution is to consider, as proposed by \textcite{ciccarello2013,ciccarello2013b}, that between system-ancilla collisions there are also ancilla-ancilla collisions [see Fig. (\ref{collisional})]. These are assumed to occur at a rate $\Gamma_c$, which can be interpreted as the memory of the environment, such that the probability that an interancilla collision occurs at time $\tau$ is given by $p=e^{-\Gamma_c \tau}$. While the system-ancilla collisions are defined by a map which acts over an operator $\rho$ as $\rho\rightarrow U_{Si}[\rho]=U_{Si}\rho U_{Si}^\dagger$, with $U_{Si}=e^{-iH_{Si}\tau}$ corresponding to a unitary evolution at the collision time $\tau$, the ancilla-ancilla collisions are defined as a nonunitary map that, with probability $p$, exchanges the ancilla states,
\bea
\rho\rightarrow {\mathcal S}_{i+1}[\rho]=(1-p)\rho+pS_{i+1,i}\rho S_{i+1,i}.
\eea
Here $S_{i+1,i}$ is the swap operator defined as $S_{i+1,i}=|\phi_j\otimes\phi_k\rangle\langle\phi_k\otimes\phi_j|$ in terms of an arbitrary orthonormal basis $\{\phi_j\}$ of the ancillas that interchanges the states of the ancillas $j$ and $j+1$. 
The sequential repetition of this system-ancilla, ancilla-ancilla collisional process gives rise in the continuum limit to a master equation of the form 
\bea
\frac{d\rho_s}{dt}=\int_0^tds e^{-\Gamma s}{\mathcal E}(t)[\dot{\rho_s}(t-s)]+e^{-\Gamma t}\dot{{\mathcal E}}(t)[\rho_s(0)].
\cr
\label{cicarello}
\eea
This equation is rather similar (but not exactly equal) to Eq. (\ref{vacc}), and it also preserves complete positivity. Here, ${\mathcal E}(t)$ is a completely positive time-dependent map related to the system-ancilla collisions. It corresponds to the continuous analog of 
\bea
{\mathcal E}_j[\rho_s]=\Ttr_B\{U^j_{Sn}[\rho_s\otimes|0\rangle_B\langle 0|]\},
\eea
where $U_{Sn}^j[\sigma]=e^{-iH_{Sn}j\tau}\sigma e^{iH_{Sn}j\tau}$ is the unitary evolution at the collision time, $j\tau$, between the system and the ancilla, $n$. A more recent work relating quantum memory effects to ancilla-ancilla collisions can be found in \cite{kretschmer2016}.
In addition, a generalization of Eq. (\ref{cicarello}) that is not restricted to the case in which the system-environment coupling is mediated via the ancillary degrees of freedom, but applies to a broader class of non-Markovian dynamics, was recently derived by \textcite{lorenzo2016}. Interestingly, this non-Markovian master equation is originated from a class of bipartite Lindblad master equations when tracing out one of the two subsystems. The idea of obtaining a non-Markovian equation by tracing a Markovian one corresponding to a larger Hilbert space is at the heart of the embedding methods presented in the following section. 

\subsubsection{Embedding methods}
\label{embeddingM}
Embedding methods consist of adding fictitious modes to the non-Markovian system in such a way as to make the enlarged hypothetical system dynamics Markovian. This idea was first proposed by \textcite{imamoglu1994}, \textcite{garraway1996,garraway1997} and \textcite{bay1997}. \textcite{garraway1996,garraway1997} described the decay of an atom strongly coupled to a reservoir by considering an enlarged system that includes a set of \textit{pseudomodes}. Such pseudomodes are related to the poles of the spectral density of the environment, and are calculated by considering its analytical continuation in the complex plane. The enlarged system obeys a Lindblad equation, and the dynamics of the OQS can be recovered by tracing out the pseudomodes. This method provides an exact solution and is particularly convenient when there is only one excitation in the total system, although generalizations to tackle the multiple-excitation case have also been developed \cite{dalton2001,dalton2003}. 

Embedding methods have been more recently extended by \textcite{breuer1999,breuer2004a}. In the most recent work, Breuer proposed an enlarged system with density operator $W$, composed of the original system and a three-level system with basis states $|a\rangle,|b\rangle,|c\rangle$ belonging to a space ${\mathcal C}^3$. The key point of the method is to consider that the enlarged density operator $W$ obeys a time-local equation of the form of Eq. (\ref{nmbreuer}). Here, the coupling or jump operators $C_k$ are chosen such that the reduced density matrix of the original system $\rho_s(t)$ can be written as a certain set of coherences $W_{ab}$ of the density matrix $W$ of the extended system. As argued in Sec. \ref{canonical}, this equation may not be completely positive, but equations with this form can be derived from first principles, for instance by considering a projection-operator method with an expansion up to second-order in the coupling parameter between the system and the environment.

Similarly, \textcite{budini2013} derives a master equation by considering that the system is combined with an ancilla, and that both system and ancilla evolve according to the Lindblad equation (\ref{Linb}). In this derivation, the reduced density matrix of the system is recovered by tracing out the ancilla's degrees of freedom from the total density operator $W$. Another important difference with respect to the former scheme is that here, the Lindblad operators $C_k$ are chosen in such a way that they lead to an uncorrelated system-ancilla state. Thanks to this condition, the system's€™ reduced density matrix evolves according to an equation that is closed, without making any further approximations. This ensures that such an equation preserves complete positivity. However, the equation does not necessarily comply with the form of any master equation obtained from a microscopic derivation, as occurs in the previous derivation. 

Finally, in the context of Anderson impurity models, \textcite{arrigoni2013,dorda2014} recently derived an approach that is based on representing the original problem (the impurity coupled to the leads) as an equivalent one, consisting of the impurity coupled to an auxiliary discrete system, which in turn is coupled to a Markovian reservoir. The parameters of the auxiliary system are chosen by optimization as those that most faithfully represent the original spectral density, written in terms of the hybridization function. 

\subsubsection{Master equations derived from variational methods}
\label{polaron}

As proposed by \textcite{mcCutcheon2011}, it is possible to derive a master equation for the spin-boson model (\ref{chapuno11}) which is valid in the strong coupling regime. 
The key ingredient of such derivation is to consider a variational polaron transformation for this Hamiltonian \cite{silbey1984,leggett1987}, $H=e^{V}H_{\ttot}e^{-V}$, with $\exp(\pm V)=|0\rangle\langle 0|+|1\rangle\langle 1| \prod_k D(\pm\alpha_\lambda)$, where $D(\pm\alpha_\lambda)=\exp(\pm\alpha_\lambda(a_\lambda^\dagger-a_\lambda))$ is a displacement operator, and $\alpha_\lambda=f_\lambda/\omega_\lambda$ is assumed to be real and dependent on certain parameters $f_\lambda$ to be determined variationally. The transformed Hamiltonian can be written as $H=H_0+H_I$, with
\bea
H_0&=&\frac{1}{2} \omega_{12} \sigma_z+\sum_{\lambda}\omega_\lambda a^{\dagger}_\lambda a_\lambda\cr
H_I&=&\frac{1}{2}\Delta_0(\sigma^+B_{-}+\sigma^- B_{+})+|1\rangle\langle 1|B_z,
\label{polaronT0}
\eea
where we defined $B_{\pm}=\exp(\pm\sum_\lambda g_\lambda(a_\lambda-a_\lambda^\dagger))$, and $B_z=\sum_\lambda (g_\lambda-f_\lambda)(a_\lambda^\dagger+a_\lambda)$. 
The variational parameters $\{f_\lambda\}$ are obtained by imposing that the free energy associated with the transformed Hamiltonian is minimized. Then, as discussed in \cite{mcCutcheon2011}, two limiting situations occur: (a) when $\Delta_0\ll \omega_\lambda$, then $f_\lambda= g_\lambda$ and the variational polaron transformation is identical to a simple polaron transformation; and (b) when $\Delta_0\gg \omega_\lambda$, $f_k$ becomes very small and the displacement produced by the transformation is almost negligible. 
A master equation can be obtained by considering the perturbative methods discussed in Secs. \ref{perturbativeM} and \ref{projectionM}, by performing an expansion up to second-order in the interaction Hamiltonian of the transformed system (\ref{polaronT0}), which is therefore considered as a small perturbation. Naturally, if the system is strongly coupled to the environment, the resulting master equation will be a particularly convenient approach in a situation close to the case (a) above discussed, \textit{i.e.} when the tunneling energy $\Delta_0$ is sufficiently small. A master equation in the simple polaron limit $f_\lambda=g_\lambda$ was previously obtained in \cite{jang2008,nazir2009} in the context of coherent resonant energy transfer between two chromophores. This problem can also be described with a spin-boson model, by interpreting the states $|0\rangle$ and $|1\rangle$ as the state corresponding to the excitation in the first and the second chromophores respectively. In this context, the analysis provided by \textcite{nazir2009} allowed to analyze the transition between a regime where energy is coherently interchanged between such states and a regime where energy is interchanged incoherently.

A closely related method based on the above discussed polaron transformation has been put forward by \textcite{diaz-camacho2015} to analyze the dynamics of a collection of quantum emitters interacting with a one dimensional EM field, and without considering the RWA (see Sec. (\ref{RWA}) for other examples where the RWA is no longer valid). This method extends upon the polaron variational ansatz originally derived to study the ground state properties of the spin-boson model. In more detail, it defines a dynamical variational ansatz (\textit{i.e.} establishes a model structure for the system wave function), by creating spin and photonic excitations over such polaron transformed ground state. The method appears to be accurate for relatively strong couplings, as shown by direct comparison with matrix product states. 

\subsection{Multiple-time correlation functions: The quantum regression theorem}
\label{QRT}

The Markovian approximation allows one to derive a formula which permits the evaluation of two-time correlations (and even $N$-time correlations) using the master equation for the reduced density operator. This result, which was first obtained by \textcite{lax1963,lax1967}, is called quantum regression theorem (QRT) \footnote{Although as noted by \textcite{statisticalmethodsquantumoptics} it would be more appropriate to use the word \textit{formula} instead of \textit{theorem}.}. It should be noted here that there is a classical hypothesis by \textcite{onsager1931,onsager1931b} which leads to the same formula as the QRT for two-time correlations (see a discussion of this by \textcite{statisticalmethodsquantumoptics}). 

We follow here the derivation of the QRT presented in \cite{quantumnoise}. Analogous derivations may be found in the original paper by Lax, and also in several books on quantum optics, for instance \cite{quantumoptics,atomphotoninteractions,statisticalmethodsquantumoptics}.

Let us consider the two-time correlation function of operators $A_1 =A$ and $A_{2}=B$ \cite{quantumnoise},
\begin{eqnarray}
&&\langle A(t_1)B(t_2)\rangle=\Ttr_{SB}[{\cal U}^{\dagger}(t_2,0){\cal U}^{\dagger}(t_1,t_2)A{\cal U}(t_1,t_2)\nonumber\\
&&\times {\cal U}(t_2,0) {\cal U}^{-1}(t_2,0)B{\cal U}(t_2,0)\rho(0)],
\label{regres1}
\end{eqnarray}
where the unitary evolution operator from $t_2$ to $t_1=t_2+\tau$, which is assumed to be in interaction picture, is
\begin{equation}
{\cal U}(t_1 ,t_2)=e^{iH_{0}t_1}e^{-iH_{tot}(t_1-t_2)}e^{-iH_{0}t_2}.
\label{evolin}
\end{equation}
Considering the unitarity of the evolution operators ${\cal U}(t,0){\cal U}^{\dagger}(t,0)=1$ and the cyclic property of the trace, we can write Eq. (\ref{regres1}) as
\bea
\langle A(t_1)B(t_2)\rangle&=&\Ttr_S\{A \Ttr_{B}\{\xi(t_1,t_2)\}\},
\label{regres2}
\eea
where 
$\xi(t_1,t_2)={\cal U}(t_1,t_2)B\rho(t_2){\cal U}^{-1}(t_1,t_2)$,
and $\rho(t_2)={\cal U}(t_2,0)\rho(0){\cal U}^{\dagger}(t_2,0)$. Let us now consider the evolution equation of $\xi(t_1,t_2)$ with respect to $t_1$ and in interaction picture,
\begin{equation}
\frac{d\xi (t_1,t_2)}{dt_1}=\frac{1}{i} [V_{t_1}H_{I},\xi(t_1,t_2)].
\label{regres3}
\end{equation}
The form of Eq. (\ref{regres3}) is identical to the von-Neumann equation for $\rho(t_2)$ in interaction picture, (\ref{total1}). Hence, in order to obtain a closed evolution equation for ${\Ttr_B}\left\{\xi(t_1,t_2)\right\}$, we follow the same procedure we used in Sec. \ref{perturbativeM} for obtaining the master equation up to the second-order in $g$ to get
\beqa
&&\frac{d\xi(t_1,t_2)}{dt_1}=-i[V_{t_1 }H_{I},\xi(t_2,t_2)]\cr
&-&\int^{t_1}_{t_2} d\tau \left[V_{t_1} H_{I},\left[V_{t_1-\tau} H_{I},\xi(t_2,t_2)\right]\right],
\eeqa
with $\xi(t_2,t_2)=B\rho(t_2)$.
We now proceed to trace out the environmental degrees of freedom, so that the final equation for ${\textmd{Tr}_B}\{\xi(t_1,t_2)\}=\xi^{S}(t_1,t_2)$ can be written in a similar way as Eq. (\ref{total4}),
\begin{eqnarray}
\frac{d\xi^{S}(t_1,t)}{dt_1}&=&-\int^{t_1}_t d\tau {\Ttr_B}\{[V_{t_1} {H}_I,[V_{t_1-\tau}{H}_I ,\xi^{B}(t,t) ]]\}\cr
&\times& \xi^{S}(t,t),
\end{eqnarray}
where we assumed an initially uncorrelated state $\xi(t_2,t_2)=\xi^{B}(t_2,t_2)\otimes \xi^{S}(t_2,t_2)$, which is equivalent to assuming the Born approximation. Assuming that the $\xi^{S}(t_1,t_2)=\xi^{S}(t_2,t_2)+{\mathcal{O}}(g)$, we can approximate the last equation up to second-order as
\begin{eqnarray}
\frac{d\xi^{S}(t_1,t_2)}{dt_1}&=&-\int^{t_1}_{t_2}d\tau {\Ttr_B}\{[V_{t_1} {H}_I,[V_{t_1-\tau}{H}_I ,\cr
&&\xi^{B}(t_2,t_2) ]]\} \xi^{S}(t_1,t_2),
\end{eqnarray}
which is not equal to the master equation (\ref{total5}), because of the limits of integration. Only in the Markovian case does the former equation become local in time and the evolution equation of $\xi^{S}(t_1,t_2)$ become equal to the Lindblad master equation (\ref{Linb}), but 
with the initial condition $\xi^{S}(t_2,t_2)=\Ttr_B \{B\rho(t_2)\}=B\rho_s (t_2)$. In other words, the evolution equation has the same form as an ordinary master equation, but considering a modified initial condition. This procedure can be repeated to show that, in general, $N$-time correlation functions are computed by considering the $N-1$-time correlations as the initial condition, and using the evolution equation of $1$-time correlations, namely the Markovian master equation.

The last derivation can be reexpressed in terms of the evolution superoperators, $\Lambda(t_1,t_2)$, which define the following mapping of the operator $\xi^{S}(t_1,t_2)$ \cite{breuerbook,quantumnoise},
\begin{eqnarray}
\xi^{S}(t_1,t_2)=\Lambda(t_1,t_2)\xi^{S}(t_2,t_2).
\end{eqnarray}
The evolution equation of $\Lambda(t_1,t_2)$ has the same form as the evolution of $\xi^{S}(t_1,t_2)$ which, as derived above, turns out to be equal to the evolution for $\rho_s (t_1)$, but with a different initial condition. 
Because of its Lindbland form, the evolution superoperators have the divisibility property $\Lambda(t_1,t_2)\Lambda(t_2,t_0)=\Lambda(t_1,t_0)$, and hence, the two-time correlation (\ref{regres2}) can be written as
\begin{equation}
\langle A(t_1)B(t_2)\rangle={\textmd{Tr}_S}\{A\Lambda(t_1,t_2){\textmd{Tr}_B}\{B\rho(t_2)\}\}.
\label{regres4}
\end{equation}
The theory of stochastic Schr\"odinger equations, initially elaborated to compute the expectation values of system observables, has been extended by many groups \cite{gisin1993,brun1996} to calculate multiple-time correlation functions (MTCFs) for the Markovian case. Such stochastic methods agree with the results expected from the QRT.

\section{Stochastic Schr\"odinger equations}
\label{SSE}

In this section, we analyze the SSEs that evolve the system wave function $|\psi_t (z^*)\rangle$, \textit{i.e.} a vector that evolves in the Hilbert space of the system following a stochastic trajectory. As shown, the reduced density matrix can be recovered as a sum of projectors of stochastic trajectories. Depending on the method used in the derivation, there are many different SSEs that recover the reduced density matrix of the OQS; these are called \textit{unravelings} of the reduced density matrix \cite{carmichael1993}. A review of quantum stochastic methods was given by \textcite{quantumnoise}.

An advantage of SSEs is that since the reduced density matrix is the result of a positive definite sum of projectors, it preserves positivity, a fundamental property discussed in Sec. \ref{nonmarkovianity}. A second advantage is that their non-Markovian version does not rely explicitly on a Born approximation that neglects the second-order system-environment correlations at all times, which allows using them to describe the evolution from initially system-environment correlated states (see Sec. \ref{expansion}). Also, non-Markovian SSE (for instance those derived with expansion methods) allow one to obtain not only system dynamical quantities, but also environment ones. Finally, the size of the wave function to be evolved grows with the system basis dimension $d$ and not with $d^2$ as the reduced density operator.

\subsection{Markovian SSEs}

Stochastic Schr{\"o}dinger equations were introduced in the context of dynamical reduction models \cite{pearle1976,pearle1989,ghirardi1986,ghirardi1990,bassi2003}. In these schemes (particularly in the \textit{continuous localization models}), a modified Schr{\"o}dinger equation is generated which, besides the standard Hamiltonian, contains stochastic terms acting at every time step of the evolution, as well as nonunitary or dissipative terms. These new terms induce a diffusion process for the state vector which is responsible for its reduction to a particular subspace in the system's Hilbert space. 
Thus, as noted by \textcite{bassi2003}, the goal of dynamical reduction models is to formally account for the wave-packet reduction process by building a modified Schr{\"o}dinger equation that describes the spontaneous suppression of the superpositions observed in a macro-system, while at the same time still accounts for all the known properties of microscopic quantum systems.

Following a different strategy, the continuously diffusive nonlinear stochastic Schr{\"o}dinger equation derived by \textcite{gisin1984} departs from the von-Neumann-L{\"u}ders postulate \cite{neumann1955}.
Thus, the resulting stochastic trajectory of the system is reduced due to a sequence of projective measurements performed by an external apparatus. 
In addition, a real-valued noise Markovian SSE was presented by \textcite{ghirardi1990}, also from a dynamical reduction model,
\begin{eqnarray}
\frac{d|\psi_t \rangle}{dt}=-iH_S |\psi_t \rangle+L \xi_t |\psi_t \rangle-\frac{dt \Gamma }{2}L^{\dagger}L |\psi_t \rangle
\label{linMreal}
\end{eqnarray}
where $\xi_t$ is a real-valued Gaussian white noise. Equation (\ref{linMreal}) is still linear, since it represents the evolution of a non-normalized state. In order to write the density operator as a mixture of pure state vectors, it has to be transformed into a nonlinear equation for normalized states $|\tilde{\psi}_t\rangle$. This transformation is formally made as 
\bea
\frac{d|\tilde{\psi}_t\rangle}{dt}=\frac{d|\psi_t \rangle}{dt}\frac{1}{\sqrt{N}}+|\psi_t \rangle \frac{d}{dt}\frac{1}{\sqrt{N}},\nonumber
\eea 
where $N=\langle\psi_t |\psi_t \rangle$.
\textcite{belavkin1989,belavkin1990} presented a new SSE that was very similar to the former one but driven by a complex white noise, $z^*_t =\zeta_t =\xi_{1,t}+i\xi_{2,t}$, where $\xi_{i,t}$ with $i=1,2$ is a real-valued Gaussian white noise process.
\begin{eqnarray}
\frac{d|\psi_t \rangle}{dt}=-iH_S |\psi_t \rangle +Lz^*_t |\psi_t \rangle -\frac{1}{2}L^{\dagger}L |\psi_t \rangle,
\label{linM}
\end{eqnarray}
Here the complex white noise has the following statistical properties, ${\mathcal M}[z_t z^*_\tau]=\Gamma \delta(t-\tau)$ and 
${\mathcal M}[z_t z_\tau]={\mathcal M}[z_t]=0$, 
where $\Gamma$ is the dissipative constant and ${\mathcal M}[\cdots]$ denotes an average over many realizations of $z_t$.
A nonlinear version of this SSE was later derived by \textcite{gisin1993a}, 
\begin{eqnarray}
\frac{d|\tilde{\psi}_t \rangle}{dt}=-iH_S |\tilde{\psi}_t \rangle+\left(L-\langle L\rangle_t \right)\left(z^*_t +\langle L^\dagger \rangle_t \right)|\tilde{\psi}_t \rangle \nonumber\\
-\frac{\Gamma}{2}\left(L^{\dagger}L-\langle L^{\dagger}L\rangle\right)|\tilde{\psi}_t \rangle+{\mathcal O}(g^3 ),\cr
\label{nolinM}
\end{eqnarray}
in Stratonovich form \cite{quantumnoise}. The mean value appearing in Eq. (\ref{nolinM}) is $\langle L^{\dagger}\rangle=\langle\tilde{\psi}_t | L^{\dagger}|\tilde{\psi}_t \rangle$.

Several models of Markovian SSE have been derived in the framework of theories of continuous observation \cite{belavkin1989,belavkin1990,belavkin1992,kampen}. In these models, contrary to those involving dynamical reduction, a particular measuring device is chosen, which determines the kind of trajectory or unraveling that will be obtained. 
Also, in order to minimize the perturbation caused to the system by the measuring device, the measurement is performed not directly on the system but on its environment. Since they are entangled, a measurement of the environment selects the particular state of the mixture compatible with the measurement result. In that way, the quantity that is continuously measured, which is not necessarily the environmental state but a combination of its eigenvalues, is related to the stochastic variable $z_t$ that drives the SSE. A sequence of measurement results $z_t$ then corresponds to a single trajectory of a Markovian SSE. In other words, the trajectory $|\psi_t (z^*)\rangle$ represents the system state conditioned to the sequence of measurements which have given the result $z_t$. 

The SSEs generated by dynamical reduction models and continuous measurement theories are of a {\bf quantum state diffusion} type, since the stochastic element acts on every time step of the trajectory. 
Particularly, in the framework of quantum optics Carmichael shows that the real noise SSE (\ref{linMreal}) derived by \textcite{ghirardi1990} corresponds to continuous homodyne detection \cite{statisticalmethodsquantumoptics}. 
In addition, \textcite{wiseman1993} showed that the complex noise linear SSE (\ref{linM}) corresponds to a continuous heterodyne detection of the environment. 
The bases chosen for homodyne and heterodyne detection are the quadrature and the coherent basis, respectively. A formal derivation of Eqs. (\ref{linMreal}) and (\ref{linM}), as well as their correspondence to homodyne and heterodyne detection, was performed by \textcite{gambetta2002} from the measurement theory. This is discussed in more detail in Sec. \ref{measurementSSE}.

Apart from diffusive trajectories, which depend on a continuous noise variable acting over the trajectory at each time step, there are also quantum trajectories in which the stochastic influence occurs in sudden jumps, interrupting a deterministic nonunitary evolution. The {\bf quantum jumps} formalism was first developed by \textcite{zoller1987} for Markovian systems, as a theory to calculate density operators conditioned to a different number of photon emissions. The density operator corresponding to the emission of $n$ photons, $\rho^{(n)}_S (t)$, is related to the total density operator by
\begin{equation}
\rho^{(n)}_S (t)=\Ttr_B\{{P_n \rho (t)}\},
\end{equation}
where $P_n$ is the projection-operator onto the state of the quantized radiation field that contains $n$ photons. A formulation of quantum jumps as a stochastic equation was later proposed by \textcite{hegerfeldt1991}, \textcite{dalibard1992,carmichael1993a,molmer1993} and \textcite{quantumnoise,zoller1987,gardiner1992} [see \cite{plenio1998} for more details]. In all these methods, a non-Hermitian term and a white noise term are added to the Schr{\"o}dinger equation. Because of the non-Hermitian term, the trace of the reduced density operator is no longer conserved, but is restored by stochastically chosen quantum jumps. 

For instance, in the algorithm by \textcite{dalibard1992}, the total wave function of an atom coupled to its environment, computed at time $t+dt$ is $|\Psi (t+dt)\rangle=|\Psi^{(0)}(t+dt)\rangle+|\Psi^{(1)}(t+dt)\rangle$, where $|\Psi^{(1)}\rangle$ represents the product state of the atom in the ground state $|g\rangle$ and a photon in the field, and $|\Psi^{(0)}\rangle=|\psi_t\rangle\otimes|0\rangle$ represents the product of a atomic state $|\psi_t\rangle$ and no photon in the field $|0\rangle$. When a photon is detected, the total state is projected into $|\Psi^{(1)}\rangle$, and when no photon is detected it remains in $|\Psi^{(0)}(t+dt)\rangle$. The probability of a spontaneous emission occurring during $dt$ is given by $dp=\langle\Psi^{(1)}|\Psi^{(1)}\rangle$. The randomness in the detection or non-detection of a photon is simulated by numerical generation of a random number $\epsilon$ chosen from the interval $[0,1]$. Thus, when $\epsilon>dp$, it is assumed that no photons are detected, so that $|\Psi (t+dt)\rangle=|\Psi^{(0)}(t+dt)\rangle=\mu (1-i dt H_{\eeff})(|\psi_t\rangle\otimes|0\rangle)$, where $\mu=(1-dp)^{-1/2}$, and $H_{\eeff}$ is a non-Hermitian Hamiltonian in ${\mathcal H}_S$. The norm of this state is no longer $1$, but is given by $1-dp$. As a consequence, the quantity $dp$ represents the loss of norm of the total state when no photon is detected. When $\epsilon<dp$, a photon is detected and the total state is projected into the normalized state $|\Psi^{(1)}\rangle$, where it is assumed that there has been no time for the atom to be re-excited after having emitted a photon. The reduced density matrix is computed as a sum of the projectors $|\psi_t\rangle\langle\psi_t|$ corresponding to a large ensemble of stochastic trajectories.

As shown by \textcite{carmichael1993}, the jump-like Markov SSE corresponds to direct photon detection, where the experimental setup consists of a photon counter and the environmental state is expressed in the number basis. An extended review of the quantum jump approach in the Markovian regime is found in \cite{plenio1998}.

\subsection{Non-Markovian SSEs}
\label{NMSSE}
Among the first proposals to describe non-Markovian effects with SSEs is that offered by \textcite{imamoglu1994}, who approximated memory effects in electron-phonon interactions by embedding the system into a larger one that could be described with a Markovian SSE. In addition, \textcite{kleinert1995} derived an exact non-Markovian quantum-Langevin equation to describe the evolution of the position operator of a harmonic oscillator. However, the first extension of a quantum state diffusion SSE to a non-Markovian environment was proposed by \textcite{diosi1997,diosi1998}, and later complemented by the works of \textcite{gaspard1999b,jack2000,cresser2000,strunz2001,alonso2005}. An extension of the quantum jump approach to non-Markovian interactions came a decade later with the proposal of \textcite{piilo2009}. In the following section we  discuss some of these equations, with an emphasis on the different derivation techniques that exist in the literature. 

\subsubsection{Expansion method}
\label{expansion}
In the last few decades, several methods have been derived for obtaining diffusive non-Markovian SSEs, where the noise acts continuously along the trajectory. 

Some methods are based on \textit{expanding the total state vector into the environmental basis}. The coefficients of such an expansion are in principle deterministic, but because the environment has a large number of degrees of freedom, it is often convenient to consider these coefficients as stochastic and compute their evolution with a stochastic Schr\"odinger equation. As shown, deriving an SSE with the expansion method provides an excellent way to understand the origin of the stochasticity in the evolution of an OQS, as well as the connection between the noise and environment states. 

The wave function corresponding to the total Hamiltonian (\ref{chapuno32}) evolves from its initial value $\mid\Psi_0 \rangle$ as $\mid\Psi_t \rangle ={{\cal U}}_I \mid\Psi_0 \rangle$, where ${{\cal U}}_I (t,0)$ is the evolution operator in interaction picture given by 
\begin{eqnarray}
{{\cal U}}_I (t,0)=e^{iH_B t}e^{-iH_{tot}t}.
\label{chapdos3}
\end{eqnarray}
The expansion method consists of representing the state $\mid\Psi_t \rangle$ in an environmental basis. Choosing the Bargmann coherent state basis \cite{bargmann1961,bargmann1962,glauber1963}, and an initial state $|\Psi_0\rangle=|\psi_0\rangle|0\rangle$, the total system state at a time $t$ can be expressed as \cite{strunz2001},
\begin{equation}
\mid \Psi_{t}\rangle=\int d\mu(z_i) G(z_i^* 0|t 0)|\psi_0\rangle| z_i \rangle.
\label{chapdos4b}
\end{equation}
In Eq. (\ref{chapdos4b}) we used the Gaussian measure 
\begin{eqnarray}
d\mu(z_i)=\prod_{\lambda} \frac {d^2 z_{i,\lambda} }{\pi}e^{-|z_{i,\lambda}|^2},
\label{Gaussianmeasure}
\end{eqnarray}
and the notation $\mid z_i \rangle=\mid z_{i,1}\rangle\mid z_{i,2}\rangle...\mid z_{i,\lambda}\rangle...$ for the state of the environment, given by a tensor product of the states of all the $\lambda$ environmental oscillators. The basis states for each oscillator are $\mid z_{i,\lambda}\rangle=\exp(z_{i,\lambda}a_{i,\lambda}^{\dagger})\mid0\rangle$. 
The system operator 
\begin{equation}
G(z_i^* 0|t 0)=\langle z_i \mid {{\cal U}}_I (t,0) \mid 0\rangle,
\label{chapdos5}
\end{equation}
with ${{\cal U}}_I (t,0)$ given by Eq. (\ref{chapdos3}), is the vacuum reduced propagator that was interpreted by \textcite{strunz2001} as a stochastic propagator. In a sense, they correspond to the Kraus operators in Eq. (\ref{map1}), considering the interaction picture and a Bargmann coherent basis€. Vacuum reduced propagators give rise to a displacement of the wave function from its initial value $\mid\psi_0\rangle$ to the value $|\psi_t (z_i^*)\rangle=G(z_i^* 0|t0)|\psi_0\rangle$ at time $t$, provided that the environment oscillators have evolved from the vacuum state $\mid 0\rangle$ to the state $\mid z_i \rangle$. 
The reduced density matrix of the system can then be recovered as 
\bea
\rho_s(t)=\int d\mu(z_i)|\psi_t(z_i^*)\rangle\langle\psi_t(z_i)|.
\label{SSEsum}
\eea
A generalized version of Eq. (\ref{chapdos5}), $G(z_i^* z_{i+1}|t_it_{i+1})=\langle z_i \mid {{\cal U}}_I (t,0) \mid z_{i+1}\rangle$, corresponding to an arbitrary initial state of the environment $z_{i+1}$, is useful to compute the OQS dynamics from any arbitrary total initial state $\rho(0)=\int d\mu(z_0) \int d\mu(z_0') |  z_0 \rangle |\psi_0 (z^*_0)\rangle\langle\psi_0 (z'_0)| \langle z'_0 |$. Its evolution can be derived as $\partial G(z_i^* z_{i+1}|t_it_{i+1})/\partial t_i=
\langle z_i | \partial {{\cal U}}_I(t_i,t_{i+1})/\partial t_i| z_{i+1} \rangle$,
where ${{\cal U}}_I (t_i,t_{i+1})$ satisfies the Schr\"odinger equation in the partial interaction picture 
\beqa
&&\frac{\partial {{\cal U}}_I(t_i,t_{i+1})}{\partial t_i}=\bigg(-i H_S -i 
\sum_{n} g_\lambda (L^\dagger a_\lambda e^{-i \omega_\lambda t_i}\nonumber\\&+& L a_\lambda^\dagger e^{i \omega_\lambda t_i})
\bigg)
{{\cal U}}_I(t_i,t_{i+1}).
\label{chapdos9}
\eeqa
Hence, the evolution equation for the reduced propagator is \cite{alonso2005}
\begin{eqnarray}
&&\frac{\partial G(z_i^* z_{i+1}|t_it_{i+1})}{\partial t_i}=
\bigg(
-i H_S\nonumber\\&-&i L \sum_\lambda g_\lambda e^{i \omega_\lambda t_i} z^*_{i,\lambda} 
\bigg) G(z_i^* z_{i+1}|t_it_{i+1}) \cr
&& -i L^\dagger \sum_\lambda g_\lambda e^{-i \omega_\lambda t_i}
\langle z_i | a_\lambda {{\cal U}}_I(t_i,t_{i+1}) | z_{i+1} \rangle,
\label{chapuno320}
\end{eqnarray}
where we used the property $\langle z|a^\dagger_\lambda=\langle z|z^*_\lambda $. 

To proceed further it is convenient to deal with the matrix element 
$\langle z_i |a_\lambda {{\cal U}}_I(t_i,t_{i+1}) | z_{i+1} \rangle$ that 
equals to $\langle z_i |{{\cal U}}_I(t_i,t_{i+1}) a_\lambda(t_i,t_{i+1}) | z_{i+1} \rangle$, with $a_\lambda(t_i,t_{i+1})={{\cal U}}^{-1}_I(t_i,t_{i+1}) a_\lambda {{\cal U}}_I(t_i,t_{i+1})$. Integrating the Heisenberg equations of motion for $a_\lambda(t_i,t_{i+1})$, $\frac{d}{dt_i }a_\lambda (t_i ,t_{i+1} )=-ig_\lambda e^{-i\omega_\lambda t_{i}}L(t_i ,t_{i+1})$,
it follows that
\begin{equation}
a_\lambda(t_i,t_{i+1})=a_\lambda(t_{i+1},t_{i+1})-i g_\lambda \int_{t_{i+1}}^{t_i} d\tau L(\tau,t_{i+1}) e^{i \omega_\lambda \tau},
\label{chapuno321}
\end{equation}
with $L(t_i,t_{i+1})={\mathcal U}^\dagger_I(t_i,t_{i+1})L{\mathcal U}(t_i,t_{i+1})$. 

Gathering the results, Eq. (\ref{chapuno320}) becomes
\begin{eqnarray}
&&\frac{\partial G(z_i^* z_{i+1}|t_it_{i+1})}{\partial t_i}=
\big(-i H_S+L z^*_{i,t_i}-L^\dagger z_{i+1,t_i}\big)\cr
&\times&G(z_i^* z_{i+1}|t_it_{i+1})-L^\dagger \int_{t_{i+1}}^{t_i} d\tau 
\alpha(t_i-\tau)
\langle z_i | {{\cal U}}_I(t_i,t_{i+1})\cr
&\times&L(\tau,t_{i+1})|z_{i+1} \rangle,
\label{chapuno323}
\end{eqnarray}
where we defined the functions
\begin{equation}
z_{i,t}=i\sum_\lambda g_\lambda z_{i,\lambda}e^{-i \omega_\lambda t},
\label{chapuno324}
\end{equation}
and
\begin{equation}
\alpha(t-\tau)={\mathcal M}[z_{i,t}z^*_{i,\tau}]=\sum_\lambda |g_\lambda|^2 e^{-i \omega_\lambda (t-\tau)}.
\label{chapuno325}
\end{equation}
Note that the last term of Eq. (\ref{chapuno323}) can also be written as 
\begin{eqnarray}
\langle z_i | {{\cal U}}_I(t_i,t_{i+1})L(\tau,t_{i+1})|z_{i+1} \rangle=\frac{\delta G(z_i^* z_{i+1}|t_i t_{i+1}) }{\delta z_{i,\tau}^* },\nonumber
\label{eqa21chap2}
\end{eqnarray}
In the above equation we defined the average ${\mathcal M}[z_{i,t}z^*_{i,\tau}]=\int d\mu(z_i)z_{i,t}z^*_{i,\tau}$, with the Gaussian measure defined in Eq. (\ref{Gaussianmeasure}), which leads to the environmental correlation function introduced in Sec. \ref{ch1sec7}, $\alpha(t-\tau)$. Since the environment is usually very large, and the coherent state variables $z_{i,\lambda}$ form a continuum, it is convenient to consider them as a complex Gaussian white noise, with the properties ${\mathcal M}[z^*_\lambda z_{\lambda'}]=\delta_{\lambda,\lambda'}$, and ${\mathcal M}[z_\lambda]=0$. In that case Eq. (\ref{chapuno324}) becomes a complex Gaussian noise with the properties,
\bea
&&{\mathcal M}[z_{i,t}]=0;\cr
&&{\mathcal M}[z_{i,t}z^*_{i,\tau}]=\alpha(t-\tau),
\label{statprop}
\eea
Hence, the only information needed about the environment is its correlation function, or equivalently its spectral density $J(\omega)$. In other words, if $\alpha(t)$ is at our disposal, we can generate a Gaussian distributed set of complex random numbers in such a way that they have the required properties (\ref{statprop}). Also, such environmental function is indeed responsible for the dependency of the evolution of the system over its past history, as it is the kernel of an integral term from the initial time $t_{i+1}$ to the actual time $t_i$. 
Note that once interpreting Eq. (\ref{chapuno324}) as a noise, Eq. (\ref{chapuno323}) leads to a SSE for the wave function $|\psi_t(z^*,z_0) \rangle=G(z^* z_0|t t_0)|\psi_0\rangle$.

Similarly, for the case $z_{i+1}=0$ the evolution equation of the system state vector can be written as \cite{strunz2001}
\begin{eqnarray}
&&\frac{d|\psi_t (z^*)\rangle}{dt}=-iH_S |\psi_t (z^*)\rangle+L z^*_t |\psi_t (z^*)\rangle\nonumber\\
&-&L^\dagger \int^t_0 d\tau \alpha(t-\tau) \frac{\delta}{\delta z_{\tau}^* } |\psi_t (z^*)\rangle.
\label{linearNM}
\end{eqnarray}
The same equation was first derived by \textcite{diosi1997} without using an expansion, but considering the equation for state vectors depending on a Wiener-stochastic process that depends on a complex colored Gaussian noise. An alternative derivation of the former equation was given by \textcite{cresser2000}. An extension of the SSE (\ref{linearNM}) for fermionic environments has recently been derived in \cite{zhao2012}. 
Such an extension can also be a useful tool for studying OQSs coupled to a spin-chain environment, when this can be transformed into an effective fermionic environment.

From Eq. (\ref{chapuno323}) [similarly (\ref{linearNM})], we could integrate the reduced propagators with the initial conditions $G(z_i^* z_{i+1}|t_i t_i)=\exp(z_i^* z_{i+1})$. However, the time dependency of the operator appearing in the last term of Eq. (\ref{chapuno323}), $L(\tau, t_{i+1})={\mathcal U}^{\dagger}_I (\tau,t_{i+1})L{\mathcal U}_I (\tau,t_{i+1})$
is over the total Hamiltonian operator, so that Eq. (\ref{chapuno323}) is still not a closed equation over the reduced Hilbert space of the system, but is merely a particular representation of the Schr{\"o}dinger equation for the system and the environment. 
In general, it is not always possible to exactly compute the last term 
and only in very exceptional cases can this be done. Particularly, when $L(\tau,t_{i+1})\propto H_S(\tau,t_{i+1})$, then $[L,H_{tot}]=0$ and therefore $H_S(\tau,t_{i+1})=H_S$, so that $\langle z_{i}|L(\tau,t_{i}){\mathcal U}_I (t_i , t_{i+1})|z_{i+1}\rangle= H_S G(t_i t_{i+1}|z^*_i z_{i+1})$. Also, as will be further discussed in Sec. \ref{brownian}, \cite{ferialdi2012} derives the exact analytical solution of an SSE similar in form to Eq. (\ref{linearNM}), for the particular case when the system is a harmonic oscillator and the environment is in a thermal state. 

In other situations, a perturbative expansion of $L(\tau,0)$ is needed, which up to the second-order leads to expressing $\langle z | {{\cal U}}_I(t,0)L(\tau,0)|z_{0} \rangle=\frac{\delta G(z^* z_{0}|\tau 0) }{\delta z_{\tau}^* }\approx V_{\tau-t}L$. 
Similarly, it can be considered, as an ansatz, that the matrix element can be written as \cite{diosi1998,yu1999}, 
$\langle z_i | {{\cal U}}_I(t_i,t_{i+1})L(\tau,t_{i+1})|z_{i+1} \rangle =
O(z_{i+1} z^*_{i},t,\tau)G(z_i^* z_{i+1}|t_i t_{i+1})$
where the operator $O$ belongs to the system's Hilbert space and shall be obtained for each case. 
For $z_{i+1}=0$, the above ansatz has been complemented with the \textit{consistency condition} \cite{diosi1998},
\begin{eqnarray}
\frac{d}{dt}\frac{\delta \mid \psi_t (z^*)\rangle}{\delta z^*_\tau}=\frac{\delta}{\delta z^*_\tau}\frac{d\mid \psi_t (z^*)\rangle }{dt},
\label{chapdos23}
\end{eqnarray}
to obtain $O(z^*_{i},t,\tau)$ systematically. Also, the ansatz and the consistency condition have been used in the many-body case to analyze the dynamics of energy transport in quantum aggregates \cite{roden2009}. In this context, the validity of the SSE approach is confirmed by comparing its solution to the one provided by the pseudomode approach discussed in Sec. \ref{embeddingM}.

An alternative to the consistency condition was recently proposed by \textcite{suess2014}, and consists of obtaining the evolution equation of $\frac{\delta \mid \psi_t (z^*)\rangle}{\delta z^*_\tau}=\psi_t^{1}$. For the case of a correlation function of the form $\alpha(t)=ge^{-\Omega t}$, this equation becomes simply
\bea
\frac{d\psi_t^{1}}{dt}=(-iH-\Omega+Lz_t^*)\psi_t^{0}+\alpha(0)L\psi_t^{0}-L^\dagger \psi_t^{2},\nonumber
\eea
where $\psi_t^{k}=\frac{\delta^k \mid \psi_t (z^*)\rangle}{\delta z^{*k}_\tau}$. In general, for exponential correlation functions the evolution equation for the $k-th$ functional derivative of the system wave-vector can be written as 
\bea
\frac{d\psi_t^{k}}{dt}=(-iH-k\Omega+Lz_t^*)\psi_t^{k-1}+\alpha(0)L\psi_t^{k-1}-L^\dagger \psi_t^{k+1},\nonumber
\eea
with $\psi^{0}_{t=0}=|\psi_0\rangle$ and $\psi^{k}_{t=0}=0$ for $k>0$. To make practical use of this hierarchy, one may truncate it at a certain order $k$, by using a terminator $\psi^{k+1}_{t}=\frac{\alpha(0)}{\Omega}L\psi^{k}_{t}$.

\subsubsection{Nonlinear SSEs}

As noted by \textcite{diosi1998}, the linear equation obtained with the previous methods has one major drawback. During the evolution of the trajectories, 
the solutions $|\psi_{t} (z^*)\rangle$ may lose their norm and therefore their statistical relevance. This problem comes from not having considered the fact that the interaction between the system and the environment not only affects the system, but also the environment itself. 

To see this more clearly, a Husimi function (or Q-function) \cite{quantumoptics} of the environment is considered \cite{strunz2001},
\begin{equation}
Q_t(z,z^{*})=\frac{e^{-| z|^2}}{\pi} \langle z|
\mbox{Tr}_s\left[|\Psi_t \rangle\langle\Psi_t|\right] | z \rangle,
\label{husi1}
\end{equation}
where $|z\rangle$ denotes a coherent state of the environment in the Bargmann basis. Since each of these states corresponds to a certain value of the noise, the function $Q_t (z,z^*)$ may be interpreted as the probability distribution of the noise. The substitution of $|\psi_t (z^*)\rangle =\int d\mu(z)|\psi_t (z^*)\rangle\langle \psi_t (z)|\otimes |z\rangle\langle z|$ into Eq.  (\ref{husi1}) gives the following expression:
\begin{equation}
Q_t(z,z^{*})=\langle\psi_t(z) |\psi_t (z^*)\rangle Q_0 (z,z^*),
\label{husi2}
\end{equation}
with $Q_0 (z,z^*)$ as the initial Gaussian distribution of coherent states $Q_0(z,z^{*})=\frac{e^{-| z |^2}}{\pi}$. In terms of Eq. (\ref{husi2}), the density operator can be defined as a mixture of pure normalized states weighted by $Q_t (z,z^*)$,
\begin{equation}
\rho_s =\int d^2 z\, Q_t (z,z^*)\frac{| \psi_t (z^*)\rangle\langle\psi_t (z)|}
{\langle\psi_t (z)|\psi_t (z^*)\rangle}.
\label{mean3}
\end{equation}
With Eq. (\ref{mean3}) it is clearer to see that once the interaction is "switched on'" and the environmental oscillators start to move away from the origin according to the distribution $Q_t (z,z^*)$, the states $|\psi_t (z^*)/\langle\psi_t (z)|\psi_t (z^*)\rangle^{1/2}$, which according to $Q_0 (z,z^*)$ correspond to small $z$, will have a decreasing weight in the sum (\ref{mean3}). 

The Husimi function shows a closed time evolution of Liouville form for the set of oscillators $z_\lambda$ composing the quantity $z_t$, corresponding to the phase space flow \cite{diosi1998} 
\begin{equation}
\dot{z}^*_\lambda =ig_\lambda e^{-i\omega_\lambda t}\langle L^{\dagger} 
\rangle_t.
\label{flow}
\end{equation}
In terms of the trajectories $z(t)$ that follow this flow, the Husimi function $Q_t (z,z^*)$ at time $t$ can be expressed as 
\begin{equation}
Q_t(z,z^{*})=\int d^2 z_0\,Q_0 (z_0,z_0^*)\delta^2 (z-z(t)),
\label{husi3}
\end{equation}
where somewhat symbolically $z(t)$ represents the set of solutions of the different trajectories of the 
oscillators starting from the set of initial values $\{z_{\lambda}^* (0)=z^* _{\lambda,0}\}$. In this way, we can now replace (\ref{mean3}) by an integral of wave functions evaluated in the dynamical 
states $z^* (t)\equiv\{z^*_\lambda (t)\}$ as
\begin{eqnarray}
&&\rho_t =\int d^2 z_0\, Q_0 (z_0,z_0^*)
\frac{| \psi_t (z^* (t))\rangle\langle\psi_t (z^* (t))|}
{\langle\psi_t (z^* (t))|\psi_t (z^* (t))\rangle} =\nonumber\\
&&\int \frac{d^2z_0}{\pi} e^{-|z_0|^2}
\frac{| \psi_t (z^* (t))\rangle\langle\psi_t (z^* (t))|}
{\langle\psi_t (z^* (t))|\psi_t (z^* (t))\rangle}.
\label{mean4}
\end{eqnarray}

Now to perform the integral (\ref{mean4}) with a Monte Carlo method, a new stochastic variable $\tilde{z}^*_t$ is defined, which corresponds to $z^* (t)$ with a random selection of the initial values for the environmental oscillators $\{z_\lambda ^* (0)\}$. From the flow equation (\ref{flow}), one obtains
\begin{eqnarray}
\tilde{z}^*_t=z^*_t+g\int d\tau\alpha^*(t-\tau)\langle L^{\dagger}\rangle_\tau.
\label{shifted}
\end{eqnarray}
Here, the variable $z^*_t$ is the noise as it appears in the linear stochastic Schr\"odinger equation,  which corresponds to the stationary statistics with distribution function $Q_0 (z,z^*)$. The last term represents a dynamical shift or displacement of each $z_t$, which depends on the history of the interaction with the system. The stochastic equation for the wave function $|\psi(z(t))\rangle$ with a shifted noise in the driving term is
\cite{diosi1998}
\begin{eqnarray}
\frac{| \psi_t(z^* (t))\rangle}{dt}&=&-iH_S | 
\psi_t(z^* (t))\rangle+gL\tilde{z}^*_t
| \psi_t(z^*(t))\rangle \nonumber \\
&-&g^2 (L^{\dagger}-\langle L^{\dagger}\rangle_t)\bar{O}(t,z^*(t))| 
\psi_t(z^*(t))\rangle \quad,
\label{nl1}
\end{eqnarray} 
with $\bar{O}=\int^t_0 d\tau \alpha(t-\tau)O(t,\tau,z^*(t))$.
By evolving Eq. (\ref{nl1}) we ensure that the wave functions $|\psi_t (z^*(t))\rangle$ correspond to those realizations that contribute with a significant probability, which is ensured by the shift term in Eq. (\ref{shifted}). This is because the equation depends on a noise (\ref{shifted}) that dynamically follows the motion of the center of the Gaussian distribution in the environment state space. 
As shown by \textcite{devega2005a}, the probability function for the noise corresponding to an environment at high temperature, evolves quite significantly in time, so that a nonlinear equation needs to be considered. 
Conversely, for low temperatures the state distribution of the environment (\textit{i.e.} the noise distribution) remains quite close to a Gaussian distribution centered at the origin during the interaction, and the linear equations provide an accurate description of the problem.

\subsubsection{Projection method}
\label{projectionSSE}
Using the \textit{Feshbach projection-operator} method, \textcite{gaspard1999b} derive a non-Markovian SSE that is identical to the one obtained with the expansion method up to the second-order in the perturbative parameter. The projection-operator method is based on the same idea as the Nakajima-Zwanzig method, but is applied to the Schr\"odinger equation instead of the master equation. 
As in Sec. \ref{expansion}, the evolution equation of the total system wave function is considered. This wave function is expressed in the coordinate representation for both the system $\{x_s\}$ and the environment $\{x_b\}$, as $\Psi_t(x_s,x_b)=\sum_n \phi_{n}(x_s,t)\chi_{n}(x_b)$, where $\{\phi_{n}(x_s,t)\}$ is the set of coefficients of this linear expansion. The $\chi_{n}(x_b)$ functions depend only on the environmental degrees of freedom, so that the dependency of the total wave function over the system degrees of freedom is entirely encoded in the coefficients $\phi_{n}(x_s,t)$ of the linear decomposition. 

The normalized version of these coefficients, 
\begin{eqnarray}
\hat{\phi}_n(x_s;t)=\phi_n(x_s;t)/\|\phi_n(x_s;t)\|,
\end{eqnarray}
can be considered as a statistical set of wave functions of the system. In terms of these, the reduced density matrix can be written as $\rho_s=\sum_n p_{n}(t)|\hat{\phi}_{n} \rangle\langle \hat{\phi}_{n}|$,
where $\hat{\phi}_n(x_s;t)=\phi_n(x_s;t)/\|\phi_n(x_s;t)\|$ and $p_n(t)=\int dx_s |\phi_{n}(x_s,t)|^2=\|\phi_n(x_s;t)\|^2$ is the probability for the environment to be observed in a certain state $\chi_{n}(x_b)$.
The statistical character of $\hat{\phi}_n(x_s;t)$ appears through its dependency on the environmental state (of index $n$). Then the probability of each system wave function is given by the probability $p_n(t)$ of observing the environment in the corresponding basis state $\chi_{n}(x_b)$. Thus, the quantum system can no longer be described through a single wave function, but through a collection of them, and the dynamics of the system is conditioned on the dynamics of its environment.

In order to obtain an evolution equation for these coefficients, the Schr{\"o}dinger equation of the total system is decomposed in two equations, using the projectors ${\mathcal P}$ and ${\mathcal Q}$ that act over the total Hilbert space, with properties (\ref{F22}), and such that 
${\mathcal P}\Psi(x_s,x_b)=\phi_{l}(x_s;t)\chi_{l}(x_b)$ and ${\mathcal Q}\Psi(x_s,x_b)=\sum_{n(\neq l)}\phi_{n}(x_s;t)\chi_{n}(x_b)$. 
The time dependency of ${\mathcal P}\Psi$ is entirely encoded in the coefficient $\phi_{l}(x_s;t)$. Its evolution in the total interaction picture is 
\begin{eqnarray}
i\frac{\varphi_l(t)}{dt}&=&f_l (t) -ig^2 \int_{0}^{t}d \tau \sum_{\eta \gamma}V_t S_{\eta}\langle l|V_t B_{\eta}V_\tau B_{\gamma} | l\rangle V_\tau S_{\gamma} \cr
&\times&\varphi_{l}(\tau)+{\mathcal{O}}(g^3).
\label{F36}
\end{eqnarray}
Here, an interaction Hamiltonian of the form (\ref{chapuno12}) has been considered, representing a sum of system $S_\eta$ and environment $B_\eta$ Hermitian operators. 
In addition, an expansion up to the second-order in the weak coupling parameter $g$ has been performed. 
Equation (\ref{F36}) has two different terms. 
The first originates from the initial condition $Q\Psi(0)$ of all the coefficients except $P\Psi$, and has the form 
\begin{eqnarray}
f_l (t)&=&g\sum_{\eta}\sum_{m(\neq l)}V_t S_{\eta} \langle l|V_t B_{\eta}| m\rangle \varphi_m (0)-ig^2 \int_{0}^{t}d \tau\cr
&\times&
\sum_{\eta \gamma}\sum_{m(\neq l)}V_t S_{\eta} 
\langle l|V_t B_{\eta}V_\tau B_{\gamma} | m\rangle V_\tau S_{\gamma} \varphi_{m}(0)+{\mathcal{O}}(g^3)\nonumber
\label{F37}
\end{eqnarray}
with the assumption that $\langle l| B_{\eta}| l\rangle=0$. This term will be identified later with the \textit{stochastic forcing} over the system due to the environmental fluctuations. The second term corresponds to the \textit{damping} of the coefficient or wave function ${\mathcal P}\Psi$ (or $\varphi_l$) due to its coupling with the other coefficients ${\mathcal Q}\Psi$, which is produced through the interaction with the environment. As it is an integral up to the actual time $t$, this term is responsible for the non-Markovian character of the equation.

In order to use the former equation to derive a stochastic Schr{\"o}dinger equation, it is necessary to assume that the coefficient $\varphi_{l}(t)$ statistically represents each of the coefficients $\varphi_{n}(t)$ of the decomposition of the total wave function. In other words, it is necessary to assume that all the coefficients evolve in a similar way, so that $\varphi_{l}(t)$ is a typical representative of the rest of the statistical ensemble. This hypothesis, known as statistical typicality, has been justified for classically chaotic systems, but is not necessarily valid for every environmental state basis $\chi_l (x_b)$ chosen. 
However, it is reasonable to assume that this hypothesis is fulfilled for most of the environmental states, since it has its origins in the fact that the typical eigenfunctions of high quantum numbers are statistically irregular. 

Thanks to statistical typicality, and following a conjecture of \textcite{berry1977}, the quantum mean value of an environmental operator $C$ over a typical eigenstate $\chi_l$ is equivalent to the quantum mean value over a representative state of the micro-canonical ensemble with the corresponding energy $e_l$.
In addition, since the environment is large, following \textcite{srednicki1994}, it can also be supposed that such a mean over the state of the micro-canonical ensemble is essentially equivalent to a mean over a typical state of the canonical ensemble. As a consequence, a quantum average of an environmental operator $B$ over a typical environmental eigenstate $\chi_l$ is approximately equal to a thermal mean,
\begin{eqnarray}
\langle l| B | l\rangle\approx \Ttr_B \bigg\{\frac{e^{-\beta H_B}}{Z_b}B\bigg\}\equiv \Ttr_B\{ \rho_B ^{eq}B\},
\label{F39}
\end{eqnarray}
where $Z_b=\Ttr_B \{\exp(-\beta H_B)\}$. The inverse temperature $\beta$ should be fixed for a given environmental eigen-energy $e_l$. Also, the variation of such environmental energy due to the interaction with the system is assumed to be negligible, since such variation is very small in comparison with its energy $e_l$. Taking Eq. (\ref{F39}), the damping term (\ref{F36}) can be written in terms of the environment correlation function
\begin{eqnarray}
&&\langle l| V_t B_{\eta}V_\tau B_{\gamma} | l\rangle\approx
\Ttr_B\{ \rho_B ^{eq}V_t B_{\eta}V_\tau B_{\gamma}\}\equiv C_{\eta \gamma}(t-\tau),\nonumber
\label{F40}
\end{eqnarray}
a form which, thanks to statistical typicality, is independent of the particular choice of the coefficient $| l\rangle$.
Also, in order to find the typical behavior of the forcing term, it is necessary to assume that the initial state is a tensor product of the system $\psi$ and the mixed canonical state of the environment. 

Assuming all these approximations over Eq. (\ref{F36}), the following stochastic differential equation is obtained for a typical coefficient $\varphi_l$:
\begin{eqnarray}
i\frac{\varphi_l(t)}{dt}&=&g\sum_\eta \zeta_\eta(t)S_\eta \varphi_l (t) -ig^2 \int_{0}^{t}d \tau \sum_{\eta \gamma}C_{\eta\gamma}(t-\tau)\cr
&\times&
V_t S_{\eta}V_\tau S_{\gamma} \varphi_{l}(\tau)+\theta(g^3).
\label{FF54}
\end{eqnarray}
Here we reexpressed the stochastic forcing as
$f_l (t)\approx g\sum_{\eta}\zeta_{\eta}(t)S_{\eta}\varphi_l (t)$,
up to the second-order in $g$, defining a term
\begin{eqnarray}
\zeta_{\eta}(t)\equiv \sum_{m(\neq l)} \langle l| V_t B_{\eta} | m\rangle e^{-\beta(e_m -e_l)/2}e^{i(\theta_m -\theta_l)}
\label{F47_}
\end{eqnarray}
which may be interpreted as the \textit{stochastic} forcing that acts on the system due to its interaction with the environment. Indeed, when the environment is large enough, the quantity defined in Eq. (\ref{F47_}) is given by a sum of a large number of oscillating complex terms that, following the central limit theorem \cite{watson1952}, gives rise to random variables of a Gaussian type. In summary, the random variables appearing in Eq. (\ref{F47_}) can be taken as Gaussian noises characterized by a zero mean value and a correlation function $C_{\eta\gamma}(t-\tau)$, \textit{i.e.} following similar properties as (\ref{statprop}).

The following step is taken to obtain an evolution from a general initial condition $\rho_s(0)=\sum_\lambda | \psi_k(0)\rangle\langle \psi_k(0)|$, 
defined in terms of the system wave functions $\psi_k$ and their probabilities $\{p_k\}$, so that $\sum_\lambda p_k=1$. In such a case, we consider the following statistical set of coefficients:
\begin{eqnarray}
\varphi_l(t)\approx|\psi_{k}(x_s;t)\rangle\sqrt{\frac{e^{-\beta w_l}}{Z_b}}e^{i\theta_l}
\label{FF56_}
\end{eqnarray}
where $l$ is the index appearing in Eq. (\ref{FF54}), $k$ specifies the member of the statistical mixture.
Replacing Eq. (\ref{FF56_}) in Eq. (\ref{FF54}), and eliminating the factor that multiplies $\psi_{k}$ on both sides, the following equation is obtained up to the second oder in $g$ :
\begin{eqnarray}
&&i\frac{d|\psi_{k}(t)\rangle}{dt}=-iH_S|\psi_{Ik}(t)\rangle+g\sum_\eta \zeta_\eta(t)S_\eta |\psi_{k}(t)\rangle\cr
&-&ig^2 \int_{0}^{t}d \tau \sum_{\eta \gamma}C_{\eta\gamma}(t-\tau)S_{\eta}V_{\tau-t} S_{\gamma} e^{-iH_S(t-\tau)}|\psi_{k}(\tau)\rangle,\cr
\label{F570}
\end{eqnarray}
where $\psi_{k}(x_s;t)=\langle x_s|\psi_{k}(t)\rangle$.
In this equation, the Gaussian noises $\eta_\beta (t)$ satisfy
\begin{eqnarray}
&&\overline{\zeta_\eta(t)}=0, \overline{\zeta_\eta (t)\zeta_\gamma (\tau)}=0, \nonumber\\
&&\overline{\zeta_\eta^{*} (t)\eta_{\gamma}(\tau)}=C_{\eta \gamma}(t-\tau)=C^{*}_{\gamma \eta}(\tau-t).
\label{F59}
\end{eqnarray}
Inserting $e^{-iH_S(t-\tau)}|\psi_{k}(\tau)\rangle=|\psi_{k}(t)\rangle+{\mathcal O}(g^2)$ in the last term of Eq. (\ref{F570}), which is already of second-order in $g$, leads to a a time-local equation in $|\psi_k\rangle$. This time-local equation is equivalent to Eq. (\ref{linearNM}) when approximating $\frac{\delta |\psi_t\rangle}{\delta z_{\tau}^* }\approx V_{\tau-t}L$, and considering the equivalences in Eq. (\ref{chapuno30}).

\subsubsection{Continuous measurement theory method, and measurement of a quantum evolution}
\label{measurementSSE}

Non-Markovian SSEs can also be derived based on \textit{continuous measurement} theories. For instance, \textcite{jack1999,jack2000} presented a formulation of non-Markovian quantum trajectories which describes the real-time spectral detection of the light emitted from a localized system. In this case, the non-Markovian behavior is not intrinsic to the interaction of the system with its environment, but arises from the uncertainty in the time of emission of particles that are later detected. 
More recently, \textcite{gambetta2002} propose a formal way to obtain non-Markovian SSEs from a continuous measurement scheme. They discussed all the mathematical ingredients to describe a continuous measurement \cite{davies1976,kraus,wiseman1996}. This includes a probability-operator-measure element, or \textit{effect},
$\tilde{F}_{\{q_\lambda \}}=|\{q_\lambda\}\rangle\langle\{q_\lambda\}|$,
where $|\{q_\lambda\}\rangle$ is the environmental basis, and $\{q_\lambda\}$ is the result of the measurement. A set of \textit{measurement operators} $\tilde{M}_{q_\lambda }$ is also necessary, with the constraint $\tilde{F}_{\{q_\lambda\}} =\tilde{M}^{\dagger}_{q_\lambda }\tilde{M}_{q_\lambda }$. For example, we can decompose the measurement operators as $\tilde{M}_{q_\lambda }=|\{n_\lambda \}\rangle\langle\{q_\lambda \}|$, 
where the final state of the environment after a measurement $\{n_\lambda \}$ can be chosen as the vacuum, since in most detection situations the measurement generally results in annihilating the detected field. A \textit{noise operator} $\hat{Z}(t)$ is also defined in such a way that
$\hat{Z}(t)|\{q_\lambda\}\rangle=\hat{z}_t |\{q_\lambda\}\rangle$,
where $\hat{z}_t$ is the noise function from which the conditioned state after a measurement depends. With these definitions at hand, two kinds of such conditioned system states can be obtained after measurement. The first state $|\psi_{q_\lambda }(t)\rangle$ is such that: a) it depends \textit{linearly} on the premeasurement state $|\psi_t \rangle$, and b) it depends on an environmental state $\{q_\lambda \}$, which is distributed according to a probability $\Lambda (\{q_\lambda \})$ that does not take into account the effects of the interaction of the environment and remains constant in time. In such terms, the linear state after the measurement of $\{q_\lambda \}$ is written as
\begin{eqnarray}
|\psi_{q_\lambda }(t)\rangle =\frac{\langle \{q_\lambda \}|\psi_t \rangle}{\sqrt{\Lambda(\{q_\lambda \})}}.
\label{appB3}
\end{eqnarray}
Because it is not normalized, they argued that the linear conditioned system state does not have a clear physical interpretation, but is useful to derive the actual probability $P(\{q_\lambda\},t)$ that the environmental states have, considering their interaction with the system as
\begin{eqnarray}
P(\{q_\lambda\},t)=\langle\psi_{q_\lambda }(t)|\psi_{q_\lambda }(t)\rangle \Lambda (\{q_\lambda \})
\label{appB4}
\end{eqnarray}
Such probability is obtained through a Girsanov transformation of the variables $\{q_\lambda \}$ \cite{gatarek1991}. 

This actual probability allows for the derivation of a second kind of conditioned state $|\tilde{\psi}_{q_\lambda }(t)\rangle$ that: a) evolves in a \textit{nonlinear} way and b) depends on an environmental state $\{q_\lambda \}$ that is sampled according to the actual distribution (\ref{appB4}),
\begin{eqnarray}
|\tilde{\psi}_{q_\lambda }(t)\rangle =\frac{\langle \{q_\lambda \}|\psi_t \rangle}{\sqrt{P(\{q_\lambda \},t)}}.
\label{appB5}
\end{eqnarray}
A linear SSE can be derived from Eq. (\ref{appB3}) as
\begin{eqnarray}
\frac{d|\psi_{\{q_\lambda\}}(t)\rangle}{dt}=\frac{\partial |\psi_{\{q_\lambda \}}(t)\rangle}{\partial t}+\sum_\lambda \frac{dq_\lambda}{dt}\frac{\partial |\psi_{\{q_\lambda\}}(t)\rangle}{\partial t},
\label{appB6}
\end{eqnarray}
and provided that a Girsanov transformation can be made, a nonlinear SSE results in
\begin{eqnarray}
\frac{d|\tilde{\psi}_{\{q_\lambda\}}(t)\rangle}{dt}&=&\frac{1}{|\psi_{\{q_\lambda\}}(t)|}\frac{d|\psi_{\{q_\lambda\}}(t)\rangle}{dt}+|\psi_{\{q_\lambda\}}(t)\rangle \cr
&\times&\frac{d}{dt}\frac{1}{|\psi_{\{q_\lambda\}}(t)|},
\label{appB7}
\end{eqnarray}
where $|\tilde{\psi}_{\{q_\lambda\}}(t)\rangle=\frac{1}{|\psi_{\{q_\lambda\}}(t)|} |\psi_{\{q_\lambda\}}(t)\rangle$, and $|\psi_{\{q_\lambda\}}(t)|=\langle\psi_{\{q_\lambda\}}(t)|\psi_{\{q_\lambda\}}(t)\rangle$.
Since it is normalized, the former state represents, with a probability equal to $1$, the conditioned state of the system after a measurement of output $\{q_\lambda \}$ has been performed at time $t$ in the environment. This statement is true whether the interaction is Markovian or non-Markovian. However, the linking of such a state with earlier states obtained by evolving Eq. (\ref{appB7}) is possible only in the first type of interaction. Once a measurement of the environmental state has been made at time $t$, a future measurement performed at time $t+\Delta t$ is altered if $\Delta t < \tau_c$. In other words, the measurement at time $t+\Delta t$ is performed before the environment has recovered from the last measurement, since the recovery time is of the order of $\tau_c$. Considering that $\Delta t \rightarrow 0$ for a continuous measurement, only in the Markovian case in which the correlation time $\tau_c =0$ do the sequences of measurements that monitor a trajectory not affect each other. Thus, according to \textcite{gambetta2002,wiseman2008}, there are no genuine non-Markovian quantum trajectories: monitoring the field feeds back into the system and this can change the average evolution of its state. The result is that an average over such a monitored trajectory would not reproduce on average the non-measurement evolution that a non-Markovian SSE does. However, \textcite{diosi2008,diosi2008erratum} has concluded that the non-Markovian SSE describes a time-continuous measurement that includes delay and \textit{retrodiction} (\textit{i.e.} an account of the past). 

As shown by \textcite{barchielli2012} (see also \cite{barchielli1995}), another way to include non-Markovian effects, but which permits to maintain at the same time the continuous measurement interpretation is to start from the linear SSE and to generalize it by considering the presence of stochastic coefficients. This allows us to describe the non-Markovian evolution of a quantum system continuously measured and controlled, thanks to a measurement-based feedback, and in a mathematically consistent way. 

In this context of measurement, a result by \textcite{galve2014,giorci2015} described how the non-Markovian character of an evolution inhibits quantum Darwinism. Such quantum Darwinism explains the emergence of a classical objective reality by the fact that a system that dissipates spreads to its environment multiple redundant copies of the same information \cite{zurek2009}. As a result, each small fraction of the environment contains almost all information classically accessible on the system, which can then be observed by multiple observers without perturbing the system. The existence of an information flow-back produced by the non-Markovianity of the system evolution (which also prevents the existence of genuine trajectories), reduces such redundancy, and hence the emergence of an objective classical reality.  

\subsubsection{Embedding methods}
\label{embeddingSSE}

Similar to the embedding methods described in Sec. \ref{embeddingM} for master equations, \textcite{breuer2004} propose a stochastic unraveling of states living in an extended space. Just as in the master equation case, such an extended state is given by a tensor product of the original system state space ${\mathcal H}$ and ${\mathcal C}^3$. Then, states $|\Phi_t \rangle$ in this extended space have the general form
\begin{eqnarray}
|\Phi_t \rangle=|\varphi_a (t)\rangle |a\rangle+|\varphi_b (t)\rangle |b\rangle+|\varphi_c (t)\rangle |c\rangle,
\end{eqnarray}
where $|\varphi_k \rangle\, \epsilon \,{\mathcal H}$ (k=a,b,c). 
Also, the coherences can be expressed as $W_{ab}={\mathcal M}\left[|\varphi_a (t)\rangle\langle\varphi_b (t)| \right]$, in terms of wave functions of the extended space $|\varphi_k \rangle$, which obey a Markovian evolution and therefore have the physical interpretation of continuous measurements. In this way, a reduced density matrix that is equivalent to the one obtained with the master equation (\ref{nmbreuer}) can be recovered by considering
\begin{eqnarray}
\rho_s (t)=\frac{{\mathcal M}\left[|\varphi_a (t)\rangle\langle\varphi_b (t)| \right]}{{\mathcal M}\langle\varphi_b (t)|\varphi_a (t)\rangle}.
\label{averagepierre}
\end{eqnarray}
Thus, an unraveling is constructed for non-Markovian dynamics, which consists of two wave functions, each of which is described by a particular Markovian SSE in the extended Hilbert space. Note that similar to the SLN method, which is presented in Sec. \ref{path}, a reduced density matrix with non-Markovian evolution is recovered with an average of two memoryless system wave functions. However, contrary to SLN, this method has the limitation that it starts from the general form of the time-convolutionless Eq. (\ref{nmbreuer}), which may not be valid for strong couplings. Similar considerations were made by \textcite{budini2013}, where a quantum jump unraveling is constructed to describe the dynamics of the OQS and an ancilla. 

\subsubsection{Quantum jumps}
\label{jumps}
One of the main obstacles to unraveling a non-Markovian master equation of the form (\ref{nmbreuer}), into a set of quantum jump trajectories, is the appearance of negative quantum jump probabilities during the evolution. 
These occur precisely at the times when the decay rates $\Delta_k(t)$ become negative. This problem was tackled by \textcite{piilo2008,piilo2009}, who realized that when the decay rates become negative, the direction of the information flow between the system and the environment is reversed. In this picture, at times when the rates are positive, the system loses its information to the environment, and quantum jumps have a similar effect and structure as for Markovian dynamics. In turn, when rates become negative, the system may regain some of the information it lost earlier, which means that the seemingly lost superpositions in the ensemble can be restored. Between jumps, the system undergoes a deterministic evolution according to the Hamiltonian 
\bea
H=H_S-\frac{i}{2}\sum_k \Delta_k(t) C_k(t)^\dagger C_k(t).
\label{heff}
\eea

It is in the jump dynamics where the non-Markovian character introduces a difference between \emph{forward jumps} that take place in channels $k_+$ when decay rates are positive, and \emph{backward jumps}, that take place in channels $k_-$ when decay rates are negative. The forward jump process occur when $\Delta_k(t)> 0$ and is very much like the Markovian case, corresponding to transitions
\begin{equation}
|\psi_{\alpha}\rangle \rightarrow |\psi_{\alpha'}(t+\delta t)\rangle\equiv \frac{C_{k}(t)}{||C^\dagger_{k}|\psi_\alpha(t)\rangle||}|\psi_\alpha(t)\rangle,
\label{eq:}
\end{equation}
with probability
\bea
\label{positivejump}
P^{k+}_\alpha(t)=\Delta_{k}(t)\delta t\langle\psi_\alpha(t)|C^\dagger_{k} (t) C_{k}(t)|\psi_\alpha (t)\rangle.
\eea
A backward jump occurs when $\Delta_k<0$ and produces the transition
\begin{equation}
|\psi_{\alpha'}(t+\delta t) \rangle \leftarrow 
|\psi_\alpha(t)\rangle \equiv 
\frac{C_{k_-}(t)}{||C_{k_-}|\psi_\alpha'(t)\rangle||}
|\psi_\alpha'(t)\rangle,
\label{eq:}
\end{equation}
with probability
\bea
P^{k-}_{\alpha}(t)=\frac{N_{\alpha'}(t)}{N_\alpha(t)}|\Delta_{k}(t)|\delta t\langle\psi_{\alpha'}(t)|C^\dagger_{k}(t) C_{k}(t)|\psi_{\alpha'} (t)\rangle,\cr
\label{negativejump0}
\eea
where $N_{\alpha}$ is the number of ensemble members in state $|\psi_\alpha(t)\rangle$ at time $t$. 
The reduced density operator of the system can be constructed as $\rho_s(t)=\sum_\alpha P_\alpha(t)|\psi_\alpha(t)\rangle\langle\psi_\alpha(t)|$, with $P_\alpha(t)=\frac{N_\alpha(t)}{N}$, and $N$ as the ensemble size.

Note that the non-Markovian quantum jump method has certain differences with respect to the Markovian quantum jumps. While fully Markovian trajectories are uncorrelated with each other, here one should in principle [although not in practice, as discussed by \textcite{piilo2009}] simultaneously propagate the ensemble of $N$ trajectories. The reason is that the quantity $N_{\alpha'}/N_\alpha$ necessary to determine the negative jump probability should be known, and this depends on the actual number of trajectories $N_\alpha(t)$ at a certain state $|\psi_\alpha\rangle$. Hence, the $N$ trajectories should be propagated in a self-consistent way, such that $N_\alpha(t)$ vary at times when one of the trajectories performs a quantum jump. As a result of this, the different realizations of the process are correlated, since the quantity $P^{\gamma -}_\alpha$ will change according to quantities that depend on the ensemble. 

As mentioned earlier, the master equation (\ref{nmbreuer}) with time-dependent rates does not guarantee positivity of the density matrix, particularly if the rates become negative at some times. The non-Markovian quantum jumps detect when positivity is about to be violated, based on the presence of a singularity in the negative jump probability (\ref{negativejump0}) \textcite{breuer2009b}. In particular, when the number of source members entering in the denominator of such a quantity becomes zero, and the rate is negative at the same time, the master equation violates positivity. This corresponds to the unphysical situation in which the environment tries to undo an event that has not happened.

In order to further understand the method, let us consider a three-level system with states $\lbrace | 0\rangle, |1 \rangle, |2 \rangle \rbrace$ and energies $E_0<E_1<E_2$, as discussed by \textcite{piilo2009}. We now assume that there are only two decay channels $k=1,2$ corresponding to the coupling operators $L_1=|0\rangle \langle 1|$ and $L_2=|1\rangle \langle 2|$. To build the state vector ensemble, we start by considering the normalized state $|\psi_0 \rangle= c_0|0\rangle+c_1|1 \rangle+c_2 |2 \rangle$. 
The two states, $|\psi_1 \rangle= |1\rangle$ and $|\psi_2 \rangle= |0 \rangle$, can be reached from $|\psi_0 \rangle$ with a forward jump. If a further forward jump occur, the state $|\psi_1 \rangle$ might jump to $|\psi_2 \rangle$. Hence the only states explored in the forward process are $\lbrace |\psi_0(t) \rangle, |\psi_1 \rangle, |\psi_2 \rangle \rbrace$, which is then the reference ensemble of states. For negative decay rates different channels open backward. If at time $t$, $\Delta_2(t)<0$ for the channel $L_2$, then the target state for $|\psi_1\rangle$ will be $|\psi_0 (t+\delta t)\rangle$ and no other jumps are allowed. However, if at time $t$ what we find is that $\Delta_1(t) < 0$, the target states for $|\psi_2\rangle$ will be either $|\psi_0 (t+\delta t)\rangle$ or $|\psi_1 \rangle$. In this case, the target state is not unique although there are different probabilities to be reached from $|\psi_2\rangle$. 

Non-Markovian quantum jumps have been successfully applied to study, for instance, exciton dynamics in photochemistry \textcite{rebentrost2009b,ai20141}.

\section{Path integral methods} \label{path}

The path integral representation, first derived by \textcite{feynman1948,feynman1963a}, constitutes a very convenient framework for performing numerical simulations of quantum dynamics and equilibrium quantum statistical mechanics, considering real and imaginary time evolution respectively \cite{weissbook}. Most of the applications are based on using a coordinate representation of the OQS, which is assumed to be coupled with one or few degrees of freedom to an environment as described by \textcite{caldeira1983,caldeira1983b,leggett1987} (see also Sec. \ref{ch1sec3}). 
In addition, the path integral approach generally considers a factorized initial condition between environment and system, and the environment in thermal equilibrium. Under these conditions, the path integral representation of the reduced density matrix of the system reads as \cite{weissbook}
\bea
\rho(x_{f},x'_{f},t)&=&\int dx_i dx'_i {\mathcal J}(x_{f},x'_{f},t;x_i,x'_i,t_i)\rho(x_i,x'_i,t_i),\nonumber
\label{path1}
\eea
where 
\bea
&&{\mathcal J}(x_{f},x'_{f},t;x_i,x'_i,t_i)=\int {\mathcal D}[x_1]{\mathcal D}[x_2]\nonumber\\&\times&e^{\frac{i}{\hbar}({\mathcal S}_S[x_1]-{\mathcal S}_S[x_2])}F[x_1,x_2],
\label{propagator}
\eea
is the propagator of the reduced density matrix, and $x$ represents the OQS's degree of freedom. The propagator presents a sum over all real-time paths $x_1$ and $x_2$ that run in time from $x_i$ and $x'_i$ at an initial time $t_i$, to $x_f$ and $x'_f$ at a final time $t_f$. The influence functional $F[x_1,x_2]$ couples these two paths, and can be written in terms of the difference and sum paths $y=(x_1-x_2)$ and $r=(x_1+x_2)/2$ \cite{weissbook}
\bea
&&{\mathcal F}[y,r]=\exp\bigg(-\frac{1}{\hbar}\int_0^t du {\mathcal W}[u,y,r]\bigg),
\label{path2}
\eea
where ${\mathcal W}[u,y,r]=\int_0^u dv y(u)[\alpha^R_T(u-v)y(v)+2i\alpha_T^I(u-v)r(v))]+i\mu\int_0^tdu y(u)r(u)$.
The functions $\alpha_T^I$ and $\alpha_T^R$ correspond respectively to the imaginary and real parts of the environment correlation function (\ref{chapuno33bb}).
Also $\mu=\frac{2}{\hbar}\int_0^\infty d\omega \frac{J(\omega)}{\omega}$ corresponds to the static susceptibility of the environment. 


The influence functional (\ref{path2}) introduces long-range non-local interaction among the system paths, so an explicit evaluation of (\ref{path1}) is possible only numerically. 
Numerical developments to evaluate the path integrals include 
the iterative tensor propagator scheme \textcite{makarov1994,makri1995}, originally introduced in terms of a quasiadiabatic propagator \cite{makri1992} and hence often referred to as the quasiadiabatic propagator path integral (QUAPI) algorithm, and the path integral Monte Carlo schemes (PIMC) derived by \textcite{egger1994,mak1996}.

The QUAPI algorithm relies on a Trotter decomposition of the evolution operator within a time slice $\Delta t$, which is   
based on the partitioning of the full Hamiltonian into a so-called adiabatic contribution $H_S$, which can be treated exactly, and a non-adiabatic reminder $H-H_S$. As a result of such decomposition, and considering also a discretization in the OQS configuration space, a discretized version of the path integral (\ref{path1}) is obtained, which includes the non-adiabatic corrections through the influence functional. The discretization is based on the choice of two parameters: a time-related parameter $K$, which settles a memory time window $\tau_k=\Delta t K$ up to which the environment correlations are included (such a window larger or of the order of the environment correlation time, $\tau_c$), and a parameter $M$ that settles the number of OQS basis states. 
After the discretization, the evolution of the reduced density operator is obtained through a temporal iterative procedure. As discussed by \textcite{nalbach2011}, the summation over all possible paths within the memory time window $\tau_k$ is exact (up to the error produced by the Trotter decomposition of the evolution operator) and deterministic. A further improvement in the implementation of iterative algorithms is the filtered propagation functional developed by \textcite{sim1996,sim2001quantum}, which takes into consideration only path segments that contribute in the path integral with significant weight.  
The QUAPI algorithm was successfully applied to study quantum transport between two particles \cite{nalbach2010}, and for such a model, its performance has been compared to that of a time-nonlocal perturbative master equation (see Sec. \ref{projectionM}) \cite{nalbach2011}, and to a variational master equation [discussed in Sec. \ref{polaron}]\cite{mcCutcheon2011}. In addition, the iterative path integral procedure has been developed for calculating equilibrium two-time correlation functions of quantum dissipative systems \cite{shao20011iterative,shao2002iterative}. Other variants of iterative algorithms were developed to compute real-time path integral expressions for quantum transport problems out of equilibrium \cite{weiss2008iterative}.


For more details of the iterative path integral algorithm see \cite{makri1995,makri1995b} and the discussion by \textcite{thorwart1998}. Also, a review of the most recent advances in the field, including a Matlab library to implement iterative tensor propagator scheme was given by \textcite{dattani2013feyndyn}.

As described by \textcite{muelbacher2004,muelbacher2005}, the PIMC algorithm is also based on a discretization of the path integral representation (\ref{path1}). 
However, as opposed to the QUAPI algorithm, PIMC relies on performing a stochastic sampling of the path integral, which is approximated by considering a finite ensemble of randomly chosen paths.  
In addition, the PIMC technique is usually focused on computing the diagonal part of the reduced density matrix, but it has recently been extended to simulate coherences as well \cite{kast2013}. In general, the PIMC is one of the most powerful means of exploring the non-perturbative range including strong coupling and high temperatures. Although the method was introduced to analyze the dynamics of spin-boson systems \cite{egger1994}, it has also been used to analyze dynamical quantities of larger systems, like single and correlated charge transfer along molecular chains \cite{muelbacher2004,muelbacher2005}, also including external driving fields \cite{muhlbacher2009}. 
The PIMC method is particularly efficient to describe quantum systems coupled to a thermal reservoir with Ohmic spectral densities, but it has also been extended to sub-Ohmic reservoirs, a situation where entanglement between the system and the environment becomes more important \cite{andre2009}. 
Motivated by the success of the PIMC algorithm, \textcite{muhlbacher2008} combined such a technique with the diagrammatic Monte Carlo approach (initially derived for the imaginary time evolution), in order to analyze the dynamics of a quantum dot coupled to two fermionic reservoirs and to a bosonic bath representing a photon environment. These diagrammatic Monte Carlo algorithms are the basis for the so-called continuous time quantum Monte Carlo methods, discussed in detail by  \textcite{gull2011}. 


One of the drawbacks of Monte Carlo methods in general is that the number of sample paths needed to achieve a sufficient signal-to-noise ratio increases exponentially with the simulated system time, which hinders their performance over long times. This \textit{dynamical sign problem} has been shown to be relieved if the sampling space is reduced by integrating out exactly large parts of the configuration space \cite{egger1994,muelbacher2005,muehlbacher2006}. Also, \textcite{cohen2015} recently presented a solution to the dynamical sign problem with a new algorithm whose computational cost scales quadratically rather than exponentially with the simulation time.

Based on the observation that for harmonic oscillator environments the Feynman path integrals have a quadratic functional form, \textcite{cao1996anovel} presented an alternative method based on performing the environment average by directly sampling paths of the discretized harmonic modes and then propagating the system under the influence of a quantum Gaussian force. While the influence functional methods are based on a cutoff in the number of discretized time slices, the method by \textcite{cao1996anovel} introduced a cutoff in the number of discretized bath frequencies, which makes it particularly amenable for environments with narrow spectral densities.

As noted in the introduction, the path integral representation is the basis of different analytical derivations and approximations that do not rely on a weak coupling approximation between the system and the environment. Three of these derivations, the noninteracting blip approximation, the stochastic Liouville von-Neumann equation, and the hierarchical equations of motion, are discussed in the following. 


\subsection{The noninteracting blip approximation}
\label{noninteracting-Blip}

Within the two-level approximation leading to a Hamiltonian of the form (\ref{chapuno11}), the variables $x_1$ and $x_2$ can take only two discrete values $|\pm\rangle=\pm\frac{1}{2}q_0$, where $q_0$ is the center of the double well. Therefore Eq. (\ref{path1}) becomes an integral over all possible pairs of paths, each of which jumps between these two states. Alternatively, it can be considered as a single path integral jumping between four states $A=\{+,+\}$, $B=\{+,-\}$, $C=\{-,+\}$, and $D=\{-,-\}$, corresponding to populations (\textit{diagonal} states $A$ and $D$) and coherences (\textit{off-diagonal} states $B$ and $C$). Periods in which the system is in a diagonal state are called \textit{sojourns}, and periods between diagonal states are called \textit{blips}. 
Within this picture, the noninteracting blip approximation (NIBA) is used to calculate the probability of the system to be at a certain state at time $t$, by assuming that the average time spent by the system in a diagonal state is very large compared to the average time spent in an off-diagonal state. This assumption leads to certain prescriptions being considered for performing the path integral (\ref{path1}) \cite{leggett1987,weissbook}, in particular to compute the functional (\ref{path2}). These prescriptions turn out to be valid at high temperatures (so that a strong decoherence suppresses the off-diagonal terms), for the super-Ohmic case, and for a situation in which the Fermi golden rule applies, \textit{i.e.} the Markovian case. With respect to the Hamiltonian (\ref{chapuno11}), the NIBA corresponds to an expansion in terms of the tunneling matrix element $\Delta_0$, which can also be performed with projection-operator techniques \cite{morrillo1991}. The result of the NIBA approximation is that the evolution of $P(t)=\langle \sigma_z \rangle$ is given by
\bea
\frac{dP(t)}{dt}=-\int_{-\infty}^t ds f(t-s)P(s),
\label{NIBA}
\eea
with $f(s)=\Delta_0^2\cos[Q_1(s)/(\pi \hbar)]e^{-Q_2(s)/(\pi\hbar)}$, and 
\bea
&&Q_1(s)=\int_0^\infty \sin(\omega s)J(\omega)d\omega/\omega^2\cr
&&Q_2(s)=\int_0^\infty (1-\cos(\omega s))\coth(\hbar\beta\omega/2)J(\omega)d\omega/\omega^2.\nonumber
\eea
\textcite{dekker1987} found a different way to obtain this expression for the evolution of the OQS population by performing a polaron transformation on the spin-boson Hamiltonian (\ref{chapuno11}). As discussed in Sec. \ref{polaron}, the polaron transformation has the form $U=\exp(-i\sigma_z\Omega/2)$, with $\Omega=\sum_\lambda(c_\lambda/m_\lambda\omega_\lambda^2)p_\lambda$, and the transformed Hamiltonian can be written as Eq. (\ref{polaronT0}) with $B_z=0$, $H'=-\frac{1}{2}\Delta_0(\sigma^+e^{-i\Omega}+\sigma^-e^{i\Omega})+H_B$, so that the Heisenberg evolution of $\sigma_z$ has the exact form 
\bea
\frac{d\sigma_z(t)}{dt}=-\frac{1}{2}\Delta_0^2\int_{-\infty}^t ds\bigg(e^{-i\Omega(t)}e^{i\Omega(s)}\sigma_z(s)+\Hhc\bigg).\nonumber
\eea
Equation (\ref{NIBA}) is recaptured simply by considering that $\Omega(t)$ evolves according to the free environment dynamics, and then assuming that the quantum average of the spin, $\sigma_z(s)$, and the environmental exponentials $e^{-i\Omega(t)}$ is decoupled. 

\textcite{orth2010,orth2013,henriet2014,henriet2016} have developed an alternative method to perform a stochastic unraveling of the influence functional similar to the one proposed by \cite{stockburger2002} discussed in the following section, but which is made after rewriting of the influence functional in the blip-sojourn language. Based on this, they have obtained a stochastic equation for the density matrix in the vector space of states A, B, C and D. 

\subsection{Stochastic Liouville von-Neumann equation}
\label{SLN2}
Path integral formulations may also give rise to the SLN equation, first proposed by \textcite{stockburger2002}. 
According to this formulation, the double time integral appearing in Eq. (\ref{path2}) can be reduced to a single time integral by introducing a Gaussian integral over two complex functions $\xi(t)$ and $\nu(t)$, and redefining the functional in a Hubbard-Stratonovich form
\bea
&&F[y,r]=\int D^2[\xi]\int D^2[\nu]W[\xi,\xi^*,\nu,\nu^*]\cr
&\times&\exp(\frac{i}{\hbar}\int_{t_0}^tdt'\xi(t')y(t')+i\nu(t')r(t'))\cr
&\times&\exp(-\frac{i\mu}{\hbar}\int_{t_0}^tdt' y(t')r(t')),
\eea
where $W[\xi,\xi^*,\nu,\nu^*]$ is a Gaussian functional. The two newly defined complex functions can be considered Gaussian noises with the following statistical properties: 
\bea
{\mathcal M}_{\xi,\nu}[\xi(t)\xi(t')]&=&\alpha_R(t-t');\nonumber\\
{\mathcal M}_{\xi,\nu}[\xi(t)\nu(t')]&=&-i\alpha_I(t-t')\theta(t-t');\nonumber\\
{\mathcal M}_{\xi,\nu}[\nu(t)\nu(t')]&=&0, 
\label{prop}
\eea
where ${\mathcal M}_{\xi,\nu}[\cdots]=\int D^2[\xi]\int D^2[\nu]W[\xi,\xi^*,\nu,\nu^*]\cdots$ is the Gaussian average over two noises $\xi$ and $\nu$.
In this definition, $\alpha_R(t)$ and $\alpha_I(t)$ correspond, respectively, to the real and imaginary parts of the correlation function given by Eq. (\ref{chapuno33bb}). 
Having decoupled the two paths of (\ref{path1}), and following a procedure similar to the one given in \cite{feynman1963a}, a stochastic differential equation can be obtained for the reduced density operator, the stochastic Liouville von-Neumann equation,
\bea
\frac{dP_{\xi,\nu}}{dt}=-\frac{i}{\hbar}[H_S,P_{\xi,\nu}]+\frac{i}{\hbar}\xi(t)[q,P_{\xi,\nu}]+\frac{i}{2}\nu(t)\{q,P_{\xi,\nu}\}\cr
\label{SLNeq}
\eea 
where considered a system coupling operator $q$ and neglected a re-normalization term. Eq. (\ref{SLNeq}) valid for environments at thermal equilibrium, allows one to compute different stochastic trajectories for the density matrix sample $P_{\xi,\nu}$, such that the reduced density operator can be obtained as $\rho_s(t)={\mathcal M}_{\xi,\nu}[P_{\xi,\nu}]$. 
Eq. (\ref{SLNeq}) can be rewritten as two stochastic equations for two different stochastic state vectors $|\psi_t^1\rangle$ and $|\psi_t^2\rangle$, 
\bea
&&\frac{d|\psi^1_t\rangle}{dt}=-iH_S|\psi^1_t\rangle+ i\xi(t)q|\psi^1_t\rangle+i\frac{1}{2}\nu(t)q|\psi^1_t\rangle,\cr
&&\frac{d|\psi^2_t\rangle}{dt}=-iH_S|\psi^2_t\rangle+ i\xi^*(t)q|\psi^2_t\rangle-i\frac{1}{2}\nu(t)^*q|\psi^2_t\rangle.
\label{SSEexact3}
\eea
such that $P_{\xi,\nu}=|\psi^1_t\rangle\langle\psi^2_t|$. 
The drawback of this method is that beyond the case of the OQS being a harmonic oscillator, the convergence of the stochastic average for relatively long times is difficult. One of the problems is that even though $\rho_s(t)$ is normalized, the individual samples $P_{\xi,\nu}$ do not stay normalized, which slows down convergence. To overcome this, \textcite{stockburger2002,stockburger2004} proposed an exact mapping of Eq. (\ref{SLNeq}) to an equation that preserves the trace of each resulting density matrix sample $\hat{P}_{\xi,\nu}$. This formulation, similar to a Girsanov transformation that leads to the shifted noise of Eq. (\ref{shifted}), results in a transformed noise
\bea
\xi\rightarrow\hat{\xi}=\xi-\int_0^tdu\chi(t-u){\hat r}_u,
\label{newnoiseSLN}
\eea
where $\hat{r}_u=\Ttr_S\{q\hat{P}_{\xi,\nu}\}$
 and $\chi(u)=-\theta(u)\alpha_I(u)/2\hbar$, with $\theta(u)$ as the Heaviside step function. Similarly as in Eq. (\ref{shifted}), this new noise improves the statistics, such that the number of stochastic trajectories needed to obtain the reduced density matrix is smaller. 
A subtle point about the shift (\ref{newnoiseSLN}) is the fact that the quantity $\hat{r}_u$ is itself defined in terms of the normalized state, which can be a source of numerical instability \cite{stockburger2004}.
As proposed in \cite{koch2008}, this limitation can be overcome if one considers that the term $\hat{r}_u$ in the shift follows a reference path given by the classical trajectory according to the classical Langevin equations of motion.

We saw previously that SSEs generally require some approximation or ansatz to handle the integral term in order to obtain a closed equation, while the SLN stochastic equations are exact. The differences between the SLN and SSE unravelings are that the former depends on two correlated noise variables and recaptures $\rho_s$ as an average of two different stochastic state vectors, while the latter depends on a single noise variable and recaptures $\rho_s$ with an average over a single stochastic vector. 
The path integral approach underlines the close connection between the path integral representation and the stochastic description of OQS. As pointed out by \textcite{strunz1996,diosi1997}, the density matrix propagator (\ref{propagator}) can also be expressed as 
\bea
{\mathcal J}(x_{f},x'_{f},t;x_i,x'_i,t_i)={\mathcal M}_z[G_z(x_f,t;x_i,t_i)G^*_z(x_f,t;x_i,t_i)],\nonumber
\eea
where the stochastic propagator in the path integral representation has the form
\bea
G_z(x_f,t;x_i,t_i)&=&\int_{x_i;t_i}^{x_f;t_f}{\mathcal D}[x_s]\exp\bigg(\frac{i}{\hbar}{\mathcal S}_S+\int_{t_i}^{t_f}ds x_sz_s\nonumber\\
&-&\int_{t_i}^{t_f}ds\int_{t_i}^s ds' x_s\alpha^*(s-s')x_{s'}\bigg)
\eea
with the noise $z_t$ obeying the statistical properties (\ref{statprop}).


\subsection{Hierarchical equations of motion}
\label{hierarchy}
A variant of path integrations in the real position space consists of using a coherent state representation, characterized by a variable $\phi$ and its conjugate $\phi'$. This is the basis to derive the HEOM for the reduced density operator, first proposed by \textcite{tanimura1989,tanimura1990}. In terms of coherent states, the path integral representation of the reduced density matrix of the system reads as $\rho_s(t)=\int d\phi_f \int d\phi'_f \rho(\phi_f,\phi'_f;t)|\phi_f\rangle\langle\phi'_f|$, with the coefficients given by 
\bea
\rho_s(\phi_f,\phi'_f;t)&=&\int D[Q(\tau)]\int D[Q'(\tau)]e^{\frac{i}{\hbar}{\mathcal S}_S(Q;t,t_i)}\cr
&\times&F(Q,Q';t,t_i)e^{-\frac{i}{\hbar}{\mathcal S}_S(Q';t,t_i)},
\label{pathc1}
\eea
where $\phi_f$ and $\phi'_f$ are the final states of the system, and $Q(t)$ represents the set of coherent state variables $\{\phi^*(t),\phi(t)\}$. Here, the functional ${\mathcal S}_S$ is an action of $H_S$ and $\int D[Q(\tau)]$ represents the functional integral of $Q(\tau)$. In addition, we defined the influence functional 
\bea
F(Q,Q';t,t_i)=\exp\bigg(\frac{-i}{\hbar}\int_{t_i}^td\tau' S^x(Q,Q';\tau')\Pi\bigg),\cr
\label{pathc2}
\eea
where we have omitted the dependencies of the function $\Pi$, which is defined as $\Pi=-\frac{i}{\hbar}\int_{t_i}^{\tau'} d\tau [\alpha_R(\tau'-\tau)S^x(Q,Q';\tau)-i\alpha_I(\tau'-\tau)S^o(Q,Q';\tau)])$, in terms of the functionals $S^x(Q,Q';\tau)=S(Q(\tau))-S(Q'(\tau))$ and $S^o(Q,Q';\tau)=S(Q(\tau))+S(Q'(\tau))$, which represent difference and sum paths similar to those appearing in Eq. (\ref{path2}). Here $S(Q(\tau))$ corresponds to the coherent state representation of the coupling operators appearing in Eq. (\ref{chapuno5}). 

For systems having Ohmic dissipation with a Lorentzian cutoff (Drude dissipation), characterized by a spectral density 
\bea
J(\omega)=\frac{\hbar\lambda\gamma^2}{2\pi}\frac{\omega}{\omega^2+\gamma^2}, \nonumber
\eea
where $\lambda$ is the reorganization energy, which is proportional to the system-environment coupling strength, the correlation function (\ref{chapuno33bb}) can be written as $\alpha(t)=\sum_{m=0}^\infty c_m \exp(-\mu_{m} t)$, in terms of the Matsubara frequencies. These are defined as $\mu_{0}=\gamma$ and $\mu_{m}=2\pi m/\hbar\beta$ when $m\ge 1$, while the coefficients are 
\bea
c_0=\frac{\hbar\gamma^2\lambda}{2}(\cot(\beta\hbar\gamma/2)-i), \nonumber
\eea
and 
\bea
c_{m\ge 0}=\frac{\gamma^2\lambda}{\beta}\frac{\mu_{m}}{\mu_{m}^2-\gamma^2}. \nonumber
\eea
For a high-temperature environment, $\beta\hbar\gamma\ll 1$, this reduces to 
\bea
\alpha(t)\approx \frac{\lambda\gamma^2}{2}(\cot\large(\frac{\hbar\beta\gamma}{2}\large)-i)\exp(-\gamma t). 
\eea
In this case, it is possible to reexpress the element $\Pi$ in the functional (\ref{pathc2}) as 
\bea
\Pi&=&-\frac{i\lambda\gamma^2}{2}\int_{t_i}^{\tau'} d\tau e^{-\gamma(\tau'-\tau)}[\cot\large(\frac{\hbar\beta\gamma}{2}\large)S^x(Q,Q';\tau)\cr
&-&iS^o(Q,Q';\tau)]).
\label{pathc3}
\eea
Then in terms of this quantity we can define the following elements
\bea
\rho_n(\phi_f,\phi'_f;t)&=&\int D[Q(\tau)]\Pi^ne^{\frac{i}{\hbar}S_S(Q;t,t_i)}\cr
&\times&F(Q,Q';t,t_i)e^{-\frac{i}{\hbar}S_S(Q;t,t_i)},
\label{pathc1}
\eea
and the corresponding operators $\rho_n(t)=\int d\phi_f \int d\phi'_f \rho_n(\phi_f,\phi'_f;t)|\phi_f\rangle\langle\phi'_f|$. The element $n=0$ corresponds to the reduced density matrix (\ref{pathc1}). The time differentiation of these operators leads to \cite{tanimura2006,tanimura20151}
\bea
&&\frac{d\rho_n}{dt}=-(\frac{i}{\hbar}H_s^X-n\gamma)\rho_n-\frac{i}{\hbar}S^X\rho_{n+1}
-i\frac{n}{\hbar}\Theta\rho_{n-1},
\label{HEOMsimple}
\eea
where $\rho_0=\rho_s$, $\Theta=(\hbar\lambda/2)[\cot(\hbar\beta\gamma/2)S^X-iS^o]$, with $A^o\rho=A\rho+\rho A$ and $A^X\rho=A\rho-\rho A$. 

Hierarchical expansions have the advantage that they allow one to obtain the reduced density matrix of the OQS while including all orders of the system-environment interactions. The fact that the different levels of the hierarchy include all orders in the coupling between the system and the environment renders the method particularly useful for strong system-environment coupling. In addition, under certain conditions, the hierarchy can be systematically truncated \cite{tanimura2006,tanimura20151}. This approach has been used, for instance, to describe the quantum dynamics of chemical and biophysical systems, in which other approaches based on the weak coupling approximation are not valid. 
An example of such systems are light-harvesting complexes, where the $N$ molecules in the complex are affected by a local Drude spectral density \cite{ishizaki2009c}, $J_j(\omega)=\hbar\lambda_j \gamma_j^2\omega/2\pi(\omega^2+\gamma_j^2)$, at each molecular site $j$. The resulting hierarchical structure describing the problem is more complex than the previous one. In this case, each member of the hierarchy $\rho_{\bf n}$ is now labeled by a set of non-negative integers ${\bf n}=(n_1,n_2,\cdots,n_N)$, each corresponding to a molecule $j$. Then, the evolution equation is given by 
\bea
\frac{\rho_{\bf n}(t)}{dt}&=&-(\frac{i}{\hbar}H_s^X-\sum_{j=1}^N n_j\gamma_j)\rho_{\bf n}(t)-\frac{i}{\hbar}\sum_{j=1}^N[S^X_j\rho_{{\bf n}^+}(t)\cr
+&&n_j\Theta_j\rho_{{\bf n}^-}(t)].
\eea
Here, ${\bf n}^\pm_j$ differs from ${\bf n}$ by changing the specified $n_j$ to $n_j+1$, \textit{i.e.} ${\bf n}_j^\pm=(n_1,n_2,\cdots,n_j\pm 1,\cdots,n_N)$. Here, $\Theta_j$ is defined similarly as $\Theta_0$, but depends on $\lambda_j$, $\gamma_j$, $S_j^X$ and $S_j^o$.
In addition, one can also consider the low-temperature case, by defining a hierarchy that depends on two indexes, one of which relates to the level $n$ of the hierarchy, and the other which is settled by the index $m$ corresponding to each of the Matsubara frequencies. Naturally, the number of Matsubara frequencies must be truncated \textcite{ishizaki2005,xu2005,han2006}. This situation was tackled in \cite{ishizaki2005} for the case of a single molecule and in \cite{li2012} for more than one molecule comprising a light-harvesting photosynthetic complex. Extending the dimension of the hierarchy, the method is able to describe a number of spectral densities leading to correlation functions that are combinations of exponentials \cite{tanaka2009,tanimura2012,ma2012}. 

A more recent proposal for low-temperature environments consist of splitting the functional into a term, $F_R$, that depends on the real part of the correlation function and thus carries the temperature dependency, and a term, $F_I$, that depends on the imaginary part of the correlation function $\alpha_I$. Then, $F_R$ is written as a function of a colored real noise $\xi(t)$ using the Hubbard-Stratonovich transformation discussed in Sec. \ref{SLN2}, and $F_I$ is used as a basis for deriving HEOM. Considering a Drude model for the spectral density, such that $\alpha_I(t)\sim e^{-\gamma t}$, the procedure results in a stochastic version of (\ref{HEOMsimple}), which depends on the real noise $\xi(t)$ \cite{moix2013ahybrid}. Previous proposals in this direction were put forward by \textcite{zhou2005,tanimura2006}). 


This description can also be extended to deal with initially correlated states between the system and the environment, and to obtain thermal equilibrium quantities of the system \cite{tanimura2014}. A similar hierarchical structure was recently derived by \textcite{devega2014}, by departing from the SLN Eq. (\ref{SLNeq}) of the previous section.

\section{Heisenberg representation}
\label{HR}
Early developments in the application of the Heisenberg representation to the OQS problem were made to describe the spontaneous emission \cite{ackerhalt1973} and strong-field resonance fluorescence \cite{kimble1975} of a two-level atom. A non-Markovian extension of the theory was proposed by \textcite{wodkiewicz1976} and \textcite{wodkiewicz1979} for the spontaneous emission and the resonance fluorescence respectively. As shown in the next section, the difficulty of solving the Heisenberg equations for OQS is that they comply with a hierarchical structure. Thus, the evolution of one-time correlations (\textit{i.e.} quantum mean values) depends on two-time correlations. Furthermore, the evolution equation of two-time correlations depends on three-time correlations, while three-time correlations show a dependency on fourth order correlations. In summary, the evolution of non-Markovian $N$-time correlations of system operators, when no approximations are made, depends on the $N+1$-time correlations. This hierarchy appears only in non-Markovian interactions, and vanishes when the environment correlation function $\alpha(t)$ is Markovian, \textit{i.e.} $\alpha(t)\approx\Gamma\delta(t)$. 

\subsection{Computing multiple-time correlation functions}
\label{HRI}

To derive MTCF with the Heisenberg equations, the idea is to express $dA_1(t_1)\cdots A_N (t_N)/dt_1$ in such a way that the environmental operators $a_\lambda (0)$ are placed on the right-hand side of the terms, while the $a^\dagger_\lambda (0)$ appear on the left-hand side. Thus, when we compute the MTCF as the quantum mean value of $A_1(t_1)\cdots A_N (t_N)$, \textit{i.e.} as $C_{{\bf A}}({\bf t}|\Psi_0)=\langle \psi_0 |\langle 0|A_1(t_1)\cdots A_N (t_N)|0\rangle|\psi_0\rangle$, where we  considered $\rho(0)=|\psi_0\rangle\langle\psi_0|\otimes|0\rangle\langle 0|$, those terms are zero, and only system operators appear in the equations.
Let us consider the Heisenberg evolution equation for a system observable $A(t,0)={\mathcal U}_I^{\dagger}(t,0)A{\mathcal U}_I(t,0)={\mathcal U}^{\dagger}(t,0)A{\mathcal U}(t,0)$, where ${\mathcal U}_I(t,0)$ is defined in Eq. (\ref{chapdos3}) and ${\mathcal U}(t_1,t_2)=\exp(-iH_{\ttot}(t_1-t_2))$, with $H_{\ttot}$ the total Hamiltonian (\ref{chapuno32}). Reexpressing $A(t,0)=A(t)$ for simplicity, we find
\begin{eqnarray}
&&\frac{dA(t_1 )}{dt_1}=i{\mathcal U}^{-1}(t_1 ,0 )[H_{tot},A]{\mathcal U}(t_1 ,0 )\cr
&=&-i [H_S (t_1 ),A(t_1 )]+i\sum_\lambda g_\lambda \large( a_\lambda^{\dagger} (t_1,0 )[L(t_1 ),A(t_1 )]\cr
&+&[L^{\dagger}(t_1 ),A(t_1 )]a_\lambda (t_1 ,0 ) \large),
\label{eq39}
\end{eqnarray}
where $L$ is a system coupling operator.
We can replace in (\ref{eq39}) the formal solution of the evolution equation of the environmental operators, $da_\lambda (t_1 ,t_2)/dt_1 =i[H_{tot}(t_1,t_2),a_\lambda (t_1,t_2 )]=-i\omega_\lambda a_\lambda (t_1,t_2)-ig_\lambda L(t_1,t_2)$,
\begin{eqnarray}
a_\lambda (t_1 ,t_2 )&=&e^{-i\omega_\lambda (t_1 -t_2 )}a_\lambda(t_2,t_2)\cr
&-&i g_\lambda \int^{t_1}_{t_2} d\tau e^{-i\omega_\lambda (t_1 -\tau)}L(\tau,t_2),
\label{eq44}
\end{eqnarray}
for $t_2=0$.
The single evolution equation (\ref{eq39}) becomes as follows:
\begin{eqnarray}
&&\frac{dA(t_1 )}{dt_1}=i [H_S (t_1 ),A(t_1 )]-\nu^{\dagger}(t_1 )[L(t_1 ),A(t_1 )]\nonumber \\
&+&\int_0^{t_1 }d\tau \alpha^*(t_1 -\tau) L^{\dagger}(\tau)[A(t_1 ),L(t_1 )]+[L^{\dagger}(t_1 ),A(t_1 )]\cr
&\times&\nu(t_1 )+\int_0^{t_1 }d\tau \alpha(t_1 -\tau)[L^{\dagger}(t_1 ),A(t_1 )]L(\tau),
\label{eq41}
\end{eqnarray}
where we used the definition (\ref{chapuno325}) of the environment correlation function. In the last expression, we also defined the environment operators
$\nu^{\dagger}(t_1 )=-i\sum_\lambda g_\lambda a_\lambda^{\dagger} (0,0)e^{i\omega_\lambda t_1 }$, and 
$\nu(t_1 )=i\sum_\lambda g_\lambda a_\lambda (0,0)e^{-i\omega_\lambda t_1}$.
In a similar way, the evolution equation of a two-time correlation can be written as 
\begin{eqnarray}
&&\frac{dA(t_1 )B(t_2 )}{dt_1}=i [H_S (t_1 ),A(t_1 )]B(t_2 )
\cr&-&\nu^{\dagger}(t_1 )[L(t_1 ),A(t_1 )]B(t_2 )+[L^{\dagger}(t_1 ),A(t_1 )]B(t_2 )\nu(t_1 )\nonumber\\
&-&\int_0^{t_1 }d\tau \alpha^* (t_1 -\tau)L^{\dagger}(\tau)[L(t_1 ),A(t_1 )]B(t_2 )\nonumber\\
&+&\int_{t_2}^{t_1 }d\tau \alpha(t_1 -\tau)[L^{\dagger}(t_1 ),A(t_1 )]L(\tau)B(t_2 )\cr
&+&\int_0^{t_2} d\tau \alpha(t_1 -\tau)[L^{\dagger}(t_1 ),A(t_1 )]B(t_2 )L(\tau).
\label{eq45}
\end{eqnarray}
From this equation, the evolution of the quantum mean value $\langle A(t_1 )B(t_2 )\rangle$ is again obtained by applying the total initial state $\mid\psi_0\rangle$ on both sides of the former expression. The generalization to an $N$-time correlation function was given by \textcite{alonso2007}. 

Note that for the quantum Brownian particle described further in Sec. \ref{brownian}, the Heisenberg equations of the form (\ref{eq39}) for system observables may be reduced to the quantum Langevin equation for the system position coordinate, $A=q$, \cite{hua1994,changpu1995}. Building on these results and formally calculating the solution of the Heisenberg equations, \textcite{yang2012} obtained the reduced density matrix for a bosonic and a fermionic open system, and analyzed the non-Markovianity of its dissipation.

\subsection{Computing multiple-time correlation functions with the weak coupling expansion}
\label{perturbativeH}
The open Hierarchy described previously can be truncated by assuming a semiclassical approximation which decouples quantum mean values of products of operators at different times. An alternative is based on assuming weak coupling between system and environment. For instance, as proposed by \textcite{wodkiewicz1976} [see also \cite{florescu2001} for a more recent application], the two-time operator product of Eq. (\ref{eq41}) can be linearized by re-writing it as an equal time product. This can be done by considering a perturbative expansion of the left Liouville operator of the system 
\bea
L(t)&=&e^{-i{\mathcal L}(t-\tau)}L(\tau)=\sum_{n=0}^\infty \frac{[-i(t-\tau)]^n}{n!}{\mathcal L}^nL(\tau),\cr
G^nL(\tau)&=&[[\cdots,[L(\tau),H_{\ttot}],H_{\ttot}],\cdots,H_{\ttot}],
\eea
where ${\mathcal L}$ is the Liouvillian associated with the total Hamiltonian. In general, it is possible to rewrite ${\mathcal L}={\mathcal L}_0+{\mathcal L}_{\iint}$, where ${\mathcal L}_0$ and ${\mathcal L}_{\iint}$ are of order $0$ and $g$ respectively in the perturbative parameter. Then, keeping contributions in the equations of motion up to order $g^2$ corresponds to replacing ${\mathcal L}\approx {\mathcal L}_0$. In general, 
a perturbative expansion in the operators $L(\tau)$ that appear in Eqs. (\ref{eq41}) and (\ref{eq45}), leads to
\begin{eqnarray}
&&L^{\dagger}(\tau)\left\{[L,A]\right\}(t_i)={\mathcal U}^{-1}_I (t_i 0)L(\tau,t_i)[L,A]{\mathcal U}_I (t_i 0 )\nonumber\\
&=&\left\{V_{\tau-t_i}L^{\dagger}[L,A]\right\}(t_i)+{\mathcal O}(g).
\label{eq61}
\end{eqnarray}
In a similar way, $L(\tau)B(t_{i+1})={\mathcal U}^{-1}_I (t_{i+1} 0)L(\tau,t_{i+1})B{\mathcal U}_I (t_{i+1} 0 )
=\left\{V_{\tau-t_{i+1}} LB\right\}(t_{i+1})+{\mathcal O}(g)$. Hence, inserting such terms in Eq. (\ref{eq41}), we find that the evolution of quantum mean values is given by a master equation of the form (\ref{master}), while 
the two-time correlation Eq. (\ref{eq45}) can be expressed as 
\begin{eqnarray}
&&\frac{d}{dt_1 }\langle A(t_1 )B(t_2 )\rangle=
i \langle\left\{[H_S ,A]\right\}(t_1 )B(t_2 )\rangle\nonumber\\
&+&\int^{t_1 }_0 d\tau \alpha^*(t_1 -\tau)\langle\left\{V_{\tau-t_1 } L^{\dagger}[A,L]\right\}(t_1 )B(t_2 )\rangle\nonumber\\
&+&\int^{t_1 }_{t_2} d\tau \alpha(t_1 -\tau)\langle\left\{[L^{\dagger},A]V_{\tau-t_1 } L \right\}(t_1 )B(t_2 )\rangle\nonumber\\
&+&\int_0^{t_2} d\tau\alpha(t_1 -\tau) \langle\left\{[L^{\dagger}, A]\right\}(t_1 )\left\{BV_{\tau-t_2} L\right\}(t_2 )\rangle.
\label{total3}
\end{eqnarray}
up to second-order in $g$. A general $N$-time correlation function can also be derived \cite{alonso2007}.

The first three terms of Eq. (\ref{total3}) are analogous to the non-Markovian evolution of the $\langle A(t_1)\rangle$, so that when the last term vanishes, \textit{i.e.} provided that $[L^{\dagger},A]=0$ or $[B,V_{\tau-t_2 } L]=0$, the QRT applies. This term is zero in the Markovian case, since the corresponding correlation function $\alpha(t_1-\tau)=\Gamma \delta(t_1-\tau)$ is zero in the domain of integration from $0$ to $t_2$. A similar result was previously given by \textcite{swain1981}, where the master equation approach is used to relate the calculation of correlation functions to the calculation of single-time expectation values. The theory of non-Markovian MTCFs was also recently analyzed by \textcite{fleming2012}.

There are particular conditions in which, even though the interaction is non-Markovian, the QRT is valid in the stationary regime. This was analyzed by \textcite{budini2008} for systems that can be described with a reduced density operator (\ref{reducedbudini}), where $\rho_R$ is obtained with a Lindblad type of equation according to a certain rate $\gamma_R$. It was determined in particular that whenever the evolution of $\rho_R$ satisfies the detailed balance condition \cite{carmichael1976}, then a QRT is valid in the asymptotic regime. This condition is automatically not satisfied when $\rho_R(\infty)$ depends on $\gamma_R$. The fact that the QRT is fulfilled in the stationary regime means that
\bea
\lim_{t_1\rightarrow\infty}\langle A(t_1)B(t_2)\rangle=\lim_{t_1\rightarrow\infty}\overline{\langle A(t_1)B(t_2)\rangle},
\eea
where $\overline{\langle A(t_1)B(t_2)\rangle}$ is computed by using the master equation with initial condition $\hat{\rho_0}=B(t_2)\rho_0$.

\subsection{Input-output formalism}
\label{inout}

The Heisenberg approach allows for the introduction of the input-output formalism, first derived by \textcite{yurke1984,gardiner1985} [see also \cite{quantumnoise}] using the Markov approximation. 
This formalism was used in the context of cavity quantum electrodynamics \cite{yurke1984,quantumnoise,koshino2008}, for systems driven by the output of another system \cite{gardiner1993}, to describe cascaded open systems \cite{carmichael1993}, or to characterize quantum memories based on atomic ensembles \cite{muschik2006,muschik2013}, to name just a few examples. Recently, it was also extended to describe few-photon transport, considering a waveguide with a single atom \cite{fan2010} and many spatially distributed atoms  \cite{caneva2015}. Although it was initially derived for bosonic fields, it has also been extended to fermion fields \cite{gardiner2004}. In general, it is particularly useful in situations where it is relevant to keep track not only of the dynamics of the OQS, but also of the environment operators. This approach was recently extended to non-Markovian systems by \textcite{diosi2012} and \textcite{zhang2012} in the context of stochastic Schr\"odinger equations and cascaded networks, respectively. 

The first step in the input-output formalism is to reexpress the environment and coupling Hamiltonians in the continuum limit, $H_B=\int_{-\infty}^\infty d\omega \omega a(\omega)^\dagger a(\omega)$, and $H_I=\int_{-\infty}^\infty d\omega G(\omega)[a(\omega)^\dagger L+L^\dagger a(\omega)]$. Here the lower limit can be extended to $-\infty$ provided that the problem is translated into a rotating frame with respect to the system resonant frequency $\omega_{s}$, which is considered to be very large $\omega_{s}\rightarrow\infty$ as is justified in quantum optics \cite{gardiner1985}. 

Then, considering the interaction picture with respect to the environment, we can write $H=H_S+i[\hat{a}^\dagger_{in}(t)L-L^\dagger \hat{a}_{in}(t)]$,
where 
\bea
\hat{a}_{in}(t)=i\int^\infty_{-\infty}d\omega G(\omega)a(\omega) e^{-i\omega t}=i\int^\infty_{-\infty}d\tau \kappa(t-\tau)a_{in}(\tau),\cr
\label{inout0}
\eea
with $\hat{a}_{in}(t)=\frac{1}{\sqrt{2\pi}}\int_{-\infty}^\infty d\omega a(\omega)e^{-i\omega t}$, and $\kappa(t)=\frac{1}{\sqrt{2\pi}}\int_{-\infty}^\infty d\omega G(\omega)e^{-i\omega t}$. Also, these operators satisfy $[\hat{a}_{in}(t),\hat{a}^\dagger_{in}(s)]=\gamma(t-s)$, where $\gamma(t-\tau)=\int_{-\infty}^\infty ds\kappa^*(t-s)\kappa(\tau-s)$.
Similar to Eq. (\ref{eq44}), it is possible to write the evolved environment operator as 
\begin{eqnarray}
\hat{a}_{out}(t)&=&\hat{a}_{in}(t)+\int^{t_1}_0 d\tau \kappa(t-\tau)L(\tau).
\label{inout1}
\end{eqnarray}
In terms of Eq. (\ref{inout1}), the evolution of an arbitrary system observable is written as \cite{zhang2012}
\begin{eqnarray}
&&\frac{dA(t)}{dt}=i [H_S (t),A(t )]+\hat{a}_{in}^{\dagger}(t)[L(t),A(t)]\nonumber \\
&+&\int_0^{t}d\tau \gamma^*(t-\tau) L^{\dagger}(\tau)[A(t),L(t)]+[L^{\dagger}(t),A(t)]\cr
&\times&\hat{a}_{in}(t)+\int_0^{t}d\tau \gamma(t-\tau)[L^{\dagger}(t),A(t)]L(\tau),
\label{inout2}
\end{eqnarray} 
which is very similar to Eq. (\ref{eq41}).
Thus, the second-order perturbative version of this equation can be derived similarly as in Sec. \ref{perturbativeH}. In addition, the traditional input-output expressions are obtained by considering the Markov limit in Eqs. (\ref{inout1}) and (\ref{inout2}). In this limit, $G(\omega)=\sqrt{\gamma}$, and hence $\alpha(t-\tau)=\sqrt{\gamma}\delta(t-\tau)$, so that Eq. (\ref{inout1}) becomes simply $\hat{a}_{out}(t)=\hat{a}_{in}(t)+i \sqrt{\gamma}L(t)$, which is the well-known expression in the Markovian input-output formalism. 
\subsection{Heisenberg equations in many-body problems}

\label{heisenbergmany}

Let us consider, for instance, a system of $M$ particles interacting with a harmonic field through a Hamiltonian of the form (\ref{ch1collective}), with $L_j=\sigma^-_j$ a spin ladder operator corresponding to the particle $j$, and $g_{\lambda}({\bf r}_j)=g_\lambda e^{i{\bf k}\cdot {\bf r}_j}$.
Then, the Heisenberg equations for some of the main quantum mean values of the system observables have the following form:
\begin{eqnarray}
&&\frac{d\langle\sigma^{-}_{i}(t)\rangle}{dt}=
\sum_{j} \int_0^t d\tau \alpha_{ij}(t-\tau)\langle \sigma^3_{i}(t)
\sigma^-_{j}(\tau)\rangle\nonumber\\
&&\frac{d\langle\sigma^{3}_{i} (t)\rangle}{dt}=-4 \Re
[\sum_{j} \int_0^t d\tau \alpha_{ij}(t-\tau)\langle \sigma^+_{i}(t)
\sigma^-_{j}(\tau)\rangle]\nonumber\\
&& \frac{d\langle \sigma^+_{i}(t) \sigma^-_{j}(t) \rangle}{dt}=
\sum_{l} \int_0^t d\tau \alpha^*_{li}(t-\tau)\langle \sigma^+_{l}(\tau)
\sigma^3_{i}(t) \sigma^-_{j}(t) \rangle\nonumber\\
&+&\int_0^t d\tau \alpha_{lj}(t-\tau)\langle
\sigma^+_{i} (t)\sigma^3_{j}(t) \sigma^-_{l}(\tau) \rangle,
\label{sistemt}
\end{eqnarray}
with $\sigma_{i}^{3}=2\sigma^+_{i} \sigma_{i}-1$, and the two-particle correlation function given by $\alpha_{lj} (t)=\sum_{\bf k}g^2_{k}e^{i{\bf r}_{lj}\cdot({\bf k}-{\bf k}_L)-i\Delta_k t }$, with ${\bf r}_{lj}={\bf r}_{l}-{\bf r}_{j}$ as the distance between the two particles. 

Indeed, for many-body problems the Heisenberg equations of system operators comply with a hierarchical structure in two different ways: first, as was explained, for non-Markovian cases one-time correlations (\textit{i.e.} quantum mean values) are dependent on two-time correlations, which are in turn dependent on three-time correlations, etc.; second, even in the Markovian case, the quantum mean value of a single operator $\langle A_j(t)\rangle$ corresponding to the particle $j$ depends on the quantum mean value of two particle operators, \textit{i.e.} $\langle B_j(t)C_l(t)\rangle$ where $B_j$ and $C_l$ are operators corresponding to the particles $j$ and $l$.

The first hierarchical structure can be removed by assuming, for instance, a weak coupling approximation up to the second-order in the system-environment coupling parameter. 
The second hierarchical structure appears because of the many-particle nature of the OQS. 
For systems with many particles, the reduced density matrix often becomes too large to be computed, and Heisenberg equations become particularly convenient. The reason is that they allow for the use of a \textit{truncation method} \cite{andreev1993,christ2007} to express correlations of three operators into correlations of two operators, enabling the calculation of a smaller set of the most relevant system quantities. As it is not based on any systematic perturbative expansion, the accuracy of this truncation has to be tested in each case, for instance by comparing it to the exact result obtained for a smaller version of the particular system under study. 
We will discuss this idea in the following section, and also the application of the \textit{mean-field} or \textit{Hartree approximation} \cite{breuerbook} to the Heisenberg equations which under certain conditions allow one to describe the dynamics of a system beyond the weak coupling approximation. 

\subsection{Relevant scales involved in the dynamics of many-body OQSs: Independent and collective limits}
\label{cooperative}
The Markovian approximation is very useful to obtain information about the relevant time scales of the problem and to derive simplified effective Hamiltonians. In addition, it allows us to define two different limits relevant to discuss the dynamics of many-body OQSs: the limit of independent emitters, where particles evolve as if they were coupled to independent reservoirs, and the collective limit where the evolution of each particle is affected by the presence of the other particles interacting with the same reservoir. To see this, let us consider that the system evolution time scale, $T_S\sim 1/\Gamma_0$ is much smaller than $\tau_c$, where $\Gamma_0$ is the dissipative rate $\Gamma_{ij}$ for ${\bf r}_i-{\bf r}_j=0$, and $\tau_c$ is given by the decaying of the correlation function $\alpha_{ij}(\tau)$ for ${\bf r}_i-{\bf r}_j=0$. The dissipative rates are defined as
\begin{eqnarray}
\Gamma_{ij}=\int_0^\infty d\tau \alpha_{ij}(\tau ).
\label{rate}
\eeqa
In this limit, the evolution Eqs. (\ref{sistemt}) can be reduced to $\frac{d\langle\sigma^{3}_{i} \rangle}{dt}=-4 \Re
[\sum_{j} \Gamma_{ji}\langle \sigma^+_{i}
\sigma^-_{j} \rangle]$ and $\frac{d\langle \sigma^+_{i} \sigma^-_{j} \rangle}{dt}=
\sum_{l} \Gamma^*_{li}\langle \sigma^+_{l}
\sigma^3_{i} \sigma^-_{j} \rangle+\Gamma_{lj}\langle
\sigma^+_{i} \sigma^3_{j} \sigma^-_{l} \rangle$.
Here, all the operators are evaluated at time $t$. The quantities $\Gamma_{ij}$ describe the dipolar interactions between the sites ${i}$ and ${j}$, and in physically realistic situations decay with the distance ${\bf r}_{ij}={\bf r}_i-{\bf r}_j$. 
A calculation of such coefficients for atoms interacting with the radiation field in the vacuum was originally given by \textcite{lehmberg1970}, where it was found that the rates can be written as a sum of three components that decay with the distance as $|{\bf r}_{ij}|^{-1}$, $|{\bf r}_{ij}|^{-2}$, and $|{\bf r}_{ij}|^{-3}$, respectively. 

A physically intuitive form for the decay rates was obtained by \textcite{devega2008,navarrete2010} for the case of atoms trapped by an optical lattice of $M$ sites and coupled to a field of nontrapped atoms. In this system, the decay rates are given by $\Gamma_{|{\bf i-j}|}\sim|\xi |\frac{e^{-|{\bf i-j}|/\xi}}{|{\bf i-j}|}$, where $\xi =1/(|k_0|d_0)$ is a parameter that quantifies the range of the interactions, with $k_0$ as the resonant wave-vector of the field, and $d_0$ the interatomic separation. Here, because the lattice is cubic, we use the notation ${\bf r}_j=d_0{\bf j}$, where ${\bf j}$ is the position of the lattice site ${\bf j}\in\mathbb{Z}^3$. 
The rate of emission in all directions, which is given by ${\mathcal R}(t)\approx -\sum_{j}d\langle
\sigma^3_{j}\rangle/dt$, depends crucially on the different values of $\xi$. If particles evolve independently, $\mathcal{R}(t)$ decays exponentially. This corresponds to the limit of independent emitters, achieved when $\xi\ll 1$. In this range, the rates $\Gamma_{|{i-j}|}\sim \delta_{ij}$, and the correlation function is such that $\alpha_{ij}(\tau)=\delta_{ij}\alpha(\tau)$.
Cooperative effects in the emission start to occur when $\xi>1$ leading to a ${\mathcal R}(t)$ that no longer decays exponentially and, furthermore, presents positive slopes at initial times. The limit where $\xi\gg 1$ gives rise to an enhanced emission rate characteristic of Dicke superrradiance, 
$\Gamma^{\ddiss}=M\Gamma_0$. This enhancement corresponds to a situation where the correlation function is site independent, \textit{i.e.} $\alpha_{ij}(\tau)=\alpha(\tau)$, and the system can be properly described by the 
effective interaction Hamiltonian
\begin{eqnarray}
H_{\eeff}=\sum_{\lambda}g_\lambda \left(a^\dagger_{\lambda}J^- +\Hhc. \right).\label{Dicke}
\end{eqnarray}
Here, $J^-=\sum_i \sigma^-_i$ ($J^+ =\sum_i \sigma^+_i$) is a collective atomic spin operator, with properties $[J^-,J^+]=2J_z=2\sum_j \sigma_z^j$. In the limit of an environment with a single mode, this Hamiltonian corresponds to the well-known multimode Dicke model, which for the single mode case was the first system in which super-radiance was described \cite{dicke1954}. 
Although the Hamiltonian (\ref{Dicke}) appears to be formally equivalent to that of a single spin coupled to a harmonic oscillator environment, its resolution is obviously more involved, since the size of the reduced system Hilbert space is not $2^2$, but rather $2^{2M}$, where $M$ is the number of atoms. 
For this type of system, it is often convenient to consider the Holstein-Primakoff approximation \cite{holstein1940}, such that 
\bea
J^z&=&S-b^\dagger b\approx S,\cr
J^-&=&\sqrt{2S}b \sqrt{1-\frac{b^\dagger b}{2S}}\approx \sqrt{2S}b^\dagger,\nonumber
\eea 
and similarly $J^+$, where $S$ is the quantum number of the operator $\hat{S}^2$.
Having transformed the original problem of $M$ two-level atoms to that of a single harmonic oscillator, the system becomes exactly tractable even when the interaction presents strong non-Markovian features as discussed in Sec. \ref{brownian}.

Although the arguments used to obtain the effective Hamiltonian (\ref{Dicke}) are based on the Markov approximation, the dynamics given by this Hamiltonian can be solved without invoking such an approximation.  
In fact, for non-Markovian couplings, the collective decay rate can vary significantly with respect to the Markovian case. This was shown by \textcite{vats1998} by considering a collection of $M$ two-level atoms coupled to the radiation field within a photonic crystal according to Eq. (\ref{Dicke}). In this work, the aforementioned Holstein-Primakoff approximation was used, and it was found that the collective decay rate scales as $M^{2/3}$ instead of $M$, as occurs for radiation in the vacuum.

In situations where the quantum fluctuations of system observables are not significant, because, for instance, there is a large number of particles $M\gg 1$ within a space of dimension $D>1$, we may use the mean-field or Hartree approximation \cite{breuerbook}. The original Eqs. (\ref{sistemt}) can be written as
$y(t)=\sum_{j} \langle\sigma^-_{j} (t)\rangle/M$ and
$z(t)=\sum_{j} \langle\sigma^3_{j} (t)\rangle/M$ can be
written as
\beqa
\frac{dy(t)}{dt}&=&M\int_0^t d\tau \alpha(t-\tau) y(\tau)z(t);\nonumber\\
\frac{dz(t)}{dt}&=&-4 M \Re\left[\int_0^t d\tau \alpha(t-\tau)y^* (\tau)y(t)\right].
\label{semic}
\end{eqnarray}
As discussed by \textcite{john1995}, the non-Markovian structure of the equations, together with the limit of a large number of atoms or particles, gives rise to a steady state where $y^{st}\neq 0$, even though $y(0)=0$. An analysis of this phenomenon,  which is very similar to the spontaneous symmetry breaking described in the semiclassical theory of the laser \cite{breuerbook}, beyond the semiclassical approximation is still an open problem. It is hard to tackle with current techniques, because it combines three different conditions that are difficult to deal with even independently: a large number of particles, a highly non-Markovian situation (with long correlation times), and strong coupling (enlarged by the collective effects of a large number of particles $M$).

An alternative to these derivations is to consider the Heisenberg equations for both system and environment operators, and then perform a mean-field approximation. Based on this idea is the cluster expansion method, introduced in \cite{gies2007} and later applied by \textcite{delvalle2011} to deal with the dynamics of quantum dots embedded in microcavities. The Heisenberg equations for system and environment operators, $\langle\sigma_j^+(t)\sigma_j(t)\rangle$ and $\langle a_\lambda^\dagger (t)a_\lambda(t)\rangle$, are found to depend on correlations of the form $\langle \sigma_j^+(t)a_\lambda(t)\rangle$, which in turn depend on higher order correlations like $\langle \sigma_l^+(t)\sigma_j(t) a_\lambda^\dagger (t)a_\lambda(t)\rangle$, and so on. Then, if the system is additionally driven by some classical source (e.g. an incoherent field) that destroys high-order quantum fluctuations, such high-order correlations can be approximated as $\langle \sigma_l^+(t)\sigma_j (t)a_\lambda^\dagger(t) a_\lambda (t)\rangle\approx \langle \sigma^+_l(t)\sigma_j(t)\rangle\langle a_\lambda^\dagger (t)a_\lambda(t)\rangle$, and the whole system of equations is truncated at lower orders. 

\section{Exact cases}
\label{Exact} 
Throughout this review many different approaches have been discussed, most of which give rise to approximated equations or equations that are somehow limited numerically. This section is dedicated to describing two different non-trivial situations in which an exact solution is known. The first is when the full dynamics can be described within one excitation sector, because only one excitation is present in the initial state, and the Hamiltonian of the total system conserves the number of particles. The second case deals with what is known as €œquantum Brownian motion, corresponding to the situation in which the OQS is a harmonic oscillator coupled to an environment of harmonic oscillators. Another exactly solvable case that will not be dealt with here, is when $L\sim H_S$, a case often referred in the literature as the purely dephasing noise \cite{breuerbook}. 

\subsection{Calculations in the one excitation sector}
\label{Exact1} 

Let us consider a single two-level quantum system with frequency $\omega_s$ coupled to a bosonic environment. The wave
function of the total system has the form
$|\Psi(t)\rangle=C_0|0,0\rangle+A(t)|1,\{0\}\rangle+\sum_{\lambda}B_{\lambda}(t)|0,{1}_{\lambda}\rangle$, where $|1,\{0\}\rangle$ describes the excitation in the two-level system and no excitations in the environment, and
$|0,{1}_{\lambda}\rangle$ represents no excitations in the two-level systems and a single excitation in the bosonic mode ${\lambda}$. The time-dependent Schr{\"o}dinger equation projected on the one-excitation sector of the Hilbert space takes the form, $dA(t)/dt=-\sum_{\lambda}g_{\lambda} B_{\lambda}(t)e^{-i \Delta_{\lambda}t}$ and $dB_{\lambda}(t)/dt=g_{\lambda} A(t)e^{i\Delta_{\lambda}t}$, with $\Delta_{\lambda}=\omega_{\lambda}-\omega_s$.

Assuming that $B_\lambda(0)=0$, and inserting the formal solution of the latter equation into the former, we have
\beqa
\frac{dA(t)}{dt}=-\int_0^t d\tau \alpha(t-\tau)A(\tau),\label{oneatomsector3}
\end{eqnarray}
where $\alpha(t)=\sum_{\lambda} g_{\lambda}^2 e^{-i\Delta_{\lambda}t}$ is the correlation function of the environment.
An analytical solution can be obtained using the Laplace transform method $A(t)={\mathcal L}^{-1}(A(s))={\mathcal L}^{-1}K(s)$,
where $K(s)=(\frac{A(0)}{s+\alpha(s)})$. Indeed, according to the residue theorem $\int_{\epsilon-i\infty}^{\epsilon+i\infty} \mathrm{d}s K\left(s\right) \mathrm{e}^{st}+\int_C K\left(s\right)\mathrm{e}^{st}= 2\pi \mathrm{i} \sum_j R_j$, 
where the sum of the two terms on the left-hand side represents a closed contour integral around the poles of the kernel, excluding its branch cuts, and $R_j$ are the residues in such poles. Therefore, the general solution of (\ref{oneatomsector3}) is
\begin{equation}
A(t)=\int_{\epsilon-i\infty}^{\epsilon+i\infty} \mathrm{d}s K\left(s\right) \mathrm{e}^{st}= 2\pi \mathrm{i} \sum_j R_j - \int_C K\left(s\right)\mathrm{e}^{st}. \label{IntegralDecompos2}
\end{equation} 
The last term of (\ref{IntegralDecompos2}) vanishes for Markovian interactions, and gives rise to an initial nonexponential decaying in the non-Markovian case. The first term gives rise to a contribution that is proportional to $e^{x_jt}$, where $x_j$ is a pole of $K(s)$. This pole is in general a complex quantity with a non-zero real part, and therefore gives rise to a decaying of the amplitude $A(t)$ at long times. However, steady state solutions of the type $A_\ss=Ae^{i\omega_0t}$, where now $A$ is an amplitude, may exist, and they correspond to imaginary poles of $K(s)$. 

From the former result, we can write the reduced density operator as \cite{breuerbook,vacchini2010}
\[ \rho_s(t)=\Lambda(t)\rho_s(0)=\left( \begin{array}{cc}
P(t) & C_0^*A(t) \\
C_0A^*(t) & 1-P(t) \end{array} \right)\rho_s(0)\] 
where $P(t)=|A(t)|^2$ and $\Lambda(t)$ represent an exact dynamical map. In terms of this map, a time-convolutionless generator for an equation of the type (\ref{timeconvolutionless}) can be defined by means of 
\bea
{\mathcal L}(t)=\frac{d\Lambda(t)}{dt}\Lambda^{-1}(t),
\eea
as long as $\Lambda(t)$ is always $\neq 0$, \textit{i.e.} as long as the map is invertible. In terms of this generator, a master equation can be written as 
\bea
&&\frac{d\rho_s(t)}{dt}=-i\Delta(t)[\sigma^+ \sigma,\rho_s(t)]+\gamma_1(t)(2\sigma\rho_s (t)\sigma^+\cr
&-&\sigma^+ a\rho_s(t)-\rho_s (t)\sigma^+ \sigma)
\label{mastersingle}
\eea
where 
\bea
\Delta(t)&=&-\Iim[\dot{u}(t)u^{-1}(t)];\cr
\gamma_1(t)&=&-\Rre[\dot{u}(t)u^{-1}(t)];
\label{constantmaster}
\eea
and $u(t)$ satisfy the following equation
\bea
&&\frac{du}{dt}+i\omega_0u(t)+\int_0^t ds \alpha(t-s)u(s)=0,
\label{evolu}
\eea
with initial condition $u(0)=1$. This equation is well-defined only when the generator is well-defined, which means that that for all times we should have $\Lambda(t)\neq 0$. Hence, a condition for the former master equation to make physical sense is that $\Lambda(t)$ shall not change from positive to negative values. 
The master equation (\ref{mastersingle}) has a form similar to Eq. (\ref{master}) obtained within the second-order in perturbation theory. 

As described by \textcite{bellomo2007}, the dynamics of $N-$independent bodies interacting with their own reservoirs (\textit{i.e.} in the limit of independent emitters discussed in Sec. \ref{cooperative}) can be expressed in terms of the dynamics of a single-body. As discussed in this section, such single body dynamics is exactly known in the one excitation sector. 

\subsection{The quantum Brownian motion model}
\label{brownian}
The dynamics of a harmonic oscillator linearly coupled to a thermal environment has been analyzed for many years \cite{feynman1963a,ullersma1966}, and it has also been known for some time that such a system exhibits Brownian motion \cite{ford1965}. Such equivalence between a damped harmonic oscillator and quantum Brownian is very nicely explained in \cite{atomphotoninteractions}. Additionally, exact solutions to the problem have been developed since the eighties with the works by \textcite{risenborough1985,haake1985}.
The system can be described with the Hamiltonian presented in Sec. (\ref{ch1sec3}) but considering that the OQS is a harmonic oscillator with Hamiltonian $H_S=\omega_sa^\dagger a$, which corresponds to choosing $V(q)=\frac{1}{2}m\omega_s^2 q^2$, with $q=\sqrt{\frac{1}{2m\omega_s}}(a+a^\dagger)$ and $p=i\sqrt{\frac{m\omega_s}{2}}(a^\dagger-a)$ being the space and the momentum coordinates of the harmonic oscillator. 
A master equation describing this quantum Brownian motion (QBM) model was derived early on by \textcite{haake1985} and \cite{talkner1986}, and later by \textcite{hu1992} (see also \cite{karrlein1997}, which recovers and discusses both results). In more detail, while the first derivation is based on Wigner functions, the one by \textcite{hu1992} is based on the Feynman-Vernon influence functional theory derived for the QBM in \cite{feynman1963a}. Following a more recent derivation by \textcite{an2007,an2007a,matisse2008,Jinshuang2010} this equation reads as
\bea
&&\frac{d\rho_s(t)}{dt}=-i\Delta(t)[a^\dagger a,\rho_s(t)]+\gamma_1(t)(2a\rho_s (t)a^\dagger\cr
&-&a^\dagger a\rho_s(t)-\rho_s (t)a^\dagger a)+\gamma_2(t)(\pm a\rho_s (t)a^\dagger+a^\dagger\rho_s (t)a\cr
&-&a^\dagger a\rho_s(t)\mp\rho_s (t)a a^\dagger),
\label{masterharmonic}
\nonumber
\eea
where $\Delta(t)$ and $\gamma_1(t)$ are given by Eq. (\ref{constantmaster}), and 
\bea
\gamma_2(t)&=&\dot{v}(t)-2v(t)\Rre[\dot{u}(t)u^{-1}(t)],
\eea
where $u(t)$ follows Eq. (\ref{evolu}), and $v(t)$ is given by
\bea
v(t)=\int_0^t ds \int_0^t ds' u(t-s)\alpha^{+*}(t-s)u^*(t-s'),\nonumber
\eea
with initial condition $u(0)=1$, and $\alpha^+(t)$ given by Eq. (\ref{icc13}). Here, the $\pm$ and $\mp$ in the above master equation correspond respectively to the reservoir being bosonic or fermionic. As shown by \textcite{Jinshuang2010,lei2012}, the coefficients $\gamma_i(t)$ can be determined exactly using nonequilibrium Green's functions, which include non-perturbatively all environment effects. 
The derivation above has recently been extended to describe the evolution of the reduced density matrix departing from an initially correlated state between the system and the environment \cite{huatang2011}. For zero temperature $\gamma_2(t)=0$, and Eq. (\ref{masterharmonic}) is identical to Eq. (\ref{mastersingle}). 
For a system of $N$ harmonic oscillators the above equation can be written as \cite{zhang2012,matisse2008}
\bea
&&\frac{d\rho_s(t)}{dt}=-i[\tilde{H}_S(t),\rho_s(t)]+\sum_{ij}\large\{\gamma^1_{ij}(t)(2a_j\rho_s (t)a_i^\dagger\cr
&-&a_i^\dagger a_j\rho_s(t)-\rho_s (t)a_i^\dagger a_j)+\gamma^2_{ij}(t)(\pm a_j\rho_s (t)a_i^\dagger+a_i^\dagger\rho_s (t)a_j\cr
&-&a_i^\dagger a_j\rho_s\mp\rho_s (t)a_j a_i^\dagger)\large\},
\label{mastermanyharmonic}
\eea
where $\tilde{H}_S(t)=\sum_{ij}\Delta_{ij}a_i^\dagger a_j$, and we have defined the $N\times N$ matrices
\bea
\bm{\Delta}(t)&=&-\Iim[\dot{\bm{u}}(t)\bm{u}^{-1}(t)];\cr
\bm{\gamma}^1(t)&=-&\Rre[\dot{\bm{u}}(t)\bm{u}^{-1}(t)];\cr
\bm{\gamma}^2(t)&=&\bm{\hat{v}}(t,t)-2\Rre[\dot{\bm{u}}(t)\bm{u}^{-1}(t)\bm{v}(t,t)],
\eea
where $\bm{u}(t)$ and $\bm{v}(t)$ are also $N\times N$ matrices with elements $u_{ij}(t)=\langle[a_i(t),a_j^\dagger(0)]_{\pm}\rangle$ and $v_{ij}(t,t)=\langle a_j^\dagger(t)a_i(t)\rangle$, respectively, which are related to the nonequilibrium Green's functions of the system in the Schwinger-Keldysh nonequilibrium theory \cite{zhang2012,schwinger1961}. They obey the equations
\bea
&&\frac{d\bm{u}}{dt}+i\bm{\omega}_s\bm{u}(t)+\int_0^t ds \alpha(t-s)\bm{u}(s)=0;\cr
&&\frac{d\bm{v}(s,t)}{ds}+i\bm{\omega}_s\bm{v}(s,t)+\int_0^t ds' \alpha(s-s')\bm{v}(s',t)\cr
&=&\int_0^t ds'\alpha^{+*}(s-s')\bm{u}^\dagger(s'),
\eea
with conditions $\bm{u}(0)=\unit$ and $\bm{v}(0,t)=0$, with $0\le s\le t$, and $\bm{\omega}_s$ as a $N\times N$ diagonal matrix with the bare single-particle energy levels of the system. The evolution of $\bm{u}(t)$ is very similar in structure as the one in Eq. (\ref{oneatomsector3}). Hence, it can be solved with the Laplace transform method, giving rise to a solution which has a similar structure as Eq. (\ref{IntegralDecompos2}), with a first term that corresponds to the exponential contribution of the residues of the Laplace transform of $\alpha(t-s)$, and a second term corresponding to a nonexponential decaying originated by the contour integral \cite{zhang2012}. 

There is also an exact stochastic Schr\"odinger equation to describe quantum Brownian motion. Indeed, as recently shown in \cite{ferialdi2012} by computing the Green's function associated with Eq. (\ref{linearNM}), the functional derivative of the last term of such an equation can be written exactly. 
Finally, the QBM master equations above discussed correspond to an interaction Hamiltonian of the form (\ref{chapuno7b}), i.e. containing a single bilinear term. Based on the results by \textcite{diosi2014}, \textcite{ferialdi2016a} has provided a generalization of the QBM master equation valid for a more general interaction Hamiltonian of the form (\ref{chapuno12}) that includes several bilinear terms. 

\section{Solving the dynamics of the full system}
\label{unitary}

Most of the approaches described above are based on calculating the reduced dynamics of the system under the assumption that the environment evolves much faster than the OQS itself. Hence, the environmental degrees of freedom are either traced out, like in the master equation approach, or their action is considered statistically through a Monte Carlo-like method, like in the stochastic Schr\"odinger equations or in the path integral methods. 
Naturally, a different approach to dealing with OQSs is to integrate the dynamics of the total system. This can be made following either standard or more elaborated exact diagonalization methods \cite{computationalmany}. 
In this regard, an important aspect to consider when describing the full system dynamics is that in general, the environment oscillators in the Hamiltonian (\ref{chapuno32}) form a quasi-continuum. Hence, the interaction and field parts of such a Hamiltonian can also be written as $\int_{0}^{1}dkg(k)(a(k)L^\dagger+a(k)^{\dagger}L)+\int_{0}^{1}dk\omega(k)a(k)^{\dagger}a(k)$, 
where $g(k)$ are the coupling strengths, and $a(k)$ ($a(k)^\dagger$) are harmonic oscillator operators with commutation relations $[a(k),a(k')^\dagger]=\delta(k-k')$. Here, the index $k$ labels the modes, which have also been re-scaled to the maximum momentum, $k_{\mmax}$, as $k=k/k_{\mmax}$. In the frequency representation, these terms can be rewritten as 
$\int_{0}^{1}d\omega D_{\ddos}(\omega)g(\omega)\,(a(\omega)L^\dagger+a(\omega)^{\dagger}L)+\int_{0}^{1}d\omega D_{\ddos}(\omega)\omega a(\omega)^{\dagger}a(\omega)$, where we have also introduced an effective upper frequency $\omega_{\mmax}$ and rescaled accordingly.  

When the environment is initially in a Gaussian state, its effect on the OQS dynamics is fully described by the spectral density $J(\omega)$.
As a consequence, a variety of Hamiltonians, given by different pairs of $g(k)$ and $\omega(k)$ that lead to the same spectral density, give rise to the same OQS dynamics. This provides us with the freedom of using an arbitrary dispersion relation, which is chosen for simplicity as $\omega(k)=\omega_c k$, with $\omega_c$ as an arbitrary coefficient is taken to be equal to one. In this case, $D_{\ddos}(\omega)=1$, and the resulting Hamiltonian can be rewritten as 
\begin{eqnarray}
H&=&H_S+\int_{0}^{1}d\omega \hat{g}(\omega)\,(a(\omega)L^{\dagger}+a(\omega)^{\dagger}L)\nonumber\\
&+&\int_{0}^{1}d\omega\omega a(\omega)^{\dagger}a(\omega),
\label{Hamil3}
\end{eqnarray}
with $\hat{g}(\omega)=\sqrt{J(\omega)}$. From now on we will just write $\hat{g}(\omega)$ as $g(\omega)$ for simplicity.
One possible approach to solving the dynamics given by Eq. (\ref{Hamil3}) is to build a finite representation of the environment in terms of a smaller set of states. This problem was tackled by \textcite{burkey1984} by choosing the relevant frequencies as those that optimally discretize the integral of the spectral density according to a 
Gaussian quadrature method. According to this, for a given measure $J(\omega)$, a set of orthogonal polynomials $\pi_{n}(\omega)$ exists, such that
\bea
\int_{0}^{1} d\omega\,J(\omega)\pi_{n}(\omega)\pi_{m}(\omega)=\rho_{n}^{2}\delta_{nm},
\label{ortho}
\eea
with $n=0,1,\cdots$, and $\rho_{n}^{2}=||\pi_{n}||^2=\int_0^1 d\omega \pi^2_{n}(\omega)$. 
Then, any integral with the weight $J(\omega)$ can be approximated by $\int_0^1 d\omega f(\omega)J(\omega)\approx \sum_{p=1}^N W_pf(\omega_p)$, where $\omega_p$ are the $N$ roots of the orthogonal polynomial, $\pi_N(\omega)$, and $W_p$ are the corresponding quadrature weights. One way to compute such roots and weights is by taking into account the recurrence relations \cite{golub1969}
\bea
\pi_{n+1}(\omega)=(\omega-\alpha_{n})\pi_n(\omega)-\beta_{n}\pi_{n-1}(\omega), 
\label{recurrence}
\eea
where $\pi_{-1}(\omega)=0$, and $\pi_0(\omega)=1$. Considering now the normalized version of the polynomials, $p_n(\omega)=\pi_n(\omega)/\rho_n$, this recurrence relation reads as $\sqrt{\beta_{n+1}}P_{n+1}(\omega)=(\omega-\alpha_{n})P_n(\omega)-\sqrt{\beta_{n}}P_{n-1}(\omega)$. The matrix representation of this relation can be written in terms of a $N\times N$ symmetric tri-diagonal matrix, $T$, where the diagonal elements are formed by the $\alpha_n$ and the off-diagonal elements are formed by the $\beta_n$. The eigenvalues of such a matrix are precisely the $N$ roots corresponding to the most representative oscillators in the environment $\omega_p$, and the Gaussian weights are $W_p=\rho_0 q_{1,j}^2$, where $q_{1,j}$ is the $j-$th element of the first eigenvector of $T$, and $\rho_0=\int_0^\infty d\omega J(\omega)$. 

As a result of the Gaussian discretization, an optimized discrete version of Eq. (\ref{Hamil3}) is obtained, $H=H_S+\sum_p \omega_pa_p^\dagger a_p+\sum_p \sqrt{W_p}(a_p^\dagger L+L^\dagger a_p)$. The recurrence coefficients (\ref{recurrence}), as well as the Gaussian quadrature parameters $\{\omega_p,W_p\}$, can be obtained numerically with standard algorithms and libraries \cite{gautschi2005}.
Similar proposals are based on using sparse polynomials \cite{alvermann2008}. An interesting variation of the latter idea is that proposed by \textcite{kazansky1997} and recently optimized by \textcite{shenvi2008}, which consists of performing analytical continuation of the integral weighted by the spectral density and then applying complex Gaussian quadrature to generate complex eigen-energies and couplings $\{\omega_p, \sqrt{W_p}\}$. The resulting non-Hermitian Hamiltonian provides a suitable representation of the continuum with an accuracy that nevertheless depends on the choice of the contour form in the complex plane. Due to the complex nature of the system eigenvalues, the OQS decays irreversibly without suffering revivals, as occurs with previous discretization methods leading to Hermitian Hamiltonians. 

Related to the aforementioned problem of discretization are the approaches used to map the original problem to that of a system coupled to a one-dimensional chain, which allows for the use of powerful numerical techniques to analyze the system ground state and dynamics. 

The numerical renormalization group (NRG) approach \cite{vojta2005,bulla2008,anders2007}, for instance, was initially derived in \cite{wilson1975} to analyze the problem of a quantum impurity coupled to a reservoir of noninteracting electrons (the Kondo problem). This method is based on first performing a coarse-grained approximation of the continuous environment spectral density in energy space (as described above), which leads to a discrete environment that can be mapped onto a semi-infinite tight-binding chain \cite{krishna1980} by using a Lanczos tri-diagonalization method. Hence, the initial problem characterized by a Hamiltonian of the form (\ref{chapuno32}), $H_{tot}=H_S +H_B +\sum_\lambda g_\lambda (L a_\lambda^{\dagger}+L^{\dagger}a_\lambda )$ is mapped onto a tight-binding linear chain
\begin{eqnarray}
&&H_{tot}=H_S +g(b_0L^\dagger+Lb_0^\dagger)\\
&+&\sum_{n=0,\cdots,M}\big[A_{n}b_n^\dagger b_n+B_{1,n+1}(b^\dagger_{n+1}b_n+b_n^\dagger b_{n+1})\big]\nonumber
\label{map}
\end{eqnarray}
which depends on new transformed modes $b_n$, and certain coefficients $A_n$ and $B_n$. 
An important aspect of the NRG approach is that the coarse graining of the continuum is made through a logarithmic discretization, such that the couplings in the resulting chain decay exponentially, and hence the chain can be truncated in a systematic way after performing an iterative diagonalization \cite{bulla2008}. Interestingly, the states obtained by iterative diagonalization can be expressed as matrix product states \cite{weichselbaum2009}.

Another renormalization group approach is the surrogate Hamiltonian method, which also consists in mapping the real Hamiltonian (which have an infinite number of environment degrees of freedom), into a simpler one that reproduces exactly the dynamics for finite times. The idea is that the evolution time induces a dynamical renormalization over the system-bath interaction, \textit{i.e.} the system interacts progressively with the environment degrees of freedom rather than with all of them at once. The surrogate Hamiltonian method was first introduced to study processes in surface science \cite{baer1997,koch2003,asplund2011} and solutions \cite{koch2002,gelman2004}. Interestingly, as discussed by \textcite{gualdi2013}, the required size of the surrogate system Hilbert space can be determined a priori by considering a Lieb-Robinson bound argument.

\begin{figure}[htbp]
\centerline{\includegraphics[width=0.45\textwidth]{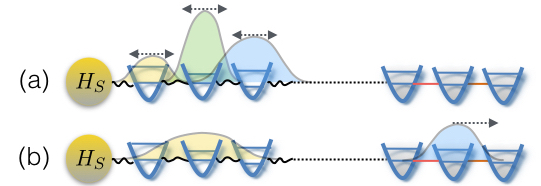}}
\caption{Partially based on \textcite{chinbook2011}. (a) When the system injects excitations into the inhomogeneous part of the chain, some backscattering occurs, reflecting the non-Markovian effects that mainly occur at initial stages of the evolution. (b) At long times the excitations penetrate into the homogeneous region and propagate away from the system irreversibly. \label{figchain}}
\end{figure}

An alternative proposal is the one by \textcite{prior2010,chinbook2011}, which considers the orthogonal polynomials $\pi_n(\omega)$ in Eq.  (\ref{ortho}) to define a unitary transformation of the environment into a new set of oscillators $a_n$, such that $b_{\omega}=\sum_n U_{n}(\omega)a_n$,
where $U_{n}(\omega)=g(\omega)\pi_{n}(\omega)/||\pi_{n}||$. Thanks to the orthogonal property (\ref{ortho}) and the normalization of the polynomials, the transformation is unitary $\int d\omega U^*_{n}(\omega)U_{m}(\omega)=\delta_{nm}$. 
The unitary transformed Hamiltonian derived from Eq. (\ref{Hamil3}) can be written as Eq. (\ref{map}), with $A_{n}=\omega_c\alpha_{n}$, and $B_{n}=\omega_c\sqrt{\beta_{n}}$, such that $\alpha_{n}$ and $\beta_{n}$ are precisely the coefficients of the recurrence relation (\ref{recurrence}). Such coefficients depend effectively on the particular set of monic polynomials that are orthogonal with respect to the weight function according to the relation (\ref{ortho}). As shown in Fig. \ref{figchain} for most spectral densities, $\alpha_{n}$ and $\beta_{n}$ are relatively small and highly dependent on $n$ for the first few sites of the chain, giving rise to an eventual backscattering of the excitation to the system, while they become large and homogeneous for higher values of $n$, leading for an irreversible loss of the excitation. 

Written as Eq. (\ref{map}), the whole system constitutes a one-dimensional structure with only nearest-neighbor interactions. This Hamiltonian can be solved with powerful numerical techniques such as matrix product states \cite{white1992,white1998,vidal2003,scholl2011}. In addition, the system is now directly coupled to the most relevant (transformed) oscillator of the environment, and since the consecutive chain oscillators only become relevant at increasingly longer times, a systematic truncation is possible. 
However, both NRG and DMRG approaches can also be used to solve the system's dynamics in its original original star configuration of Eq.  (\ref{Hamil3}).  This configuration has been proven by \textcite{wolf2014b} to be more convenient for tackling the Anderson impurity model. Indeed, in contrast to what had been commonly believed, the star configuration can become much less entangled during the dynamics than the chain representation, which favors its numerical implementation.

Similar in spirit is the effective modes approach by \textcite{hughes2009}, also based on the construction of a hierarchy of coupled effective environmental modes that is terminated by coupling the final member of the hierarchy to a Markovian bath. Closely related to this is the derivation in \cite{iles-smith2014}, where the OQS is enlarged by including the first mode of the chain, such mode being in turn coupled to a new bath conformed by the remainder oscillators. The enlarged system Hamiltonian is thus  $\tilde{H}_S=H_S +g(a_0L^\dagger+La_0^\dagger)$, and a perturbative master equation can then be derived for its reduced density operator $\rho_{\tilde S}$. To test the accuracy of this approach, \textcite{iles-smith2014} consider a Drude model for the spectral density of the original bath, and compare the results of the perturbative master equation for $\rho_{\tilde S}$ to those obtained by the HEOM described in Sec. \ref{hierarchy}, obtaining an almost perfect fit between both results. The idea of enlarging the system with the first few oscillators of the chain is further discussed by \textcite{woods2014}, who derive the general expression of the \textit{residual} spectral density $J_m(\omega)$ describing the reminder environment when an increasing number of chain oscillators $m$ are included in the system. 

Recently, a multiple-chain-bath model has been derived by \textcite{huh2014}, to transform the noninteracting star-bath into a set of weakly coupled multiple parallel chains. The transformation is based on a partitioning strategy of the bath modes that leads to the multiple parallel chains in such a way that as the number of chains is increased, the coupling strengths between the OQS and the first (primary) mode of each chain is reduced, and the lengths of each chain is shortened.  Finally, a general analysis on how different connectivities in a chain (and more generally in a network) of environment oscillators may give rise to different shapes of the corresponding spectral density can be found in  \cite{nokkala2015}.

\section{Perspectives}

The theory of OQSs presents many challenges to be further developed in the future, which we summarize in the following. 

Recent advances in experimental techniques and in the fabrication of novel materials allow us to access regimes where non-Markovian effects become crucial, leading to new arenas for further exploration. Strongly correlated systems, for instance, constitute a broad class of electronic materials that display unusual and intriguing properties. Their intrinsic interest, as well as their many applications, have led to the development of powerful numerical tools to analyze them, such as the density matrix renormalization group (DMRG), time-dependent DMRG, or matrix product states \cite{white1998,calzadilla2002,scholl2005,perez2007,scholl2011}, and of advanced experimental techniques to measure, control and observe their properties. Extending the theory of OQS to describe the interaction of impurities with such strongly correlated environments, is an interesting research topic.

In addition, the possibility of using non-Markovianity as a resource to build the desired quantum state may open new avenues for developing further the concept of dissipative quantum computation and state preparation proposed in \cite{verstraete2008,diehl2008,diehl2010} and experimentally realized in \cite{subasi2012}. 

\begin{figure}[htbp]
\centerline{\includegraphics[width=0.45\textwidth]{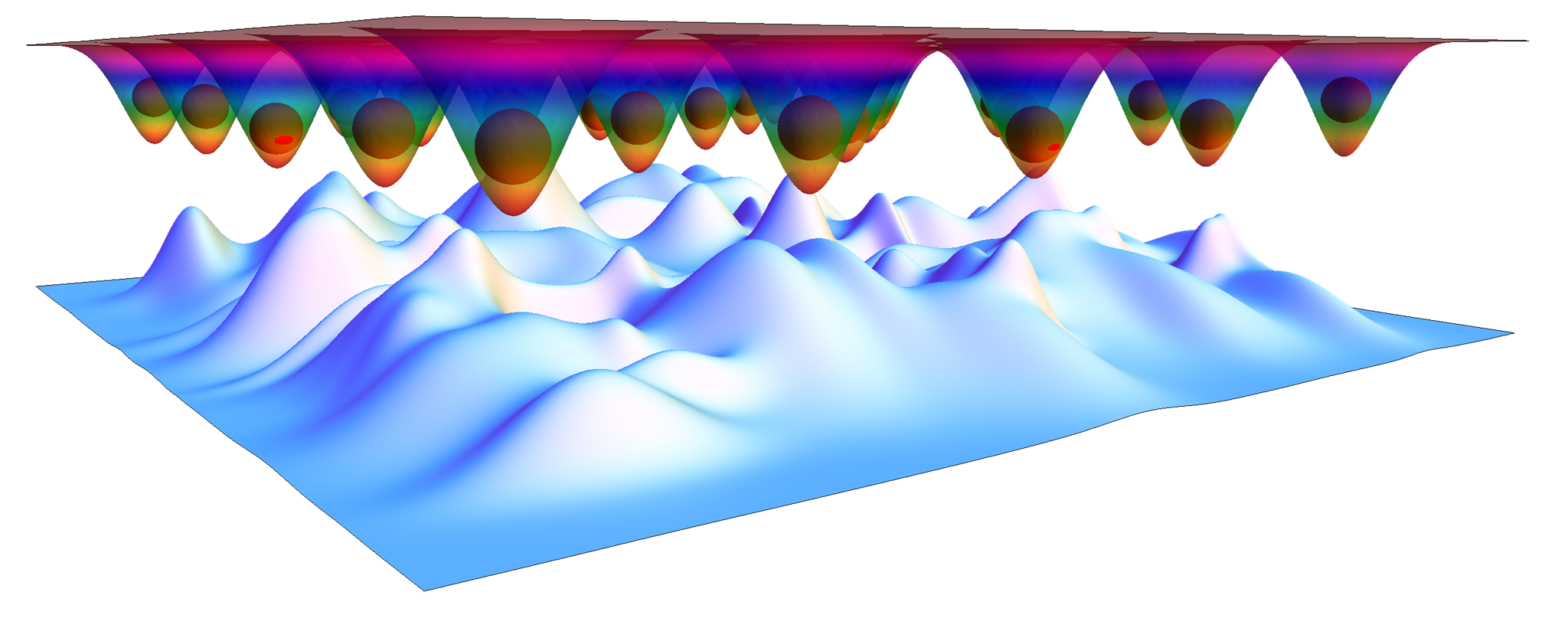}}
\caption{An idealized view of a many-body open quantum system consisting of a set of confined ions or atoms interacting with a thermal field. \label{fig8}}
\end{figure}

Another almost unexplored topic is to understand the role of non-Markovianity in the relaxation and thermalization in few or many-body quantum systems (see Fig. \ref{fig8}). Moreover, as briefly discussed in Sec. \ref{BBcorrelationsNM} in some realistic scenarios the open system is coupled to more than one environment. In this regard, for small systems interacting with several thermal baths it is possible to give a full characterization of their dynamics as quantum thermal machines within the Markov approximation \cite{kosloff2013}. A thermal machine obtained by coupling alternatively a confined ion to hot and cold reservoirs has been experimentally realized in \cite{rossnagel2015}. However, the study of the quantum thermodynamics when non-Markovian effects are present raises fundamental questions that are still subject of debate (for recent progress see \cite{bylicka2015thermodynamic}).

An interesting possibility is that of characterizing the relatively unknown environment properties of quantum artificial devices such as nano-resonators or quantum circuits and junctions, by measuring OQS dynamics. This idea has been developed in the pioneering experiment by \textcite{groblacher2015} that monitors the motion of an opto-mechanical resonator to determine the spectral density of its condensed-matter heat bath. Furthermore, this type of analysis may be useful to characterize the environment of certain biological systems and molecular ensembles like those found in photosynthetic complexes. Indeed, the ability to understand energy transport in the presence of a well-characterized environment and beyond the Markov limit may lead to important insights in the analysis of photosynthetic systems, which may also pave the way to the design of artificial light-harvesting devices \cite{schorles2011}. 

From a methodological point of view, deriving a computable form for the coefficients of the formally exact master equation (\ref{nmbreuer}) beyond the one excitation sector and the Brownian particle case discussed in Sec. \ref{brownian} is an interesting research goal. Another challenging task is extending the hierarchical equations of motion of Sec. \ref{hierarchy} to arbitrary spectral densities and beyond the low temperature regime. As we have discussed, some advances for tackling finite temperatures have been made by increasing the dimensionality of the hierarchy as in \cite{ishizaki2005,xu2005,han2006}, or by using an hybrid method \cite{moix2013ahybrid}. In addition, a very interesting research problem would be to import some of the numerical advances and achievements in Monte-Carlo methods (for instance the taming of the dynamical sign problem as reported in \cite{cohen2015}) to improve the sampling of stochastic Schr\"odinger and SLN equations. 

From a more fundamental perspective, a full mathematical characterization of non-Markovian quantum dynamical maps is still an open problem, although promising advances in this direction has been recently made by \textcite{ferialdi2016a,ferialdi2016b}. Similarly, it would be interesting to analyze whether there is a connection between the non-Markovianity measures in Sec. \ref{nonmarkovianity} and the computational complexity for solving the OQS dynamics. In addition, despite the advances reviewed in Sec. \ref{initialstate}, it is still a challenge to understand further the relationship between the structure of the initial system-bath states (which may include system-environment correlations) and the nature of the resulting dynamics, which includes addressing the question of whether or not such dynamics are completely positive and can be described by a map. 


\acknowledgements
We thank A. Bermudez, L. Di\'osi, G. Giedke, J. Stockburger, A. Rivas, G. Nikolopoulos, C. Navarrete, J. Piilo, R. Lo Franco, C. Koch, H. Lo\"ic, K. Le Hur, and J. Liu for their constructive criticisms and comments in reading preliminary versions of this work. IDV would like to express her gratitude to U. Schollw\"ock for encouragement and support. Our enlightening discussions and collaborations with M.C. Ba\~{n}uls, H.P-Breuer, S. Brouard, J.I. Cirac, H. Carmichael, L.A. Correa,  A. del Campo, L. Di\'osi, J. Eisert, M. Esposito, L. Ferialdi, P. Gaspard, G. Giedke, S. Kholer, R. Kosloff, F. Nori, J.G. Muga, C. Navarrete, A. Ruiz, J.P. Palao,  L. Pollet, D. Porras, A. Rivas, J. Halimeh, G. Hegerfeldt, S.F. Huelga, U. Schollw\"ock, N.O. Linden, L.S. Schulman, D. Sokolovski, W.T. Strunz, M. Tanaka, Y. Tanimura, A.A. Valido, and A. Wolf, have played a key role in shaping our understanding of various aspects of the dynamics of quantum systems. 
IDV acknowledges funding from Nanosystems Initiative Munich (NIM) (project No. 862050-2), and we also acknowledge funding from the Spanish MINECO through project FIS2013-41352-P and EU through COST Action MP1209.

%
%
\bibliographystyle{apsrmp4-1}
\bibliography{Bibtexelesdrop_d1}



\end{document}